\documentclass[a4paper,11pt]{article}	% Choose this for a physics paper
\pdfoutput=1 % if your are submitting a pdflatex to arXiv
\usepackage{jheppub} % If you decide to use the JHEP style
\newcommand{\blue}[1]{\textcolor{blue}{#1}}
\newcommand{\dualCox}{h_G^{\vee}}

\def\SL{SL(2,\IZ)}
\def\GN{\Gamma_1(N)}

\usepackage[table]{xcolor}                  % assign color to table cells via \cellcolor{}; allow alternate coloring
\usepackage{colortbl}

\usepackage{dynkin-diagrams}
\usepackage{tabu}

\newcommand{\cyan}[1]{\textcolor{cyan}{#1}}

\usepackage{booktabs}                       % better tables: \toprule,\midrule,bottomerule
\usepackage{multirow,bigdelim}              % things span over rows of a table: \rdelm, \multirow
\usepackage{longtable}                      % enable longtable environment
\usepackage{arydshln}                       % enable dashed rules in tables. It has to be loaded after longtable,array,colortbl!

\usepackage{units}                          % \unit[val]{dim}, \unitfrac[val]{num}{denom}, \nicefrac[fontcmd]{num}{denom}
\usepackage{amsmath,amssymb,mathtools}      % to enable some mathematical symbols	
\allowdisplaybreaks[4]
\usepackage{slashed}                        % for p slash
\usepackage{amsthm}
\usepackage{bbm,dsfont}                     % provide \mathbbm{1}(serif) and \mathds{1}(san serif) respectively

\usepackage{bm}                             % provide \bm{} for bold greek letters and other math symbols
\usepackage{upgreek}                        % upright normal greek letters \upalpha\Upalpha
\usepackage[vcentermath]{youngtab}          % a second solution to Young diagrams as well as Young tableaux
\usepackage{ytableau}                       % the best solution for Young tableaux
\usepackage{empheq}                         % put boxes around align and other math environments
\usepackage{suffix}	                        % used to define Meijer's G funciton
\usepackage{lscape}	                        % enable the ``landscape'' environment

\usepackage{graphicx}                       % to enable the inserting of graphics
\usepackage{subfig}	                        % for \subfloat commands
\usepackage{float}
\usepackage[export]{adjustbox}              % align figures vertically with \includegraphics[valign=c]{}

\usepackage{url}                            % for url reference
\usepackage{makeidx}                       % macro package for index making
\usepackage[numbers,sort&compress]{natbib}  % bibliography
\usepackage[colorlinks=true,urlcolor=blue,anchorcolor=blue,citecolor=blue,filecolor=blue,linkcolor=blue,menucolor=blue,pagecolor=blue,linktocpage=true,pdfproducer=medialab,pdfa=true]{hyperref}	
\usepackage{ifthen}	                        % enables
\usepackage{tikz}                           % the almighty PGF/TikZ! Useful TikZ packages: positioning,scopes,chains,matrix,arrows,calc,trees,intersections.
\usetikzlibrary{calc,positioning,intersections,arrows,shapes,fit,tikzmark}
\tikzstyle{roundbox}=[rectangle, draw=black, rounded corners, text
centered, text width=3.4cm, rectangle split part align={center}]
\tikzstyle{arrow}=[->,>=stealth]
\tikzstyle{split3box} = [rectangle,rounded
corners,draw=black,rectangle split,rectangle split parts=3]
\def\e{\epsilon}

\def\md{\mathbf}
\def\mf{\mathfrak}

\newcommand{\red}[1]{\textcolor{red}{#1}}

\def\IC{\mathbb{C}}
\def\IE{\mathbb{E}}

\def\IP{\mathbb{P}}

\def\IR{\mathbb{R}}
\def\IZ{\mathbb{Z}}

\def\cO{\mathcal{O}}

\def\fn{\mathfrak{n}}

\newcommand{\cirG}{\mathring{G}}
\newcommand{\cirF}{\mathring{F}}
\newcommand{\cirDelta}{\mathring{\Delta}}
\newcommand{\cirR}{\mathring{R}}
\newcommand{\lF}{\lambda_{\cirF}}
\newcommand{\lG}{\lambda_{\cirG}}
\newcommand{\mF}{m_{\cirF}}
\newcommand{\mG}{m_{\cirG}}
\newcommand{\hg}{h_G^\vee}
\newcommand{\hgc}{h_{\cirG}^\vee}

\newcommand{\Ind}{\mathrm{Ind}}

\DeclarePairedDelimiterX\opm[3]{\langle}{\rangle}{#1 \delimsize\vert #2 \delimsize\vert #3}
\DeclarePairedDelimiterX\ip[2]{\langle}{\rangle}{#1 \delimsize\vert #2}
\def\({\left(}
\def\){\right)}
\def\[{\left[}
\def\]{\right]}

\newcommand{\bfi}{\begin{figure}[h]}
\newcommand{\efi}{\end{figure}}
\newcommand{\be}{\begin{equation}}
\newcommand{\ee}{\end{equation}}
\newcommand{\ba}{\begin{aligned}}
\newcommand{\ea}{\end{aligned}}
\newcommand{\ben}{\begin{eqnarray}\displaystyle}
\newcommand{\een}{\end{eqnarray}}

\newcommand{\ri}{{\mathsf{i}}}
\newcommand{\rd}{{\rm d}}

\newcommand{\nn}{\nonumber \\}

\newcommand{\wb}[1]{\overline{#1}}

\newcommand{\eq}{\epsilon_1}
\newcommand{\et}{\epsilon_2}

\newcommand{\und}[1]{\underline{#1}}
\newcommand{\w}{\omega}

\newcommand{\so}{\mf{so}}
\newcommand{\su}{\mf{su}}
\newcommand{\SO}{\mf{so}}
\newcommand{\SU}{\mf{su}}
\newcommand{\Sp}{\mf{sp}}

\theoremstyle{definition}

\title{Twisted Elliptic Genera}
\author[a]{Kimyeong Lee,}
\author[b]{Kaiwen Sun,}
\author[b]{Xin Wang}

\affiliation[a]{School of Physics, Korea Institute for Advanced Study,\\Hoegiro 85, Seoul 02455, Korea}
\affiliation[b]{Quantum Universe Center, Korea Institute for Advanced Study,\\Hoegiro 85, Seoul 02455, Korea}
\emailAdd{klee@kias.re.kr, ksun@kias.re.kr, wxin@kias.re.kr}
\preprint{KIAS-Q22012}
\abstract{We study the twisted elliptic genera of 2d $(0,4)$ SCFTs associated with the BPS strings in the twisted circle compactification of 6d rank-one $(1,0)$ SCFTs. Such objects can arise when the 6d gauge algebra allows outer automorphism, thus are classified by twisted affine Lie algebras. We study several fascinating aspects of the twisted elliptic genera including 2d localization, twisted elliptic blowup equations, Higgsing and spectral flow symmetry. We derive a recursion formula with respect to the number of strings to exactly compute the twisted elliptic genera. We also investigate the modular bootstrap of twisted one-string elliptic genera and find the modularity of congruence subgroups $\Gamma_1(N)$  naturally appears with possible $N=2,3,4$. Geometrically, our study solves the refined BPS partition functions of the underlying genus-one fibered Calabi-Yau threefolds with $N$-section.}
\begin{document}

\maketitle
\section{Introduction}
The classification of superconformal field theories (SCFTs) and supersymmetric gauge theories in various dimensions has been a major subject in the study of supersymmetry in the recent decades. Given Nahm's classification on superconformal algebras \cite{Nahm:1977tg}, the highest dimension in which a nontrivial SCFT can exist is six. It is now well-known that the 6d $(2,0)$ SCFTs admit an ADE classification \cite{Witten:1995zh, Strominger:1995ac}, while the 6d $(1,0)$ SCFTs admit an ``atomic classification" with certain generalized quiver structures \cite{Heckman:2015bfa}. See an excellent review \cite{Heckman:2018jxk}. More recently, much progress has been made in the understanding of the compactification of 6d SCFTs to lower dimensions. In particular, the compactification to 5d Kaluza-Klein (KK) theories is mostly understood and results in a conjectural classification on 5d SCFTs by decoupling 5d hypermultiplets from the KK theories \cite{Jefferson:2017ahm,Jefferson:2018irk,Bhardwaj:2018yhy,Bhardwaj:2018vuu,Bhardwaj:2019fzv}. 

Besides the straightforward circle compactification, it was observed in \cite{Bhardwaj:2019fzv} that when a 6d SCFT has a discrete global symmetry, one can do {\it twisted circle compactification} to a new 5d KK theory. This process brings in a large class of highly nontrivial 5d KK theories, many of which have fascinating 5d Lagrangian descriptions \cite{Bhardwaj:2020gyu}. There exist two possibilities for the 6d discrete global symmetry: the first kind comes when the gauge algebra allows outer automorphism, in which case the twisted circle compactification means to fold the 6d vector multiplets; the second kind comes when the quiver structure has a discrete symmetry, in which case one folds the 6d tensor multiplets. In the current work, we are interested in the first kind of twisted circle compactification of 6d $(1,0)$ SCFTs as it is classified by {\it twisted affine Lie algebras} and related to the modularity of congruence subgroups $\Gamma_1(N)$. The second kind of twisted circle compactification for 6d $(2,0)$ SCFTs has been studied in \cite{Duan:2021ges}.

The 6d $(1,0)$ SCFTs can be geometrically engineered by F-theory compactified on non-compact elliptic Calabi-Yau threefolds. More precisely, such Calabi-Yau geometry $M$ is an elliptic fibration over a non-compact base surface $B$ in which all curves are simultaneously shrinkable to zero volume. The $\dim H^{1,1}(B,\IZ)$ gives the dimension of the tensor branch, i.e., the rank of the 6d $(1,0)$ SCFT. We are particularly interested in the rank one case, where the non-compact base $B$ is just $\mathcal{O}_{\mathbb{P}^1}(-\fn)$ for $\fn=1,2,3,\dots,8,12$. The geometric engineering suggests that when a 6d $(1,0)$ SCFT is put on the 6d Nekrasov Omega background $\IC^2_{\eq,\et}\times T^2$, the partition function should be equal to the refined topological string partition function on Calabi-Yau $M$. On the other hand,  6d $(1,0)$ SCFTs contain the BPS strings or the so-called self-dual strings with worldsheet theory as 2d $(0,4)$ SCFTs. From the viewpoint of worldsheet, the elliptic genera of the 2d theories should also be equal to the 6d partition function upon suitable expansion by the number of strings. Along with the circle compactification to 5d KK theory, these yield the well established relation chain
\be\label{eq:chain1}
\IE^{\textrm{2d } (0,4) \textrm{ SCFT}}=Z_{\IR^4\times T^2}^{\textrm{6d } (1,0) \textrm{ SCFT}}=Z_{\IR^4\times S^1}^{\textrm{5d KK}}=Z^{\textrm{ref. top.}}_{\textrm{non-compact elliptic CY3}}.
\ee

In the case of twisted circle compactification, the non-compact elliptic Calabi-Yau threefolds are generalized to non-compact genus-one fibered Calabi-Yau threefolds, with additional rational sections or just $N$-sections. The genus-one fibered Calabi-Yau geometries and their role in F-theory compactification have drawn lots of attention in recent years \cite{Morrison:2014era,Braun:2014oya,Anderson:2018heq,Cota:2019cjx,Anderson:2019kmx,Oehlmann:2019ohh,Kimura:2019bzv,Knapp:2021vkm,Schimannek:2021pau,Dierigl:2022zll}.  In our situation, the number of sections is naturally reflected by the type of twists:
\be
N=\left\{
\begin{array}{cl}
   2,\phantom{=}  & A_{2r-1}^{(2)},D_r^{(2)},E_6^{(2)}, \\
   3,\phantom{=} & D_4^{(3)}, \\
   4,\phantom{=}  & A_{2r}^{(2)}.
\end{array}
\right.
\ee
The geometric engineering of these special genus-one fibered Calabi-Yau threefolds produces the twisted compactification of 6d $(1,0)$ SCFTs. The BPS strings of the 6d SCFTs on the torus $T^2=S^1\times S^1$ also get twisted along one $S^1$ in the sixth direction such that the elliptic genera now have fractional $q^{1/N}$ orders.  Although both the 6d SCFTs and the 2d worldsheet theories are in twisted circle compactification, one can still look from 6d and 2d to keep the modularity manifest.  Notably, the modularity is no longer the $\SL$ of ordinary elliptic genera, but the congruence subgroups $\Gamma_1(N)$ of $\SL$. In summary, we expect the following twisted version of the relation chain \eqref{eq:chain1} as
\be
{\IE^{\textrm{2d } (0,4) \textrm{ SCFT}}_{\rm twisted}}=Z_{\IR^4\times S^1\times S^1, \textrm{twisted }}^{\textrm{6d } (1,0) \textrm{ SCFT}}=Z_{\IR^4\times S^1}^{\textrm{5d KK}}=Z^{\textrm{ref. top.}}_{\textrm{non-compact genus-one fibered CY3}}.
\ee
We are interested in the \emph{twisted elliptic genera} arising here. It should be noted that this term actually has already been used in a different context for  2d $\mathcal{N}=(2,2)$ SCFTs such as the twisted elliptic genus of K3 surface, see e.g. \cite{Kawai:2009ci,Eguchi:2010fg}. The twist in our term comes from the same twist as in twisted affine Lie algebras and is specifically for 2d $\mathcal{N}=(0,4)$ SCFTs, thus should not be confused. For the second kind of twist, one can also study the associated twisted elliptic genera, which for the 6d $(2,0)$ cases have been studied in \cite{Duan:2021ges}. However, those are not related to twisted affine Lie algebras, thus are not the concern of the current paper. 

Some typical examples of twisted circle compactification include the $\IZ_2$ twist of 6d $(1,0)$ pure $\su(3)$ SCFT which compactifies to the 5d $\SU(3)$ gauge theory with Chern-Simons level 9, and the $\IZ_3$ twist of 6d $(1,0)$ pure $\so(8)$ SCFT which compactifies to 5d $\SU(4)$ gauge theory with Chern-Simons level 8. The 5d KK theories are not necessarily gauge theories. For example, the $\IZ_2$ twist of 6d $(1,0)$ pure $\so(8)$ SCFT is known to be a 5d non-Lagrangian theory. We will study the twisted elliptic genera for all these examples in this paper.

In the following, we summarize several approaches to the twisted elliptic genera.
\begin{itemize}
    \item Localization of 2d $(0,4)$ SCFTs. Elliptic genera of some special 2d theories can be exactly computed  by 2d localization, i.e., Jeffrey-Kirwan residue, analogous to the ADHM construction in 0d \cite{Benini:2013nda,Benini:2013xpa}. This approach can be generalized to some special $\IZ_2$ twisted theories \cite{Kim:2021cua}. In general, this method when applicable is the most efficient way to compute (twisted) elliptic genera to arbitrary number of strings. Two known examples are the $\fn=3,$ $\su(3)^{(2)}$ theory and the $\fn=4,$ $\so({2r+8})^{(2)}$ theories which we will review in Section \ref{sec:2dquiver}.
    \item $\GN$ modular ansatz. The modular ansatz method exploits the modularity of elliptic genera \cite{DelZotto:2016pvm,Gu:2017ccq,DelZotto:2017mee,DelZotto:2018tcj,Kim:2018gak}, inspired by the early work on compact elliptic Calabi-Yau threefolds \cite{Huang:2015sta}. In particular, the $SL(2,\mathbb{Z})$ modular ansatz for the one-string elliptic genera of rank-one 6d $(1,0)$ SCFTs has been extensively studied in \cite{DelZotto:2018tcj}.  On the other hand, the modular ansatz of congruence subgroups $\Gamma_0(N)$ and $\Gamma_1(N)$ has been investigated in  \cite{Cota:2019cjx,Knapp:2021vkm,Schimannek:2021pau} for topological strings on genus-one fibered Calabi-Yau threefolds with $N$-section. Recently, the genus-one fibered Calabi-Yau threefolds associated with twisted affine Lie algebras were discussed in \cite{paultalk}. We combine these results together to propose a suitable $\Gamma_1(N)$ modular ansatz for the one-string twisted elliptic genera and successfully fix the ansatz for almost all rank-one twisted theories. Interestingly, we find that for almost all rank-one twisted theories, the modular group is enhanced from $\Gamma_1(N)$ to $\Gamma_0(N)$.
\item Brane webs and topological vertex. Many 6d $(1,0)$ SCFTs allow brane-web interpretations. When the brane-web construction exists, one can use the (refined) topological vertex to compute the 6d partition function. This method can be cleverly extended to some $\IZ_2$ twisted theories by including O5-planes \cite{Hayashi:2020hhb,Kim:2021cua,Kim:2022dbr}. For example, the 5d $\SU(3)_9$ partition function can be computed by an extension of topological vertex called O-vertex
\cite{Hayashi:2020hhb,Nawata:2021dlk}.
\item Twisting from Higgsing. It has been noticed that some 6d twisted theories can be obtained by the Higgsing from 6d untwisted theories  \cite{Kim:2019dqn,Hayashi:2021pcj}. This generalizes the known Higgsing trees for 6d $(1,0)$ SCFTs, see e.g. \cite{DelZotto:2018tcj}. Most known examples of twisting from Higgsing come from $\so({2r})^{(2)}$ type of theories with $\fn=2,3,4$ \cite{Kim:2019dqn,Hayashi:2021pcj}. We will systematically study the   twisting from Higgsing phenomenon for all rank-one theories, including the Higgsing from both untwisted theories and twisted ones. We will also propose a simple algebraic method to determine the precise Higgsing conditions based on representation decompositions.

\item Twisted elliptic blowup equations and the recursion formula. This will be the main computational method in the current paper. In a series of works \cite{Gu:2018gmy,Gu:2019dan,Gu:2019pqj,Gu:2020fem}, the elliptic blowup equations have been established for the elliptic genera of all 6d $(1,0)$ SCFTs, which generalized Nakajima-Yoshioka's blowup equations in 4d and 5d \cite{Nakajima:2003pg,Nakajima:2005fg,Gottsche:2006bm,Nakajima:2009qjc,Keller:2012da,Kim:2019uqw}. Now we further generalize these very efficient and universal functional equations to all 6d rank-one twisted theories, both unity and vanishing blowup equations. We focus on the elliptic form, rather than the 5d form or geometric form of blowup equations, such that the twisted elliptic genera can be directly computed. Some unity blowup equations for some simple twisted theories have been studied in \cite{Kim:2020hhh}.
\end{itemize}

This paper is organized as follows. In Section \ref{sec:gen}, we give an overview on some salient features of 6d $(1,0)$ SCFTs and their twisted circle compactification. We will define twisted elliptic genera of the BPS strings and describe a fundamental quantity -- the modular index. In Section \ref{sec:blowup}, we generalize the elliptic blowup equations in previous works to the twisted cases and derive a recursive formula to compute the twisted elliptic genera efficiently. In Section \ref{sec:spectralflow}, we focus on twisted one-string elliptic genera and study their universal features. In particular, we generalize the spectral flow symmetry of the one-string elliptic genera of 2d $(0,4)$ theories discovered by Del Zotto and Lockhart \cite{DelZotto:2016pvm,DelZotto:2018tcj} to the twisted cases.  In Section \ref{sec:MA}, we exploit the modularity of twisted elliptic genera, and propose the proper modular ansatz with modular group $\Gamma_1(N)$ with $N=2,3,4$. In Section \ref{sec:higgsing}, we discuss the Higgsing relations among untwisted/twisted 6d $(1,0)$ SCFTs. In Section \ref{sec:outlook}, we summarize our results and point out some further directions.

Our conventions for simple Lie algebras and affine Lie algebras are the same with \cite{Gu:2020fem}, see Appendix A therein. We often use $\su,\so$ and $\mf{sp}$ for classical gauge algebras where $\mf{sp}(r) $ is equivalent to $C_r$. We will also use $A,B,C,D$ for more algebraic contexts. We frequently use $q=e^{2\pi i \tau}$, $v=e^{2\pi i \epsilon_+}$ and $x=e^{2\pi i \epsilon_-} $ as multiplicative variables following \cite{DelZotto:2018tcj}, where $\epsilon_\pm=(\epsilon_1\pm \epsilon_2)/2$.

\section{General theory}\label{sec:gen}
\subsection{Review of rank one 6d $(1,0)$ SCFTs}
In this section, we briefly review some known results on the rank-one 6d $(1,0)$ SCFTs established in \cite{Morrison:2012np,Heckman:2013pva,Heckman:2015bfa}. More extensive reviews can be found in e.g. \cite{Heckman:2018jxk,DelZotto:2018tcj,Gu:2020fem}. 

The rank-one 6d $(1,0)$ SCFTs on the tensor branch are the natural elliptic lift of 4d $\mathcal{N}=2$ and 5d $\mathcal{N}=1$ gauge theories with 8 supercharges. They are geometrically engineered by compactifying F-theory on certain elliptic non-compact Calabi-Yau threefolds \cite{Morrison:1996na,Morrison:1996pp}. To be precise, these Calabi-Yau threefolds can be realized as elliptic fibration over some non-compact base surfaces $\mathcal{O}(-{\fn})\to \IP^1$. The self-intersection number $-\fn$ of the base curve $\IP^1$ can only take values from 1 to 8 and 12. This $\fn$ is also called the tensor coefficient. The Kodaira and Tate singularity type of the elliptic fibration gives the 6d gauge algebra ${G}$ which is supported on the $(-\fn)$ base curve. The 6d global flavor symmetry ${F}$ on the other hand is supported on a non-compact curve intersecting with the $(-\fn)$-curve. The 6d matters, i.e., hypermultiplets in representation ${{R}}$ sit at the intersection point between curves \cite{Katz:1996xe}. The ${(\fn,G,{R})}$ together are highly constrained by the {Calabi-Yau condition}. All possibilities have been classified in \cite{Morrison:2012np,Heckman:2013pva,Heckman:2015bfa}. For example, there exist in total six pure gauge theories shortly denoted by $\fn_G$ as $3_{\su(3)},4_{\so(8)},5_{F_4},6_{E_6},8_{E_7}$ and $12_{E_{8}}$. Together with a $(7,{E_7},\frac12\bf56)$ theory, these are the rank-one non-Higgsable clusters which serve as the ``atoms" to be linked together to build higher rank 6d $(1,0)$ SCFTs such as conformal matter theories \cite{DelZotto:2014hpa}. Notably, the $4_{\so(8)}$ theory can be extended to an infinite series of 6d $(1,0)$ SCFTs with $\fn=4$ which are the $\so(r+8)+r\bf V$ theories. The full list of rank-one 6d $(1,0)$ SCFTs with ${(\fn,G,F,{R})}$ can be found in e.g. \cite[Tables 20, 21]{DelZotto:2018tcj}.  We collect those whose gauge algebra allows an outer automorphism i.e. type $A,D$ and $E_6$ in Table \ref{tb:rk1-1}, which will be our starting theories to perform twisted circle compactification.

Interestingly, it was noticed in \cite{Gu:2020fem} that all rank-one 6d $(1,0)$ SCFTs with tensor coefficient $\fn$,  gauge algebra $G$ and $N_i$ number of matters in representation $R_i^G$ satisfy the following constraint
\be\label{eq:c}
c= \frac{\dualCox}{6}-\sum_i\frac{N_i}{12}\mathrm{Ind}(R_i^G)=\frac{\fn-2}{2}.
\ee
Here $\mathrm{Ind}$ is the quadratic index of a representation. 
The identity shows that the combination in the left hand side depends only on the self-intersection number $-\fn$ of the base curve. This value will be called  {\it the $c$ constant} in the current paper. The physical meaning of $-c$ is the  left-moving Casimir energy in the Ramond sector of one BPS string. It also has a geometric meaning in elliptic non-compact Calabi-Yau threefolds and can be computed from the intersection theory \cite{Cota:2019cjx}. Later we will also propose an interesting generalization of the $c$ constant for the twisted theories.

\begin{table}[h]
\centering
%\resizebox{\linewidth}{!}{
\begin{tabular}{|c|l|l|c|}\hline
  $\fn$
  &$G$
  &$F$
  & $(R_G,R_F)$
  \\ \hline
  6&$E_6$&$-$&$-$
  \\\hline
  \rowcolor{gray!10}
  5&$E_6$&$\mf{u}(1)_6$
  & $\md{27}_{-1}\oplus c.c.$
  \\\hline
  4&$\mf{so}(8)$&$-$&$-$
  \\
  4&$\mf{so}(N\geq 9)$&$\mf{sp}(N-8)_1$
  & $(\md{N},\md{2(N-8)})$
  \\
  4 &$E_6$&$\mf{su}(2)_{6}\times \mf{u}(1)_{12}$
  & $({\bf 27},\wb{\md 2})_{-1}\oplus c.c.$
  \\\hline
  3&$\mf{su}(3)$&$-$& $-$
  \\
  3&$\mf{so}(8)$&$\mf{sp}(1)_1^a\times \mf{sp}(1)_1^b\times \mf{sp}(1)_1^c$
  & $(\md 8_v\oplus\md 8_c\oplus\md 8_s,\md 2)$
  \\
   \rowcolor{gray!10}
  3&$\mf{so}(10)$&$\mf{sp}(3)_1^a\times (\mf{su}(1)_4\times \mf{u}(1)_4)^b$
  & $(\md{10},\md 6^a)\oplus[(\md{16}_s)^b_1\oplus c.c.]$
  \\
  \rowcolor{gray!10}
  3&$\mf{so}(12)$&$\mf{sp}(5)_1$
  & $(\md{12},\md{10})\oplus(\md{32}_s,\md{1})$
  \\
   \rowcolor{gray!10}
  3 &$E_6$&$\mf{su}(3)_{6}\times \mf{u}(1)_{18}$
  & $({\bf 27},\wb{\md 3})_{-1}\oplus c.c.$
  \\
  \hline
   2&$\mf{su}(1)$&$\mf{su}(2)_1$&$-$
    \\
    2&$\mf{su}(2)$&$\mf{so}(7)_1\times\text{Ising}$
    & $(\md 2,\md 8_s\times \md 1_s)$
    \\
    2&$\mf{su}(N\geq 3)$&$\mf{su}(2N)_1$
    & $(\md N,\wb{2\md N})\oplus c.c.$
    \\
    2&$\mf{so}(8)$&$\mf{sp}(2)_1^a\times \mf{sp}(2)_1^b\times \mf{sp}(2)_1^c$
    & $(\md 8_v,\md 4^a)\oplus (\md 8_s,\md 4^b)\oplus (\md
      8_c,\md 4^c)$
    \\
    2&$\mf{so}(10)$&$\mf{sp}(4)_1^a\times (\mf{su}(2)_4\times \mf{u}(1)_{8})^b$
    & $(\md{10},\md 8^a)\oplus[(\md{16}_s,\md 2^b)_1\oplus c.c.]$
    \\
    \rowcolor{gray!10}
    2&$\mf{so}(12)_a$&$\mf{sp}(6)_1^a\times \mf{so}(2)_8 $
    & $(\md{12},\md{12}^a)\oplus(\md{32}_s,\md 2^b)$
    \\
    2&$\mf{so}(12)_b$&$\mf{sp}(6)_1^a\times \text{Ising}^b\times \text{Ising}^c$
    & $(\md{12},\md{12}^a)\oplus(\md{32}_s,\md
      1^b_s)\oplus(\md{32}_c,\md 1_s^c)$
    \\
    2 &$E_6$&$\mf{su}(4)_{6}\times \mf{u}(1)_{24}$
    & $(\md{27},\wb{\md 4})_{-1}\oplus c.c.$
    \\\hline
    1&$\mf{sp}(0)$&$(E_8)_1$
    & $-$
    \\
    1&$\mf{su}(2)$&$\mf{so}(20)_1$
    & $(\md{2},\md{20})$
    \\
    1&$\mf{su}(3)$&$\mf{su}(12)_1$
    & $(\md 3,\wb{\md{12}})_1\oplus c.c.$
    \\
     \rowcolor{gray!10}
    1&$\mf{su}(4)$&$\mf{su}(12)_1^a\times \mf{su}(2)_1^b$
    & $[(\md 4,\wb{\md{12}}_1^a)\oplus c.c.]\oplus (\md 6,\md
      2^b)$
    \\
     \rowcolor{gray!10}
    1 &$\mf{su}(N\geq 5)$&$\mf{su}(N\!+\!8)_1\!\times\! \mf{u}(1)_{2N(N-1)(N+8)}$
    &$[(\md{N},\wb{\md{N+8}})_{-N+4}\oplus(\md{\Lambda}^2,\md{1})_{N+8}]\oplus
      c.c.$
    \\
     \rowcolor{gray!10}
    1&$\mf{su}(6)_*$&$\mf{su}(15)_1$
    & $[(\md{6},\overline{\md{15}})\oplus c.c.]\oplus
      (\md{20},\md{1})$
    \\
    1&$\mf{so}(8)$&$\mf{sp}(3)_1^a\times \mf{sp}(3)_1^b\times \mf{sp}(3)_1^c$
    & $(\md{8}_v,\md{6}^a)\oplus(\md{8}_s,\md{6}^b)\oplus(\md{8}_c,\md{6}^c)$
    \\
     \rowcolor{gray!10}
    1&$\mf{so}(10)$&$\mf{sp}(5)_1^a\times (\mf{su}(3)_4\times \mf{u}(1)_{12})^b$
    & $(\md{10},\md{10}^a)\oplus[(\md{16}_s,\md{3}^b)_1\oplus c.c.]$
    \\
    \rowcolor{gray!10}
    1&$\mf{so}(12)_a$&$\mf{sp}(7)_1^a\times \mf{so}(3)_8^b$
    & $(\md{12},\md{14}^a)\oplus(\md{32}_s,\md{3}^b)$
    \\
    \rowcolor{gray!10}
    1&$\mf{so}(12)_b$&$\mf{sp}(7)_1^a\times ?^b\times ?^c$
    & $(\md{12},\md{14}^a)\oplus(\md{32}_s,\md{2}^b)\oplus(\md{32}_c,\md{1}^c)$
    \\
    \rowcolor{gray!10}
    1 &$E_6$&$\mf{su}(5)_{6}\times \mf{u}(1)_{30}$
    & $(\md{27},\overline{\md{5}})_{-1}\oplus c.c.$
    \\\hline
\end{tabular}
\caption{Gauge, flavor symmetries and matter contents of all rank-one 6d SCFTs \cite{DelZotto:2018tcj} with gauge algebra allowing an outer automorphism. The subscript in a flavor symmetry indicates the level $k_F$ of the associated current algebra $F$. Matters are presented as the gauge and  flavor representations of the half-hypermultiplets. The theories in gray have unpaired matter content upon $\IZ_2$ twist.}
\label{tb:rk1-1}
\end{table}

The 6d $(1,0)$ SCFTs contain tensionless strings which we call BPS strings resembling the features of instantons in 5d $\mathcal{N}=1$ and 4d $\mathcal{N}=2$ gauge theories. Simple examples include E-strings which are realized by M2-branes stretched between a M5-brane and a M9-brane in the Horava-Witten picture of M-theory on $S^1/\IZ_2$. The worldsheet theories on $k$ E-strings can be realized as a series of 2d $(0,4)$ $O(k)$ quiver gauge theories \cite{Kim:2014dza}. The E-string theory is the simplest 6d $(1,0)$ SCFTs, which has no gauge symmetry but a $E_8$ flavor symmetry. It can be geometrically engineered by compactifying F-theory on local half-K3 Calabi-Yau threefolds. The E-string theory can be extended to two infinite series of 6d $(1,0)$ SCFTs with $\fn=1$ which are $\Sp(N)+(2N+8)\bf F$ and $\su(N)+(N+8)\mathbf{F}+\bf\Lambda^2$ theories.  
Another simple example of BPS strings is the M-strings which are realized by M2-branes stretched between two M5-branes \cite{Haghighat:2013gba}. In this case, the 6d supersymmetry is enhanced to $(2,0)$ and the worldsheet theories become $(0,8)$. The worldsheet theories of $k$ M-strings can also be realized as a series of 2d quiver gauge theories \cite{Haghighat:2013gba}. The M-string theory can be extended to an infinite series of 6d $(1,0)$ SCFTs with $\fn=2$ which are $\su(N)+2N\bf F$ theories. In general, the BPS strings in a 6d $(1,0)$ SCFT have worldsheet theories as rather nontrivial 2d $(0,4)$ SCFTs. Only in some special cases, these exist known 2d quiver gauge theory constructions. These include the four infinite series of 6d $(1,0)$ SCFTs with $\fn=1,2,4$ mentioned earlier.

The 6d $(1,0)$ SCFTs usually contain three types of supermultiplets: the tensor, vector and hyper multiplets. As we are only interested in rank-one 6d $(1,0)$ SCFTs in the current work, there is always only one tensor multiplet with tensor coefficients $\fn$. For all rank-one 6d $(1,0)$ SCFTs except E-string theory, there is a nontrivial gauge algebra $G$ and the vector multiplets are in the adjoint representation $\bf Adj$ of $G$. When there is nontrivial matter content, the hypermultiplets are in the representation $R$ of $G$. If we take into consideration the flavor symmetry $F$, the matter representations are often denoted as $(R^G,R^F)$ of the half-hypermultiplets.  The partition function of 6d $(1,0)$ SCFTs contains three parts: the classical part, the one-loop part and the elliptic genera of BPS strings. The first two parts can be directly written down once the data $(n,G,R)$ are known, while the elliptic genera are much more nontrivial and difficult to compute.

One main task in the study of 6d SCFTs is to compute the elliptic genera of the 2d $(0,4)$ worldsheet theories of the BPS strings. In the past decade, a huge amount of effort has been made and plenty of methods have been developed, see the summaries in e.g. \cite{Gu:2019dan,Gu:2020fem}. Nevertheless, there still remain quite some rank-one 6d SCFTs whose elliptic genera we could not compute to the extent we would like.  The existing major approaches include 2d localization, modular bootstrap, elliptic blowup equations, topological vertex and brane webs and so on. Each has their merits and limitations. For example, 2d localization is the most efficient method for arbitrary number of strings yet only applicable to those cases with known 2d quiver gauge theory descriptions.  Elliptic blowup equations imply recursion formulas for elliptic genera or BPS invariants yet only applicable to the cases without half-hyper. 
All these approaches have counterparts for twisted elliptic genera, which we will discuss in detail later.

The circle reduction of 6d $(1,0)$ SCFTs produces the 5d $\mathcal{N}=1$ gauge theories with the same gauge group and matter content. For 6d $(1,0)$ SCFTs with $\fn\ge 3$, by sending $\tau\to i\infty$, i.e., $q\to 0$, the $k$ string elliptic genera of 6d SCFTs exactly reduce to the $k$ instanton Nekrasov partition function of the 5d gauge theories. 
For the $\fn=1,2$ cases, there are some subtitles for the circle reduction such that the limit of elliptic genera and 5d Nekrasov partition function can have some small differences, which we refer to the discussion in \cite[Section 7]{DelZotto:2018tcj}. The circle reduction should not be confused with the circle compactification where all KK modes are kept. The circle compactification leads to 5d KK theories which are not necessarily gauge theories,  while circle reduction leads to 5d gauge theories as deformation of 5d SCFTs.

\subsection{Twisted circle compactification}
\begin{figure}[h]
\centering
\begin{tikzpicture}
\centering
\node (b) at (3,1.5){};
\draw (0,2) node{\underline{$A_{2}^{(1)},\mathcal{O}_2$}:};
 \dynkin[edge length=.75cm, involution/.style={red,stealth-stealth},
involutions={12},extended,at=(b)]{A}{2}; 
\draw (8,2) node{\underline{$A_{2}^{(2)}$}:};
\node (a) at (9+1.8,2){};
\dynkin[edge length=.75cm,involutions={},extended,at=(a)]A[2]2;
\node (b) at (1.1,0){};
\draw (0,0) node{\underline{$A_{2l}^{(1)},\mathcal{O}_2$}:};
 \dynkin[edge length=.75cm, involution/.style={red,stealth-stealth},
involutions={16;25;34},extended,at=(b)]{A}{*.****.*}; 
\draw (8,0) node{\underline{$A_{2l}^{(2)}$}:};
\node (a) at (9,0){};
\dynkin[edge length=.75cm,involutions={},extended,at=(a)]A[2]{even};
\node (b) at (1.4,-2){};
\draw (8,-2) node{\underline{$A_{2l-1}^{(2)}$}:};
\draw (0,-2) node{\underline{$A_{2l-1}^{(1)},\mathcal{O}_2$}:};
 \dynkin[edge length=.75cm, involution/.style={red,stealth-stealth},
involutions={15;24},extended,at=(b)]{A}{*.***.*};
\draw (8,-4) node{\underline{$D_{l+1}^{(2)}$}:};
\node (a) at (9.2,-2){};
\dynkin[edge length=.75cm,involutions={},extended,at=(a)]A[2]{odd};
\node (b) at (2.5,-4){};
\draw (0,-4) node{\underline{$D_{l+1}^{(1)},\mathcal{O}_2$}:};
 \dynkin[edge length=.75cm, involution/.style={red,stealth-stealth},
involutions={54},extended,at=(b)]{D}{**.***};
\node (a) at (9+0.3,-4){};
\dynkin[edge length=.75cm,involutions={},extended,at=(a)]D[2]{**.***};
\draw (8,-6) node{\underline{$D_{4}^{(3)}$}:};
\node (b) at (3.0,-6){};
\draw (0,-6) node{\underline{$D_{4}^{(1)},\mathcal{O}_3$}:};
 \dynkin[edge length=.75cm, involution/.style={red,stealth-stealth},
involutions={13;43;14},extended,at=(b)]{D}{4};
\node (a) at (9+1.5,-6){};
\dynkin[edge length=.75cm,involutions={},extended,at=(a)]D[3]4;
\draw (8,-8.5) node{\underline{$E_{6}^{(2)}$}:};
\node (b) at (2.,-8.5){};
\draw (0,-8.5) node{\underline{$E_{6}^{(1)},\mathcal{O}_2$}:};
 \dynkin[edge length=.75cm, involution/.style={red,stealth-stealth},
involutions={16;35},extended,at=(b)]{E}{6};
\node (a) at (9+1,-8.5){};
\dynkin[edge length=.75cm,involutions={},extended,at=(a)]E[2]6;
\end{tikzpicture}
\caption{Dynkin diagrams}\label{fig:dynkins}
\end{figure}
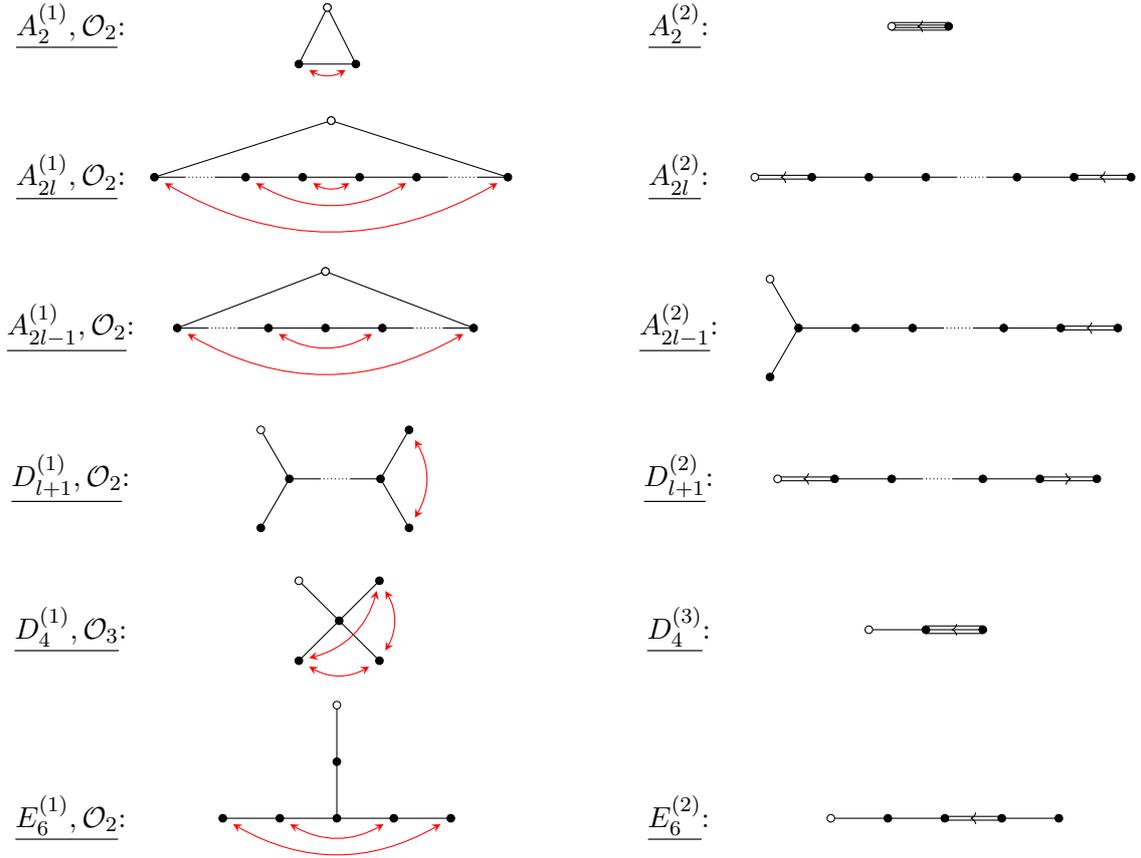

It was conjectured in \cite{Jefferson:2018irk} that all 5d SCFTs can be obtained by RG flows from 5d Kaluza-Klein (KK) theories, while the KK theories can be obtained from the compactifications of 6d SCFTs. When a 6d SCFT is compactified on a circle, one can turn on non-trivial discrete holonomies for the gauge fields from the discrete global symmetries, which are realized in \cite{Bhardwaj:2019fzv} as combinations of the outer automorphisms of gauge algebras $G$ and permutations of tensor multiplets. Such a circle compactification is usually called the twisted circle compactification \cite{Bhardwaj:2019fzv}.  In this paper, we are interested in the twisted circle compactification of rank-one 6d $(1,0)$ SCFTs, with a single tensor multiplet which is invariants under any kind of discrete global symmetries, so the outer automorphism of the gauge algebra $G$ determines the possible types of twist. In fact, such automorphisms have been classified in the study of twisted affine Lie algebras \cite{kac1990infinite}. Denote $\mathcal{O}^{(r_{\mathrm{out}})}$ as the order $r_{\mathrm{out}}$ outer automorphism. Then $G=A_{2r-1},D_r,E_6$ admit order two outer automorphisms while $G=D_4$ admits an order-three outer automorphism which act on the simple roots in patterns as described in Figure \ref{fig:dynkins}. 

Upon twisted circle compactification, the gauge algebra of affine Lie type $G^{(1)}$ in the tensor branch effectively becomes the twisted affine Lie algebra $G^{(r_{\mathrm{out}})}$. The vector multiplets are in the representation described by the roots of the twisted affine algebra, which are described by the representations of the finite dimensional subalgebra $\cirG$ obtained from outer automorphism folding of $G$ with KK charge $\tau$. The Dynkin diagram of $\cirG$ is obtained by simply removing the affine node of the Dynkin diagram of $G^{(r_{\mathrm{out}})}$. Denote $n_G$ as the rank of the automorphism of $G$, for $G=A_{2r-1},D_{r},E_6$, the inner automorphism is trivial, such that $n_G$ is equal to the rank of outer automorphism. The roots of $G^{(n_G)}$ are
\begin{align}
    \Delta=\Delta^{\mathrm{im}}\cup \Delta^{\mathrm{re}},
\end{align}
where 
\begin{align}
    \Delta^{\mathrm{im}}&=\left\{\frac{k}{n_G}\tau|k\in\mathbb{Z},k\neq 0\right\},\\
    \Delta^{\mathrm{re}}&=\left\{\alpha+\frac{k}{n_G}\tau|k\in\mathbb{Z},\alpha\in \cirDelta_s\right\}\cup \left\{\alpha+{k}\tau|k\in\mathbb{Z},\alpha\in \cirDelta_l\right\},
\end{align}
where $\cirDelta_s$ and $\cirDelta_l$ are the sets of short roots and long roots of $\cirG$. When $G=A_{2l}$, there is a nontrivial inner automorphism such that the whole rank of the automorphism is $n_G=4$, the roots of $A_{2r}^{(2)}$ are
\begin{align}
    \Delta=\Delta^{\mathrm{im}}\cup \Delta^{\mathrm{re}},
\end{align}
where 
\begin{align}\label{A2r_root}
    \Delta^{\mathrm{im}}&=\left\{\frac{k}{2}\tau|k\in\mathbb{Z},k\neq 0\right\},\\
    \Delta^{\mathrm{re}}&=\left\{\frac{1}{2}\alpha+\frac{1}{4}(2k-1)\tau|k\in\mathbb{Z},\alpha\in \cirDelta_l\right\}\cup \left\{\alpha+\frac{1}{2}{k}\tau|k\in\mathbb{Z},\alpha\in \cirDelta_s\right\}\nonumber\\
    &\quad\quad\quad\quad\quad\quad\quad\quad\quad\quad\quad\quad\quad\quad\quad\quad\ \,\cup \left\{\alpha+{k}\tau|k\in\mathbb{Z},\alpha\in \cirDelta_l\right\},
\end{align}
where $\cirDelta_s$ and $\cirDelta_l$ are the sets of short roots and long roots of $C_r$. 
We summarize all KK-momentum shifts of adjoint representations under twist in Table \ref{tb:adj}.
\begin{table}[h]
\def\arraystretch{1.1}
	\centering
	\begin{tabular}{|c|c|c|c|rcl|} \hline
		$ G $ & $ \cirG $ & $r_{\mathrm{out}}$ & $n_G$ & $ R $ & $ \to $ & $ \mathring{R} $  \\ \hline
		$ A_{2r}^{(2)} $ & $ C_r $ & $2$ & $ 4$ & $ \mathbf{Adj} $ & $ \to $ & $ \mathbf{Adj}_0 \oplus \mathbf{F}_{1/4} \oplus \mathbf{\Lambda}^2_{1/2} \oplus \mathbf{1}_{1/2} \oplus \mathbf{F}_{3/4}  $ \\ 
\hline
		$ A_{2r-1}^{(2)} $ & $ C_r $ & $2$ &$2$ &  $ \mathbf{Adj} $ & $ \to $ & $ \mathbf{Adj}_0 \oplus \mathbf{\Lambda}^2_{1/2} $ \\
\hline
		$ D_{r+1}^{(2)} $ & $ B_r $ & $2$ &$2$ & $ \mathbf{Adj} $ & $ \to $ & $ \mathbf{Adj}_0 \oplus \mathbf{F}_{1/2} $ \\
\hline
		$ E_6^{(2)} $ & $ F_4 $ & $2$ & $2$ & $ \mathbf{Adj} $ & $ \to $ & $ \mathbf{Adj}_0 \oplus \mathbf{F}_{1/2} $ \\
\hline
		$ D_4^{(3)} $ & $ G_2 $ & $3$ & $3$ & $ \mathbf{Adj} $ & $ \to $ & $ \mathbf{Adj}_0 \oplus \mathbf{F}_{1/3} \oplus \mathbf{F}_{2/3} $ \\
	\hline
	\end{tabular}
\caption{Data of twisted affine Lie algebras and the KK-momentum shifts of the adjoint representations.}
\label{tb:adj}
\end{table}

We collect all 6d $(1,0)$ rank-one twisted theories with {\it{paired matter}} content in Table \ref{tb:main0a}. Here by paired we mean that the twisted matter content is invariant upon KK-charge shift by $1/r_{\mathrm{out}}$. For example, consider the $\fn=4,$ $E_6+2\bf F$ theory upon $\IZ_2$ twist, twisted matter content is $\mathbf{F}_0\oplus\mathbf{F}_{1/2}\oplus\mathbf{1}_0\oplus\mathbf{1}_{1/2}$ for $F_4$, which apparently is invariant under KK-charge shift by $1/2$. However, for $\fn =5,$ $E_6+\bf F$ theory upon $\IZ_2$ twist, obviously one cannot make such an arrangement for the twisted matter content, thus we say unpaired. For the cases with unpaired matter content, we refer to the discussion in \cite{Bhardwaj:2020kim}. Those cases normally result in 5d KK theories with half-hypermultiplets.

\begin{table}[h]
\centering
%\resizebox{\linewidth}{!}{
\begin{tabular}{|c|l|l|l|l|c|}\hline
  $\fn$
  &$G$ & $\mathring{G}$
  &$\mathring{R}$ & $\mathring{F}$ & $c$
  \\ \hline
  6&$E_6^{(2)}$& $F_4$ & $-$ & $-$ & 5/4
   \\
\hline
  4&$D_4^{(3)}$& $G_2$ &  $-$ & $-$ & 5/9
 \\
  4&$D_{r+4}^{(2)}$& $\!\!B_{r+3}\!\!$  &$2r( \mathbf{V}_0\oplus\mathbf{1}_{1/2})$ & $\mf{sp}(2r)$ & 3/4
 \\
  4 &$E_6^{(2)}$& $F_4$ &  $\mathbf{F}_0\oplus\mathbf{F}_{1/2}\oplus\mathbf{1}_0\oplus\mathbf{1}_{1/2}$ & $\mf{sp}(1)$ & 1
   \\
\hline
  3&$A_2^{(2)}$  & $C_1$ &$-$ & $-$ & 5/16
   \\
  3&$D_4^{(2)}$& $B_3$ &   $\mathbf{V}_0\oplus\mathbf{1}_{1/2}\oplus\mathbf{S}_0\oplus\mathbf{S}_{1/2}$ & $\mf{sp}(1)\times\mf{sp}(1)$ & 1/2
\\
  3& $D_4^{(3)}$& $G_2$ &$\mathbf{F}_0\oplus\mathbf{F}_{1/3}\oplus\mathbf{F}_{2/3}\oplus\mathbf{1}_0\oplus\mathbf{1}_{1/3}\oplus\mathbf{1}_{2/3} $ & $\mf{sp}(1)$ & 1/2
  \\
 \hline
          2& $A_{2r}^{(2)}$&$C_r$ & $(2r+1)(\mathbf{F}_{0}\oplus \mathbf{F}_{1/2}\oplus \mathbf{1}_{1/4}\oplus \mathbf{1}_{3/4})$ & $\mf{so}(4r+2)$ & $3/16$
     \\
    2&$A_{2r-1}^{(2)}$&$C_r$ & $2r(\mathbf{F}_{0}\oplus \mathbf{F}_{1/2})$ & $\mf{so}(4r)$ & 1/4
    \\
    2&$D_4^{(2)}$& $B_3$ &   $2(\mathbf{V}_0\oplus\mathbf{1}_{1/2}\oplus\mathbf{S}_0\oplus\mathbf{S}_{1/2})$ & $\mf{sp}(2)\times\mf{sp}(2)$ & 1/4
  \\
    2& $D_4^{(3)}$& $G_2$ &$2(\mathbf{F}_0\oplus\mathbf{F}_{1/3}\oplus\mathbf{F}_{2/3}\oplus\mathbf{1}_0\oplus\mathbf{1}_{1/3}\oplus\mathbf{1}_{2/3}) $ & $\mf{sp}(2)$ & 4/9
   \\
    2&$D_5^{(2)}$ & $B_4$ & $4(\mathbf{V}_0\oplus\mathbf{1}_{1/2})\oplus \mathbf{S}_0\oplus \mathbf{S}_{1/2}$ & $\mf{sp}(4)\times \mf{sp}(1)$ & 1/4
    \\
       2&$D_6^{(2)}$& $B_5$   & $6( \mathbf{V}_0\oplus\mathbf{1}_{1/2})\oplus{ \frac12\mathbf{S}_0\oplus \frac12\mathbf{S}_{1/2}}$ & $\mf{sp}(6)$ & 1/4 \\
    2 &$E_6^{(2)}$& $F_4$ &  $2(\mathbf{F}_0\oplus\mathbf{F}_{1/2}\oplus\mathbf{1}_0\oplus\mathbf{1}_{1/2})$ & $\mf{sp}(2)$ & $3/4$
    \\
   \hline
        1& $A_2^{(2)}$ & $C_1$ & $6(\mathbf{F}_{0}\oplus \mathbf{F}_{1/2}\oplus \mathbf{1}_{1/4}\oplus \mathbf{1}_{3/4})$ & $\mf{so}(12)$  & 1/16 \\
    1&$D_4^{(2)}$& $B_3$ &   $3(\mathbf{V}_0\oplus\mathbf{1}_{1/2}\oplus\mathbf{S}_0\oplus\mathbf{S}_{1/2})$ & $\mf{sp}(3)\times\mf{sp}(3)$ & 0
 \\
    1&$D_4^{(3)}$& $G_2$ &$3(\mathbf{F}_0\oplus\mathbf{F}_{1/3}\oplus\mathbf{F}_{2/3}\oplus\mathbf{1}_0\oplus\mathbf{1}_{1/3}\oplus\mathbf{1}_{2/3}) $ & $\mf{sp}(3)$& 7/18 \\ 
    \hline
\end{tabular}
%}
\caption{Gauge algebra, matter representation and effective flavor symmetry for all twisted 6d $(1,0)$ rank-one theories with paired matter contents. Here $\mathring{R}=(\mathring{R}_{\cirG},\mathring{R}_{\cirF})$ and we only write down $\mathring{R}_{\cirG}$ for short. The important value $c$ will be discussed later. 
  }
\label{tb:main0a}
\end{table}

For 6d SCFTs with hypermultiplets in the representations of the flavor algebra $F$, the representations have integer KK charges for a circle compactification without twist. For twisted circle compactifications, the hypermultiplets are in the representations of the sub flavor algebra $\cirF$, with fractional KK charges. We refer to the Reference A of \cite{Kim:2020hhh} for a detailed discussion. The twisted matter representations for all twisted 6d $(1,0)$ rank-one theories with paired matter contents are summarized in Table \ref{tb:main0a}. For a twisted theory with gauge group $G$ and non-trivial matter representations, they always take the form
\begin{align}
   \bigoplus_i N_i(\cirR_{i,0}\oplus \cirR_{i,1/n^G_i}\oplus\cdots\oplus \cirR_{i,(n^G_i-1)/n^G_i}),
\end{align}
where $n^G_i$ is either the rank of outer automorphism $r_{\mathrm{out}}$ or $1$. We denote a representation with zero KK charge $R_{i,0}$ as $R_i$, and $N_i$ is the number of matters in the representation $R_i$.

\subsection{Twisted prepotential and one-loop free energy}
The prepotential, which is also the classical part of the genus zero free energy of the corresponding topological strings, contains the tree-level and one-loop level contributions
\begin{align}\label{eq:F00}
\mathcal{F}^{\mathrm{cls}}_{(0,0)}=\mathcal{F}^{\text{tree}}+\mathcal{F}^{\text{1-loop}}.
\end{align}
The tree level contribution contains the information of tensor multiplet, which should not be changed for the twisted circle compactification of rank-one theories. Under $\mathcal{O}^{(r_{\mathrm{out}})}$ action, we expect that
\begin{align}
    \mathcal{F}^{\text{tree}}&=\frac{t_{\text{ell}}}{2\fn}\bigg(-\fn \mathcal{O}^{(r_{\mathrm{out}})}(m_G\cdot m_G)+ \mathcal{O}^{(r_{\mathrm{out}})}\sum_i(k_{F_i}m_{F_i}\cdot m_{F_i})\bigg)-\frac{1}{2\fn}t_{\text{ell}}^2\tau\nonumber\\ 
    &=\frac{t_{\text{ell}}}{2\fn}\bigg(-\fn m_{\cirG}\cdot  m_{\cirG}+\sum_i k_{\cirF_i}  m_{\cirF_i}\cdot  m_{\cirF_i}\bigg)-\frac{1}{2\fn}t_{\text{ell}}^2\tau,
\end{align}
where the $\mathcal{O}^{(r_{\mathrm{out}})}$ acts on the bilinear form $m\cdot m$ by identifying the masses in the orbit of $\mathcal{O}^{(r_{\mathrm{out}})}$, such that it becomes the bilinear form of the algebra $\cirG/\cirF$. The $\mathcal{O}^{(r_{\mathrm{out}})}$ acts on $k_{F_i}$ by multiplying it to $k_{\cirF_i}=n^G_ik_{F_i}$, which can be understood as follows. For the case without twist, $k_F$ is the intersection number of the non-compact curve and the base curve \cite{Gu:2020fem}. When we do the outer automorphism twist, the non-compact curves that along the obit of the action $\mathcal{O}^{(r_{\mathrm{out}})}$ should be identified, which naturally changes the intersection number from $k_{F_i}$ to $k_{\cirF_i}=n^G_ik_{F_i}$.

As suggested in \cite{Kim:2020hhh}, the one-loop prepotential can be evaluated from the zeta regularization of the summation over infinite roots or weights of the twisted affine algebra. For ${G}^{(r_{\mathrm{out}})}$  not of type $A_{2r}^{(2)}$, we have $n_G=r_{\mathrm{out}}$, and the one loop prepotential can be formally written as 
\begin{align}
\mathcal{F}^{1\text{-loop}}=&-\frac{1}{12}\sum_{k\in \mathbb{Z}}\sum_{\alpha\in \cirDelta_l}|\alpha\cdot m_{\cirG}+k\tau|^3-\frac{1}{12}\sum_{k\in \mathbb{Z}}\sum_{\alpha\in \cirDelta_s}|\alpha\cdot m_{\cirG}+k\tau/{n_G}|^3\nonumber\\
&+\frac{1}{12}\sum_{k\in\mathbb{Z}}\sum_{i,f_i}\sum_{w_{\cirG}\in \mathring{R}_{i}^{\cirG}}|w_{\cirG}\cdot m_{\cirG}+k\tau/n^G_i+m_{f_i}|^3,
\end{align}
where $\mathring{R}_{i}^{\cirG}$ are the representations of the matter contents with KK charge zero. Using the zeta regularization\footnote{The formal definition of zeta function $\zeta(s)=\sum_{k=1}^{\infty}\frac{1}{n^s}$ for $\mathrm{Re}(s)>1$ can be extended to the whole complex plane via analytic continuation.}
\begin{align}
 \sum_{k=1}^{\infty}1=\zeta(0)=-\frac{1}{2},   \quad\quad \sum_{k=1}^{\infty}k=\zeta(-1)=-\frac{1}{12},   \quad\quad \sum_{k=1}^{\infty}k^2=\zeta(-2)=0, 
\end{align}
the relevant parts in the one-loop prepotential become
\begin{align}
&-\frac{1}{12}\sum_{k\in \mathbb{Z}}\sum_{\alpha\in \cirDelta_l}|\alpha\cdot m_{\cirG}+k\tau|^3\sim-\frac{1}{6}\sum_{\alpha\in \cirDelta_l^+}(\alpha\cdot m_{\cirG})^3+\frac{1}{24}\tau\sum_{\alpha\in \cirDelta_l}(\alpha\cdot m_{\cirG})^2,\\
&-\frac{1}{12}\sum_{k\in \mathbb{Z}}\sum_{\alpha\in \cirDelta_s}|\alpha\cdot m_{\cirG}+k
\tau/{n_G}|^3\sim-\frac{1}{6}\sum_{\alpha\in \cirDelta_s^+}(\alpha\cdot m_{\cirG})^3+\frac{1}{24n_G}\tau\sum_{\alpha\in \cirDelta_s}(\alpha\cdot m_{\cirG})^2,\\
&\frac{1}{12}\sum_{k\in\mathbb{Z}}\sum_{i,f_i}\sum_{w_{\cirG}\in \mathring{R}_{i}^{\cirG}}|w_{\cirG}\cdot m_{\cirG}+k\tau/n^G_i+m_{f_i}|^3 \nonumber\\ &\quad\quad\sim\frac{1}{12}\sum_{k\in\mathbb{Z}}\sum_{i}\sum_{w\in \cirR_{i}^{+}}(w_{\cirG}\cdot m_{\cirG}+w_{\cirF}\cdot m_{\cirF})^3-\sum_i\frac{N_i}{24 n^G_i}\tau\cdot\sum_{w_{\cirG}\in \cirR_{i,r_i}^{\cirG}}(w_{\cirG}\cdot m_{\cirG})^2.
\end{align}
Then the total one-loop prepotential becomes
\be\label{eq:Foneloop}
\ba
\mathcal{F}^{1\text{-loop}}&=-\frac{1}{6}\sum_{\alpha\in \cirDelta^+}(\alpha\cdot m_{\cirG})^3+\frac{1}{12}\sum_{k\in\mathbb{Z}}\sum_i\sum_{w\in \cirR_{i}^{+}}(w_{\cirG}\cdot m_{\cirG}+w_{\cirF}\cdot m_{\cirF})^3\\
&\phantom{=}+\frac{1}{24}\tau\sum_{\alpha\in \cirDelta}(\alpha\cdot m_{\cirG})^2+\frac{1-n_G}{24n_G}\tau\sum_{\alpha\in \Delta_s}(\alpha\cdot m_{\cirG})^2-\sum_i\frac{N_i}{24 n^G_i}\tau\cdot\sum_{w_{\cirG}\in \cirR_{i,r_i}^G}(w_{\cirG}\cdot m_{\cirG})^2\\
&=-\frac{1}{6}\sum_{\alpha\in \cirDelta^+}(\alpha\cdot m_{\cirG})^3+\frac{1}{12}\sum_{k\in\mathbb{Z}}\sum_{i}\sum_{w\in \cirR_{i}^{+}}(w_{\cirG}\cdot m_{\cirG}+w_{\cirF}\cdot m_{\cirF})^3+c\tau m_{\cirG}\cdot m_{\cirG}.
\ea
\ee
Here we define the positive weights as
\begin{align}
    \cirR_{i}^{+}=\{w_{\cirG}\in \cirR_{i}^{\cirG},w_{\cirF}\in \cirR_{i}^{\cirF}\,\big|\,w_{\cirG}\cdot m_{\cirG}+w_{\cirF}\cdot m_{\cirF}> 0\},
\end{align}
and the constant $c$ as
\begin{align}
    c=\frac{\hgc}{6}-\frac{n_G-1}{12 n_G}\mathrm{Ind}(R_i^{\mathring{G}})-\sum_i\frac{N_i}{12 n^G_i}\mathrm{Ind}(\cirR_{i}^{\cirG}).
\end{align}
For the $A_{2r}^{(2)}$ cases, as described in \eqref{A2r_root}, there are additional ``shortest" roots taking the KK-charge $n+\frac{1}{4}$ and $n+\frac{3}{4}$. They contribute an additional term
\begin{align}\label{eq:A2lcterm}
-\frac{1}{12}\times 2\times 3 \times \left(\zeta(-1,\frac{1}{4})+\zeta(-1,\frac{3}{4})\right)\tau\sum_{\alpha\in \Delta_l^+}\left(\frac{\alpha\cdot t}{2}\right)^2=-\frac{1}{96}\tau\sum_{\alpha\in \Delta_l^+}\left(\frac{\alpha\cdot t}{2}\right)^2,
\end{align}
which  contributes  to $c$ by $-\frac{1}{96}\times2\,\Ind(\mathbf{F})=-\frac{1}{48}$. Therefore, for $A_{2r}^{(2)}$ theory with $N$ KK zero mode matter in the fundamental representation, the one-loop prepotential takes the form of the last line in \eqref{eq:Foneloop} with 
\begin{align}
    c=\frac{r+1}{6}-\frac{1}{12}(r-1)-\frac{1}{48}-\frac{N}{24}\times 1
    =\frac{1}{48} (11 + 4 r) - \frac{1}{24} N.
\end{align}
Here we have used that for $G=C_r$, $\hg=r+1$, $\Ind(\mathbf{F})=1$ and $\Ind(\mathbf{\Lambda}^2)=2(r-1)$.

By explicit computation for all twisted theories, we observe the following interesting identities of the $c$ constant. For all the cases with twist coefficients $n_G=2,3,4$ in Table \ref{tb:main0a},
\be\label{goldall1}
\boxed{
c=
\left\{\begin{array}{l}
 \frac{\fn}{2n_G}-\frac{1}{(n_G)^2}, \qquad \cirG={B_r,C_r}, \\
  \frac{\fn}{2(n_G)^2}+\frac{1}{n_G},\qquad \cirG={G_2,F_4}.
\end{array}
\right.}
\ee
The physical meaning of the constant  $c$ is (the negative of) the left-moving Casimir energy in the twisted Ramond sector which will be discussed later.  In geometry, $c$ can be absorbed by the base K\"ahler parameter. A similar shift of the base K\"ahler parameters has been discussed in \cite{Cota:2019cjx}, which is closely related to the intersection number of the divisors in the genus-one fibered Calabi-Yau threefolds. It is interesting to construct the geometry explicitly and reproduce the identity \eqref{goldall1}.

In order to make connection to the blowup equation in Section \ref{sec:blowup}, we need the information about the classical part of the genus one free energies $\mathcal{F}^{\mathrm{cls}}_{(0,1)}$ and $\mathcal{F}^{\mathrm{cls}}_{(1,0)}$. As described in Section 2 of \cite{Gu:2020fem}, these two functions only contain bilinear forms of the gauge and flavor algebra and the part related to the tensor multiplet. The action of the automorphism twist maps the bilinear forms to the bilinear forms of the folded algebras directly and leaves the tensor part unchanged. We expect that
\begin{align}
    \mathcal{F}^{\mathrm{cls}}_{(0,1)}&=-\frac{1}{12}\sum_{\alpha\in\cirDelta^+}\alpha\cdot m_{\cirG}+\frac{1}{24}\sum_i\sum_{w\in \cirR_{i}^{+}}(w_{\cirG}\cdot m_{\cirG}+w_{\cirF}\cdot m_{\cirF})+\frac{\fn-2}{2\fn}t_{\mathrm{ell}},\label{eq:F01}\\
    \mathcal{F}^{\mathrm{cls}}_{(1,0)}&=\frac{1}{12}\sum_{\alpha\in\cirDelta^+}\alpha\cdot m_{\cirG}+\frac{1}{24}\sum_i\sum_{w\in \cirR_{i}^{+}}(w_{\cirG}\cdot m_{\cirG}+w_{\cirF}\cdot m_{\cirF})+\frac{\fn-2-h^{\vee}_{G^{(r)}}}{4\fn}t_{\mathrm{ell}}.\label{eq:F10}
\end{align}
Notice that the dual Coxeter number $\hg$ in \eqref{eq:F10} is the number for the original group $G$ or the twisted group $G^{(r)}$, according to the identity
\begin{align}\label{dualCox}
    h^{\vee}_{G^{(r)}}=h^{\vee}_{G^{(1)}}=\hg.
\end{align}
This is easy to understand because the dual Coxeter number is the summation of all comarks of the affine Dynkin diagrams and the comark of each node of twisted affine Lie algebra just comes from the summation of the comarks of the nodes folded together.

\subsection{Twisted elliptic genera of BPS strings}\label{sec:2dquiver}
The Ramond-Ramond (R-R) elliptic genus of the BPS strings in 6d $(1,0)$ SCFTs is defined as the trace over the Hilbert space of the R-R sector of the 2d $(0,4)$ worldsheet theory:
\be\label{RRdef}
\IE(q,v,x,m)=\mathrm{Tr}_{\rm RR}(-1)^{F_L+F_R}e^{2\pi i \tau H_L} e^{2\pi i \bar{\tau} H_R}x^{J_L^3}v^{J_R^3+J_I^3}\prod e^{2\pi i m\cdot K} ,
\ee
where $m$ and $K$ are the fugacities and the Cartan generators associated to the 2d global symmetry, which is composed of 6d gauge and flavor symmetries. The $J_L^3,J_R^3$ and $J_I^3$ are the usual Cartan generators of the little group $\SO(4)=\SU(2)_L\times \SU(2)_R$ and the R-symmetry group $\SU(2)_I$ of the 6d $(1,0)$ SCFT. It is common to refer to the R-R elliptic genera as just \emph{elliptic genera}. One can also consider the trace over the Hilbert space of the Neveu-Schwarz-Ramond (NS-R) sector of the 2d theory, in which case the above definition gives the NS-R elliptic genera. This type of elliptic genera will also be used in later sections when we discuss the spectral flow symmetry. 

In the case of twisted circle compactification of 6d SCFTs, the 2d worldsheet theories also get twisted circle compactified on one circle of the torus. We can formally define the twisted elliptic genera by replacing the Hilbert space in  \eqref{RRdef} to the Hilbert space of the 2d states appearing in the twisted circle compactification. The Hamiltonians $H_L$ now have fractional eigenvalues due to the twist. For the twist coefficient $n_G=2,3,4$, the gap between energy eigenvalues  becomes $1/n_G$. This brings in many new fascinating features for the twisted elliptic genera. We find the primary feature of $d$-string twisted elliptic genera is
\be\label{eq:dc}
\IE_d(q,v,x,m_{\cirG},m_{\cirF})= q^{-dc}\Big(Z_d^{5d}(v,x,m_{\cirG},m_{\cirF})+ \mathcal{O}(q^{1/n_G}) \Big).
\ee
Here $ Z_d^{5d}$ is the $d$-instanton Nekrasov partition function of the circle reduction, i.e., the 5d low energy theory of the twisted 6d $(1,0)$ SCFT. Such a 5d theory has gauge group $\cirG$ and the 5d matter content taken from the KK-charge 0 part of the 6d twisted matter content. 

When the 2d $(0,4)$ worldsheet theory has a known 2d quiver gauge theory construction, the above definition can be explicitly computed by 2d localization, also known as  Jeffrey-Kirwan residue  \cite{Benini:2013nda,Benini:2013xpa}. However, this good situation happens very rarely. In most cases, it is very hard or impossible to construct a 2d gauge theory for the worldsheet theory of BPS strings. For the twisted cases, this is even hard and rare. 
In this section, we review the 2d localization formulas for 6d $\so({2r+8})^{(2)}$ theories with $\fn=4$ and pure $\su(3)^{(2)}$ theory with $\fn=3$, which were first established in \cite{Kim:2021cua}. As far as we know, these are the only two types of twisted theories, for which the twisted elliptic genera can be computed directly. 

\paragraph{\underline{$\so({2r+8})$ theory with $\mathbb{Z}_2$ twist}}
The instanton partition function of the $\so({2r+8})^{(2)}$ theory can be obtained from the $\mathbb{Z}_2$ twist of 6d $\so({2r+8})$ theory with $2r$ fundamental flavors. In the 2d $\mathcal{N}=(0,4)$ quiver gauge theory description of the 6d $\so({2r+8})$ theory, at $k$-strings, the resulting theory is a $\Sp(k)$ gauge theory coupled to the bulk $\SO(2r+8)$ gauge group and the $\Sp(2r)$ flavor group as follows
\begin{align}
    \raisebox{-.5\height}{
    \begin{tikzpicture}
  \tikzstyle{gauge} = [circle, draw,inner sep=3pt];
  \tikzstyle{flavour} = [regular polygon,regular polygon sides=4,inner sep=3pt, draw];
  \node (g1) [gauge,label=below:{$\Sp(k)$}] {};
  \node (f1) [flavour,above of=g1, label=above:{$\SO(2r+8)$}] {};
  \node (f2) [flavour,right of=g1, label=right:{$\Sp(2r)$}] {};
  \draw (g1)--(f1);
  \draw[dashed] (g1)--(f2);
  \draw (g1) to [out=140,in=220,looseness=10] (g1);
  \node at (-1,0) {\small{asym}\phantom{1} };
  \end{tikzpicture}
    }
    \label{eq:2d_quiver}
\end{align}
Here solid/dashed lines denote 2d hypermultiplets/Fermi multiplets respectively. Under the $\mathbb{Z}_2$ outer automorphism twist, the $\Sp(k)$ gauge algebra and $\Sp(2r)$ flavor algebra keep invariant, so we only need to take care of the bulk gauge group $\SO(2r+8)$. In the 2d description, the bulk gauge group serves as a flavor group and it couples with the gauge group $\Sp(k)$ in the fundamental representation. Under $\mathbb{Z}_2$ twist, the fundamental representation of $\SO(2r+8)$ is split into the fundamental representation of $\SO(2r+7)$ with KK charge zero and a trivial representation of $\SO(2r+7)$ with KK charge one half. The 6d partition function turns out to be \cite{Kim:2021cua,Chen:2021rek}
\begin{align}
    Z_{\text{str}}^{\rm{6d}}=\sum_{k=0}^{\infty}{q}_{\phi}^k \,\IE_k,
\end{align}
where $\IE_k$ is the $k$-string elliptic genus 
\begin{align}
    \IE_k=\frac{1}{2^kk!}\oint \prod_{I=1}^k\frac{\rd u_I}{2\pi i}\cdot Z_{k,1-\text{loop}}^{\rm{6d}}(u_I),
\end{align}
with the integrand to be
\be
\ba
Z_{k,1-\text{loop}}^{\rm{6d}}(u_I)=
&\left(-\frac{\vartheta _1\left(2 \epsilon _+\right)}{\vartheta _1\left(\epsilon_{1,2}\right)}\right)^k 
\cdot \prod _{I=1}^k \vartheta_1\left(\pm 2   u_I\right) \vartheta _1\left(2 \epsilon _+\pm 2 u_I\right)\\
   &\cdot\prod _{I<J}^k \frac{\vartheta _1\left(\pm
     u_I\pm   u_J\right) \vartheta _1\left(2 \epsilon _+\pm   u_I\pm
     u_J\right)}{\vartheta _1\left(\epsilon _1\pm   u_I\pm u_J\right) \vartheta _1\left(\epsilon _2\pm   u_I\pm u_J\right)}\\
&\cdot \left(\prod _{I=1}^k \frac{\prod _{l=1}^{2 r} \vartheta_1\left(\pm   u_I+m_l\right)}{\vartheta _1\left(\epsilon _+\pm
     u_I\right)\vartheta _4\left(\epsilon _+\pm
     u_I\right)\prod _{i=1}^{r+3} \vartheta _1\left(\epsilon _+\pm
     u_I\pm \alpha_i\right)}\right)\, ,
\ea
\ee
where we define the variant theta functions $\vartheta_i(z)= {\theta_i(z,\tau)}/{\eta(\tau)},i=1,\cdots,4$. Here the $\vartheta_4$ in the last line comes from the trivial representation of $\SO(2r+7)$ with $\frac{1}{2}$ KK charge.

\paragraph{\underline{$\su(3)$ theory with $\mathbb{Z}_2$ twist}}
It was realized in \cite{Kim:2021cua} from brane webs that the $\mathbb{Z}_2$ twisted circle compactification of 6d pure $\mf{su}(3)$ can be obtained from Higgsing of 6d $G_2$ gauge theory with one fundamental flavor. The 2d description of 6d $G_2$ theory with one fundamental has been constructed in \cite{Kim:2018gjo}, where the theory is written from the $G_2$ subalgebra $\SU(3)$. Note the $G_2$ algebra contains a subalgebra $\SU(2)\times \SU(2)$. Denote $\alpha,\alpha^{\prime}$ the Coulomb parameters for the $\SU(2)$'s, and $m$ the mass for the fundamental flavor in the 6d $G_2$ theory, we have the following tuning of parameters for the Higgsing
\begin{align}\label{eq:higgsG2A2}
    \alpha\rightarrow \alpha,\quad \alpha^{\prime}\rightarrow \frac{\tau}{4},\quad m\rightarrow \epsilon_+-\frac{\tau}{2}.
\end{align}
Apply the reparametrization \eqref{eq:higgsG2A2}, we have the $k$-string elliptic genus of 6d $\SU(3)$ theory with $\mathbb{Z}_2$ twist as
\begin{align}
    Z_{\text{str}}^{\rm{6d}}=\sum_{k=0}^{\infty}{q}_{\phi}^k \,\IE_k,
\end{align}
where $\IE_k$ is the $k$-string elliptic genus 
\begin{align}
    \IE_k=\frac{1}{k!}\oint \prod_{I=1}^k\frac{\rd u_I}{2\pi i}\cdot Z_{k,1-\text{loop}}^{\rm{6d}}(u_I),
\end{align}
with the integrand in the elliptic form to be
\be
\ba
  &Z_{k,1-\text{loop}}^{\rm{6d}}(u_I)=\prod_{I=1}^k\frac{d{u}_I}{2\pi i}\cdot
  \frac{\prod_{I\neq J}\vartheta_1({u_{IJ}})\cdot
  \prod_{I,J}\vartheta_1({2\epsilon_+-{u}_{IJ}})}
  {\prod_{I=1}^k\prod_{i=1}^2 {\vartheta_1}({\epsilon_+\pm({u}_I-\alpha_i)})
  \cdot\prod_{I,J}\vartheta_1({\epsilon_{1,2}+{u}_{IJ}})}\\
  &\quad\times\frac{\prod_I\!\big(  \vartheta_4^{[\frac{3}{4}]}({\epsilon_++{u}_I})\cdot\prod_{I\leq J}\big(\vartheta_4({{u}_I+{u}_J})
  \cdot \vartheta_4({{u}_I+{u}_J-2\epsilon_+})\big)}
  {\prod_I\!\big( \vartheta_4^{[\frac{3}{4}]}(\epsilon_+-{u}_I) \vartheta_4^{[\mp\frac{1}{4}]}({\epsilon_+\pm{u}_I})\prod_{i=1}^2 \vartheta_4({\epsilon_+-{u}_I-\alpha_i})
 \big)
  \cdot\prod_{I<J}\vartheta_4({\epsilon_{1,2}-{u}_I-{u}_J})},
  \\
\ea
\ee
where $\alpha_1=-\alpha_2=\alpha$.
The residue sum can be labeled by $\SU(2)$ colored Young diagrams, then we have
\be\label{su3z2Ed}
\ba
  \IE_k =&\,\sum_{\vec{Y};|\vec{Y}|=k}\prod_{i=1}^2\prod_{s\in Y_i}
  \frac{\vartheta_4(2{u}(s))\cdot \vartheta_4(2\epsilon_+-2{u}(s))}
  {\prod_{j=1}^2\left(\vartheta_1({E_{ij}(s)})\cdot
  \vartheta_1({E_{ij}(s)-2\epsilon_+})\right)}\\
&\prod_{i=1}^2\prod_{s\in Y_i} \frac{\vartheta_4^{[\frac{3}{4}]}(\epsilon_++u(s))}{\vartheta_4^{[\frac{3}{4}]}(\epsilon_+-u(s))\vartheta_4^{[\mp\frac{1}{4}]}(\epsilon_+\pm u(s))\cdot\prod_{j=1}^2 \vartheta_4({\epsilon_+-{u}(s)-\alpha_j})}  \\
&\cdot\prod_{i\leq j}^2\prod_{s_{i,j}\in Y_{i,j};s_i<s_j}
  \frac{\vartheta_4({{u}(s_i)+{u}(s_j)})\cdot
  \vartheta_4({{u}(s_i)+{u}(s_j)-2\epsilon_+})}
  {\vartheta_4({\epsilon_{1,2}-{u}(s_i)-{u}(s_j)})},
 \ea
 \ee
where
\be
{u}(s)=\alpha_i-\epsilon_+-(n-1)\epsilon_1-(m-1)\epsilon_2,\quad s=(m,n)\in Y_i,
\ee
and 
\be
E_{ij}(s)= \alpha_i-\alpha_j-\epsilon_1 h_i(s)+\epsilon_2(v_j(s)+1),
\ee
Here $s_i\!<\!s_j$ means ($i<j$) or ($i\!=\!j$ and $m_i\!<\!m_j$) or
($i\!=\!j$ and $m_i\!=\!m_j$ and $n_i\!<\!n_j$).
The $h_i(s)$ denotes the distance from $s$ to
the right end of the diagram $Y_i$ by moving right and the $v_j(s)$ denotes the distance from $s$ to the bottom of the diagram $Y_j$ by moving down.

\subsection{Anomaly and modular index}
Elliptic genera $\IE$ of 2d $(0,4)$ SCFTs in general are weight zero Jacobi forms of multi elliptic variables on $SL(2,\IZ)$. They are equipped with an important quantity -- the modular index $\mathrm{Ind}\, \IE$, as a quadratic form of the  elliptic variables. Under modular transformation of the torus, the elliptic genera transform as
\be
\IE_d\Big(-\frac{1}{\tau},\frac{\eq}{\tau},\frac{\et}{\tau},\frac{m_G}{\tau},\frac{m_F}{\tau}\Big)=\exp{\Big(\frac{2\pi i}{\tau} \mathrm{Ind}\,\IE  \Big)} \IE(\tau,\eq,\et,m_G,m_F). 
\ee
Equivalently, one can write as
\begin{equation}
\partial_{E_2} \log\IE = -\frac{1}{12}(\mathrm{Ind}\, \IE).
\end{equation}
Here $E_2(\tau)$ is the weight-two Eisenstein series that measures the modular anomaly.

For 6d rank-one $(1,0)$ theories, the modular index of the elliptic genera can be derived from the anomaly polynomials for the 2d $(0,4)$ SCFTs on the worldsheet of the BPS strings \cite{Ohmori:2014pca,Ohmori:2014kda,Shimizu:2016lbw}. To be precise, for a 6d $(1,0)$ theory with tensor coefficient $\fn$, gauge group $G$ and flavor group $F$, the modular index of the $d$-string elliptic genus can be uniformly written as
\be \label{eq:Ed-index}
\ba
  \mathrm{Ind}\, \IE_d(\eq,\et,m_G,m_F) =
  &-\Big(\frac{\eq+\et}{2}\Big)^2(2-\fn+\hg)d +
    \frac{\eq\et}{2}(\fn d^2 +(2-\fn)d) \\
  & +\frac{d}{2} (- \fn\,m_G\cdot m_G +
    k_F\,m_F\cdot m_F ) \ .
\ea
\ee

The modularity for twisted elliptic genera is more complicated. Though from the viewpoint of twisted circle compactification, it is convenient to discuss the fractional KK-charges, which brings in the fractional $q^{1/n_G}$ orders in twisted elliptic genera, where $n_G=2,3,4$. From the geometric viewpoint, it is more suitable to scale $q$ to $q^{n_G}$ such that we are still in the integral bases of curves. After such scaling, it is natural to expect that the twisted elliptic genera behave as weight zero Jacobi forms of some congruence subgroups of $SL(2,\IZ)$. We propose that in general the congruence subgroup should be $\Gamma_1(n_G)$, though in many cases, we notice that the twisted elliptic genera satisfy a bigger modular symmetry that is $\Gamma_0(n_G)$. These are consistent with the geometric observation in \cite{Cota:2019cjx} for multi-section Calabi-Yau threefolds. We collect the definition of various congruence subgroups of $SL(2,\IZ)$ in Appendix \ref{app:B}.

The index quadratic form does not really depend on the modular groups. For $\mathcal{O}^{(r_{\mathrm{out}})}$ twisted compactification of rank-one SCFTs, the tensor multiplet is invariant under the twist, the gauge and flavor algebras become their twisted versions due to the identification of the masses in the same orbits of $\mathcal{O}^{(r_{\mathrm{out}})}$. This indicates that the modular index of twisted elliptic genus can be directly deduced from its untwisted origin by identifying the masses of the particles of the hyper and vector multiplets in the same orbits of $\mathcal{O}^{(r_{\mathrm{out}})}$. Thus we have the following uniform formula for the index of $d$-string twisted elliptic genus:
\be\label{eq:twistedEd-index}
\ba
  \mathrm{Ind}\, \IE_d(\eq,\et,\mG,\mF) =
  &-\Big(\frac{\eq+\et}{2}\Big)^2(2-\fn+\hg)d +
    \frac{\eq\et}{2}(\fn d^2 +(2-\fn)d) \\
  & +\frac{d}{2} (- \fn\,\mG\cdot \mG +
    k_{\cirF}\,\mF\cdot \mF ) .
\ea
\ee
One can see that only the second line changes from \eqref{eq:Ed-index}. It should be emphasized that the dual Coxeter number $\hg$ here does not change upon the twist, as was explained in the identity \eqref{dualCox}.

\section{Twisted elliptic blowup equations}\label{sec:blowup}
\subsection{Elliptic blowup equations and their twists}\label{sec:ellipticblowup}
For refined topological strings on a non-compact Calabi-Yau threefold $X$, the general form of blowup equations was proposed in \cite{Huang:2017mis}:
\be\label{blowupgeom}
\ba
    \Lambda(\epsilon_1,&\,\epsilon_2,m_i) Z\left(\epsilon_1,\epsilon_2,t_i+\pi \ri B_i\right)\\
    =\sum_{\mathbf{k}\in \mathbb{Z}^{b_4}} &(-1)^{|\mathbf{k}|}Z\left(\epsilon_1,\epsilon_2-\epsilon_1,t_i+(C_{ij}k_j+B_i/2)\epsilon_1+\pi \ri B_i\right)\\[-3.6mm]
    &\phantom{--,}\times Z\left(\epsilon_1-\epsilon_2,\epsilon_2,t_i+(C_{ij}k_j+B_i/2)\epsilon_2+\pi \ri B_i\right),
\ea
\ee
as a generalization of the K-theoretic blowup equations for 5d gauge theories in \cite{Nakajima:2005fg}. Here $b_4$ is the Betti number which counts the number of compact divisors of $X$. The $t_i$ is the K\"ahler parameter which is the volume of the curve class in $H_2(X;\mathbb{Z})$ and $C_{ij}$ is the intersection numbers of divisors and curves in $X$. The $B_i$ is the flux for the corresponding K\"ahler parameter, where the value can be determined from the classical prepotential of the topological strings on $X$ \cite{Huang:2017mis}. The $\Lambda(\epsilon_1,\epsilon_2,m_i)$ is a function that can be fixed from the fluxes $B_i$ and the classical prepotential, and it only depends on the mass parameters of the theory.

In a series of works \cite{Gu:2018gmy,Gu:2019dan,Gu:2019pqj,Gu:2020fem,Duan:2021ges}, elliptic blowup equations have been established for all 6d $(1,0)$ and $(2,0)$ SCFTs. These functional equations serve as very efficient tools to solve the elliptic genera of the BPS strings associated to 6d SCFTs. For example, for rank-one 6d $(1,0)$ SCFTs with $\fn\ge 3$ and no half-hyper, there exist a universal recursive formula for the elliptic genera with respect to the number of strings derived from unity elliptic blowup equations \cite{Gu:2020fem}. In this section, we generalize both the elliptic blowup equations and the recursive formulas to the twisted cases. The partition functions of twisted circle compactification of 6d SCFTs can be written in terms of twisted elliptic genera as
\begin{align}
Z\left(\epsilon_1,\epsilon_2,t_i\right)=e^{\mathcal{F}^{\mathrm{cls}}}Z_0(\epsilon_1,\epsilon_2,\mG,\mF)\bigg(1+\sum_{d=1}^{\infty}e^{d\cdot t_{\mathrm{ell}}}  \IE_d(\eq,\et,\mG,\mF)\bigg),
\end{align}
where $\mathcal{F}^{\mathrm{cls}}$ is the classical prepotential $\mathcal{F}^{\mathrm{cls}}=\mathcal{F}^{\mathrm{cls}}_{(0,0)}+\mathcal{F}^{\mathrm{cls}}_{(0,1)}+\mathcal{F}^{\mathrm{cls}}_{(1,0)}$ defined by \eqref{eq:F00}, \eqref{eq:F01} and \eqref{eq:F10}. 
The derivation of twisted elliptic blowup equations from \eqref{blowupgeom} to the elliptic form is exactly parallel to the untwisted cases, by a ``de-affinization procedure", thus we only present the results here. We refer readers interested in the derivation to Section 2.4 of \cite{Gu:2019dan} and Section 3.1 and Appendix D of \cite{Gu:2020fem}.

Consider a rank one 6d SCFT with tensor branch coefficient $\fn$,
gauge symmetry $G$, flavor symmetry $F$, and half-hypermultiplets
transforming in the representations $(R_{G},R_{F})$.  The flavor
symmetry induces a current algebra of level $k_F$ on the worldsheet of
BPS strings. Upon discrete twist, the twisted affine Lie algebra $\widehat{G}^{(n)}$ has truncated part $\cirG$ as low energy 5d gauge symmetry. In the meantime, matter representation becomes $\mathring{R}$ and reduced flavor symmetry becomes $\mathring{F}$. Let $\lambda_0$ be a coweight base of $G$ that is \emph{invariant} upon the twist.\footnote{For example for $\so(2r)^{(2)}$, the $\lambda_0$ associated to the vector representation is invariant upon the twist, while the one associated to the spinor representation is not.} If  $(\lambda_0,\lambda_F)$ gives admissible elliptic blowup equation for the original 6d SCFT, and let $\lambda_{\cirF}$ be the reduction of $\lambda_F$, then we propose the twisted elliptic genera $\IE_d(\tau,m_{\cirG},m_{\cirF},\e_{1,2})$
satisfy the following \emph{twisted elliptic blowup equations}:
\be\label{eq:ebeq}
\ba
\,&\sum_{\lG\in\phi_{\lambda_0}(Q^\vee(\cirG))}^{\frac{1}{2}||\lG||^2+d'+d''=d+\delta}
    (-1)^{|\lG|}\\
  &\phantom{===}\times\theta_{i}^{[a]}(\fn\tau,-\fn\lG\cdot\mG +
    k_{\cirF}\lF\cdot\mF+(y-\small{\frac{\fn}{2}}||\lG||^2)(\eq+\et)
    -\fn d'\eq-\fn d''\et) \\[-1mm]
  &\phantom{===}\times
    A_{{V}}^{\cirG}(\tau,\mG,\lG)   A_{{V}}^{frac}(\tau,\mG,\lG)
    A_{{H}}^{\mathring{R}}(\tau,\mG,\mF,\lG,\lF)\\
  &\phantom{===}\times
    \IE_{d'}(\tau,\mG+\eq\lG,\mF+\eq\lF,\eq,\et-\eq)
    \IE_{d''}(\tau,\mG+\et\lG,\mF+\et\lF,\eq-\et,\et)\\
  &\phantom{==}=\Lambda(\delta)\,
    \theta_{i}^{[a]}(\fn\tau,k_F\lF\cdot\mF+\fn y(\eq+\et))
    \IE_d(\tau,\mG,\mF,\eq,\et),\qquad d=0,1,2,\ldots
        \ea
\ee
where
\begin{equation}
  \Lambda(\delta) =
  \begin{cases}
    1, & \delta =0,\\
    0, & \delta >0.
  \end{cases}
\end{equation}
The $\Lambda(\delta)=1$ case is called \emph{unity} blowup equations, while the $\Lambda(\delta)=0$ case \emph{vanishing}. The parameter $y$ is determined in \cite{Gu:2020fem} as
\begin{equation}\label{eq:y}
  y = \frac{\fn-2+\hg}{4}+\frac{k_F}{2}(\lambda_F\cdot\lambda_F)=\frac{\fn-2+\hg}{4}+\frac{k_{\cirF}}{2}(\lambda_{\cirF}\cdot\lambda_{\cirF}),
\end{equation}
which does not change upon twist. 
The Jacobi theta function $\theta_{i}^{[a]}$ with characteristics $a$ are defined in Appendix \ref{app:A}.  
It comes from the contribution of the prepotential of 6d twisted theories. The subscript $i$ is 3 if $ \fn$ is even and 4 if $\fn$ is odd. For twist coefficient $n_G=2,3$, the characteristic $a$ of the theta function can be one of the following
$\fn$ numbers
\begin{equation}\label{eq:a}
  2\fn a = h^\vee_{\cirG}-\frac12\sum_i {N_i}\mathrm{Ind}(R_i^{\mathring{G}})    - 2k,\quad k =0,1,\ldots,\fn-1.
\end{equation}
Recall $N_i$ are the number of hypers in $R_i^{\mathring{G}}$ with KK-charge zero. This formula means that only the KK-charge 0 part is relevant for the characteristic $a$, which is required for the 5d limit of twisted elliptic blowup equations to be consistent with the 5d blowup equations with matters \cite{Kim:2019uqw}. For twist coefficient $n_G=4$, the low energy gauge group $\cirG=C_r$ which needs special care. When there is no  KK-charge 0 matter, the $C_r$ allows another possibility called the $\theta=\pi$ theory, in which case the $h^\vee_{\cirG}$ in \eqref{eq:a} needs to be replaced as $h^\vee_{\cirG}-1$. This is also consistent with the 5d blowup equations in \cite{Kim:2019uqw}. The one-loop vector-multiplet contribution $A_{{V}}^{\cirG}(\tau,\mG,\lG)$ is defined as Equation $(3.7)$ of \cite{Gu:2020fem}, while the fractional part $A_{{V}}^{frac}(\tau,\mG,\lG)$ is defined similarly with the fractional KK-charge taken into account.  The one-loop hypermultiplet contribution $A_{{H}}^{\mathring{R}}(\tau,\mG,\mF,\lG,\lF)$ is defined similar with Equation $(3.8)$ of \cite{Gu:2020fem} with KK-charges taken into account. We collect the relevant formulas in Appendix \ref{app:A}. We summarize all admissible unity twisted elliptic blowup equations in Table \ref{tb:u}. We also collect some simple vanishing ones in Table \ref{tb:v}. It can be expected that there will be more vanishing blowup equations for some theories. Besides, we will focus on the twisted theories with paired matter content here. The blowup equations and twisted elliptic genera for theories with unpaired matter content are more subtle, we leave the discussion in Appendix \ref{app:D}.

\begin{table}[h]
\centering
%\resizebox{\linewidth}{!}{
\begin{tabular}{|c|l|l|l|l|c|c|c|c|}\hline
  $\fn$
  &$G$ & $\mathring{G}$
   & $\mathring{F}$
  &$\#$
  &$y$
  &$\lF$ 
  \\ \hline
  6&$E_6^{(2)}$& $F_4$ & $-$
  & 1 & 4 & $\emptyset$ \\
\hline
  4&$D_4^{(3)}$& $G_2$ &  $-$
  & 1  & 2 & $\emptyset$ \\
  4&$D_{r+4}^{(2)}$& $\!B_{r+3}\!$   & $\mf{sp}(2r)$
  & $2^{2r}$ & $r+2$ & $(0\ldots01)$ \\
  4 &$E_6^{(2)}$& $F_4$  & $\mf{sp}(1)$
  & 2 & 5 & $(1)$   \\
\hline
  3&$A_2^{(2)}$  & $C_1$ & $-$
  & 1 & 1 & $\emptyset$ \\
  3&$D_4^{(2)}$& $B_3$  & $\mf{sp}(1)\times\mf{sp}(1)$
  & 4 & 5/2 & (1),(1) \\
  3& $D_4^{(3)}$& $G_2$  & $\mf{sp}(1)$
  & 2 & 5/2 & (1) \\
 \hline
     2& $A_{2r}^{(2)}$&$C_r$  & $\mf{so}(4r+2)$
    & $2^{2r+1}$ & $r+\frac12$ & $(0\ldots01)$ \\
    2&$A_{2r-1}^{(2)}$&$C_r$  & $\mf{so}(4r)$
    & $2^{2r}$ & $r$ & $(0\ldots01)$ \\
    2&$D_4^{(2)}$& $B_3$ &    $\mf{sp}(2)\times\mf{sp}(2)$
  & 16 & 3 & (01),(01) \\
    2& $D_4^{(3)}$& $G_2$  & $\mf{sp}(2)$
    & 4 & 3 & (01) \\
    2&$D_5^{(2)}$ & $B_4$ &  $\mf{sp}(4)\times\mf{sp}(1)$
    & 32 & 4 & (0001), (1) \\
    2 &$E_6^{(2)}$& $F_4$ & $\mf{sp}(2)$ 
    & 4 & 6 & $(01)$ \\
\hline
    1& $A_2^{(2)}$ & $C_1$  & $\mf{so}(12)$ & 64
    & 2 & $(0\ldots01)$ \\
    1&$A_3^{(2)}$ & $C_2$  &  $\mf{so}(12)\times \mf{sp}(1)$
    & 128 & 5/2 & $\!(0\ldots01),(1)\!$ \\
    1&$D_4^{(2)}$& $B_3$ &    $\mf{sp}(3)\times\mf{sp}(3)$
  & 64 & 7/2 & (001),(001) \\
    1&$D_4^{(3)}$& $G_2$  & $\mf{sp}(3)$ & 8 & 7/2 & (001) \\ \hline
\end{tabular}
%}
\caption{The parameters $y, \lF$ of unity twisted elliptic blowup equations for rank
  one models. The \# is the number of unity equations
  with a fixed characteristic $a$.}
\label{tb:u}
\end{table}

\begin{table}[h]
\centering
%\resizebox{\linewidth}{!}{
\begin{tabular}{|c|l|l|l|l|c|c|c|}\hline
  $\fn$
  &$G$ & $\mathring{G}$
   & $\mathring{F}$
  &$\#$
  &$y$
  &$\lF$
  \\ \hline
  4&$D_{r+4}^{(2)}$& $\!B_{r+3}\!$   & $\mf{sp}(2r)$
  & $2^{2r}$ & $r+2$ & $(0\ldots01)$ \\
\hline
  3&$D_4^{(2)}$& $B_3$  & $\mf{sp}(1)\times\mf{sp}(1)$
  & 2 & $3$ & (1),(0) \\
 \hline
    2&$A_{2r-1}^{(2)}$&$C_r$  & $\mf{so}(4r)$
    & $1$ & $1 $ & $(0\ldots0)$ \\
    2&$D_4^{(2)}$& $B_3$ &    $\mf{sp}(2)\times\mf{sp}(2)$
  & 4 & $4$  & (01),(00) \\
    2&$D_5^{(2)}$ & $B_4$ &  $\mf{sp}(4)\times\mf{sp}(1)$
    & 16 &$5$& (0001), (0) \\
       2&$D_6^{(2)}$ & $B_5$ &  $\mf{sp}(6)$
    & 64 & $6$  & (000001) \\
\hline
    1&$D_4^{(2)}$& $B_3$ &    $\mf{sp}(3)\times\mf{sp}(3)$
  & 8 &  $5$ & (001),(000) \\
 \hline
\end{tabular}
%}
\caption{The parameters $y, \lF$ of vanishing twisted elliptic blowup equations for rank-one models.  The \# is the number of vanishing equations with fixed characteristics $a$.}
\label{tb:v}
\end{table}

To solve blowup equations, four methods have been presented in Section 4 of \cite{Gu:2020fem}, which are recursion formula and Weyl orbit expansion for elliptic blowup equations, and refined BPS expansion and $\epsilon_1,\epsilon_2$ expansion for general local Calabi-Yau threefolds. All four methods apply to the current twisted cases as well. 
In Section \ref{sec:recursion}, we will explicitly present the recursion formula for twisted elliptic genera, which is the most efficient approach but only works for $\fn\ge 3$ theories. We also use the Weyl orbit expansion to solve the twisted elliptic genera of many $\fn=1,2$ theories. Both methods utilize the unity part of \eqref{eq:ebeq}. 
The vanishing part of \eqref{eq:ebeq} is less useful in solving twisted elliptic genera, but still gives some interesting vanishing theta identities which will be discussed in Section \ref{sec:vanish}.

Blowup equations can even shed new light on the structure of 6d $(1,0)$ SCFTs itself. As mentioned earlier, it was found in \cite{Gu:2020fem} that for the modularity of elliptic blowup equations for 6d $(1,0)$ SCFTs to hold, the following constraint on tensor coefficient $\fn$, gauge algebra $G$ and matter representations $R$ need to be satisfied:
\be
\frac{\dualCox}{6}-\sum_i\frac{N_i}{12}\mathrm{Ind}(R_i^G)=\frac{\fn-2}{2}.
\ee
Simply speaking, this constraint comes from the consistency requirement when converting the blowup equations of refined topological strings on non-compact Calabi-Yau threefolds to the elliptic form. 
For twisted elliptic blowup equations, we find there exists similar constraints depending on the type of twists. For $\IZ_2$ twist with $n_G=2$, we find\footnote{The $\fn=1,\mf{su}(2)+10\bf F$ theory is not included.}
\be\label{gold}
\frac{h^\vee_{\mathring{G}}}{6}-\frac{1}{24}\mathrm{Ind}(V^{frac}_{1/2})-\sum_i\frac{N_i}{12n^G_i}\mathrm{Ind}(R_i^{\mathring{G}})=\left\{\begin{array}{cc}
 \frac{\fn}{4}-\frac14,    & \quad \so(2r)^{(2)}, \\
  \frac{\fn}{8}+\frac{1}{2},   & \quad E_6^{(2)}.
\end{array}
\right.
\ee
Here $N_i$ is the number of hypers $R_i^{\mathring{G}}$ with KK-charge 0, with a twist coefficient $n^G_i$. We checked (\ref{gold}) is true for all $\fn=4,\mf{so}(2N)^{(2)}$ theories\footnote{Since $\mathring{G}=\mf{so}(2N-1)$, we have $\frac{2N-3}{6}-\frac{2}{24}-\frac{2N-8}{12}\times 2=\frac{3}{4} $.}, $\fn=1,2,3,4,G=\mf{so}(8)^{(2)}$ theories, $\fn=2,G=\mf{so}(10),\mf{so}(12)$ theories,  $\fn=2,G=\mf{su}(2N)$ theories\footnote{Since $\mathring{G}=\mf{sp}(N)$, we have $\frac{N+1}{6}-\frac{2N-2}{24}-\frac{2N\times 1}{24}=\frac{1}{4}.$} and $\fn=1,G=\mf{su}(4)$ theories.
For $\IZ_3$ twist, we have
\be\label{gold2}
\frac{h^\vee_{\mathring{G}}}{6}-\frac{1}{18}\mathrm{Ind}(V^{frac}_{1/3})-\frac{N}{36}\mathrm{Ind}(R^{\mathring{G}})=\frac{\fn}{18}+\frac{1}{3}.
\ee
Here $N$ is the number of hypers $R^{G_2}$ with twist coefficient 3. 
We checked (\ref{gold2}) is true for $\fn=1,2,3,4,G=\so(8)^{(3)}$ theories.
For $n^G=4$, we find
\be\label{gold3}
\frac{h^\vee_{\mathring{G}}}{6}-\frac{1}{24}\mathrm{Ind}(V^{frac}_{1/2})-\frac{1}{48}\mathrm{Ind}(V^{frac}_{1/4})-\frac{N}{24}\mathrm{Ind}(R^{\mathring{G}})=\frac{\fn}{8}-\frac{1}{16}.
\ee
Here $N$ is the number of fundamental hypers $R=\bf F$ of $\cirG=\Sp(r)$. We checked (\ref{gold3}) is true for $\fn=1,2,3,G=\su(3)^{(2)}$ theories and $\fn=2,G=\su(2r+1)^{(2)}$ theories\footnote{Since $\mathring{G}=\mf{sp}(N)$, we have $\frac{N+1}{6}-\frac{2N-2}{24}-\frac{1}{48}-\frac{(2N+1)\times 1}{24}=\frac{3}{16}.$}. Notice that these interesting constraints exactly give the $c$ constants discussed in \eqref{goldall1}. In summary, denote $s$ as all fractional KK charges, we have for all paired theories,
\be\label{goldall}
c=\frac{h^\vee_{\mathring{G}}}{6}-\sum_{s}\zeta(-1,s)\mathrm{Ind}(V^{frac}_{s})-\sum_i\frac{N_i}{12r_i^{\rm out}}\mathrm{Ind}(R_i^{\mathring{G}})=
\left\{\begin{array}{l}
 \frac{\fn}{2n_G}-\frac{1}{(n_G)^2}, \\
  \frac{\fn}{2(n_G)^2}+\frac{1}{n_G}.
\end{array}
\right.
\ee
Here $n_G=1,2,3,4$. The two possibilities here perhaps are related to the different types of multi-section Calabi-Yau geometries.

\subsection{Modularity}
The modularity of all elliptic blowup equations for rank-one 6d SCFTs has been proved in the Section 3.2 of \cite{Gu:2020fem}. The modularity of twisted elliptic blowup equations actually inherits from its untwisted origin, that is the index of each part is directly reduced from its untwisted origin. For example, as we have mentioned, the modular index of twisted elliptic genera can be reduced from the modular index of untwisted ones. Therefore, we will be brief and only show one example here. Consider the pure $\so(8)^{(3)}$ theory with $\fn=4$. From the general formula \eqref{eq:ebeq}, we have the following unity twisted elliptic blowup equations for the twisted elliptic genera:
\be
\ba
  \sum_{\frac{1}{2}||\alpha^\vee||^2_{G_2}+d'+d''=d}&
    (-1)^{|\alpha^\vee|} \theta_{3}^{[a]}(4\tau,-4m_{\alpha^\vee} +(2-2||\alpha^\vee||^2)(\eq+\et)
    -4 d'\eq-4 d''\et)\nonumber \\[-3.3mm]
  &\times
    A_{{V}}^{G_2}(\tau,m,\alpha^\vee)
    A^{\mathbf{7}_{1/3} \oplus \mathbf{7}_{2/3}}_V(\tau,m,\alpha^\vee)\nn
  &\times
    \IE_{d'}(\tau,m+\eq\alpha^\vee,\eq,\et-\eq)
    \IE_{d''}(\tau,m+\et\alpha^\vee,\eq-\et,\et)\nn
  =&\,
    \theta_{3}^{[a]}(4\tau, 2(\eq+\et))
    \IE_d(\tau,m,\eq,\et),\qquad d=0,1,2,\ldots
        \label{eq:ebeq2}
\ea
\ee
Here the characteristic $a=0,\pm\frac14,\frac12$.
Note the $B$ field of the base $(-4)$-curve is even, which was already observed in \cite[Section 6.1]{Kim:2020hhh}.

We now show how to prove the modularity of twisted elliptic blowup equations (\ref{eq:ebeq2}). Denote $d_0=\frac{1}{2}||\alpha^\vee||^2_{G_2}$.
The classical part contributes to the index quadratic form as
\be
\mathrm{Ind}_{\rm cl}=2(-\alpha^{\vee}\cdot m_{G_2} + (1/2 - d_0) (\e_1 + \e_2) - d' \e_1 - d'' \e_2)^2.
\ee
The $\Lambda$ factor $ \theta_{3}^{[a]}(4\tau, 2(\eq+\et))$ contributes index with $\mathrm{Ind}_{\Lambda}=\frac{1}{2}(\e_1 + \e_2)^2$.
The following identities for the coroot lattice of $G_2$ are useful when computing the index from vector multiplets contribution: $\forall \alpha^{\vee}\in Q^{\vee}_{G_2}$,
\begin{align}
\sum_{\beta\in \Delta_+(G_2)}(m\cdot \beta)(\alpha^{\vee}\cdot \beta)&=4 m\cdot\alpha^{\vee},\\
\sum_{\beta\in \Delta_+(G_2)}(m\cdot \beta)(\alpha^{\vee}\cdot \beta)^3&=10d_0m\cdot\alpha^{\vee}=5(\alpha^{\vee}\cdot\alpha^{\vee})(m\cdot\alpha^{\vee}),\\
\sum_{\beta\in \Delta_+(G_2)}(m\cdot \beta)^2(\alpha^{\vee}\cdot \beta)^2&= \frac{10}{3}(m\cdot\alpha^{\vee})^2+\frac{5}{3}(m\cdot m) (\alpha^{\vee}\cdot\alpha^{\vee}).
\end{align}
Then by the definition in Appendix \ref{app:B}, we find the $G_2$ vector multiplet $A_{{V}}^{G_2}(\tau,m,\alpha^\vee)$ contributes to index as
\be
\ba
\mathrm{Ind}_{V}^{G_2}=&-\frac{5}{3}\((\alpha^{\vee}\cdot m_{G_2})^2+d_0( m_{G_2}\cdot m_{G_2})\)-\frac{10d_0-4}{3}(\eq+\et)\alpha^{\vee}\cdot m_{G_2}\\&
-\frac{d_0}{3}\(5d_0-2\)(\e_1^2+\eq\et+\e_2^2).
\ea
\ee
To compute the index contribution from the fractional vector multiplets, we still need the following useful formulas for $G_2$ orbit $\mathcal{O}_{1,6}\subset \bf 7$: $\forall \alpha^{\vee}\in Q^{\vee}_{G_2}$,
\begin{align}
\sum_{w\in \mathcal{O}_{1,6}}(m\cdot w)(\alpha^{\vee}\cdot w)&=2 m\cdot\alpha^{\vee},\\
\sum_{w\in \mathcal{O}_{1,6}}(m\cdot w)(\alpha^{\vee}\cdot w)^3&=2d_0m\cdot\alpha^{\vee}=(\alpha^{\vee}\cdot\alpha^{\vee})(m\cdot\alpha^{\vee}),\\
\sum_{w\in \mathcal{O}_{1,6}}(m\cdot w)^2(\alpha^{\vee}\cdot w)^2&= \frac{2}{3}(m\cdot\alpha^{\vee})^2+\frac{1}{3}(m\cdot m) (\alpha^{\vee}\cdot\alpha^{\vee}).
\end{align}
Then by the definition in Appendix \ref{app:B}, we find the KK charge $1/3$ and $2/3$ parts $A^{\mathbf{7}_{1/3} \oplus \mathbf{7}_{2/3}}_V(\tau,m,\alpha^\vee)$ contribute to index as
\be
\ba
\mathrm{Ind}^{\mathbf{7}_{1/3} \oplus \mathbf{7}_{2/3}}_V= &-\frac{1}{3}\Big((\alpha^{\vee}\cdot m_{G_2})^2+d_0( m_{G_2}\cdot m_{G_2})\Big)-\frac{2d_0-2}{3}(\eq+\et)\alpha^{\vee}\cdot m_{G_2}\\&
-\frac{d_0(d_0-1)}{3}(\e_1^2+\eq\et+\e_2^2).
\ea
\ee

From \eqref{eq:twistedEd-index} we know the $d$-string twisted elliptic genus has index
\be
\mathrm{Ind}_{\IE}(d,m_{G_2},\eq,\et)=-d({\eq+\et})^2+(2d^2-d)\eq\et-2dm_{G_2}\cdot m_{G_2}.
\ee
All together, we checked that given $d=d_0+d'+d''$, the modularity of twisted elliptic blowup equation (\ref{eq:ebeq2}) holds, i.e.,
\be
\ba
\,&\mathrm{Ind}_{\rm cl}+\mathrm{Ind}_V^{G_2}+\mathrm{Ind}^{\mathbf{7}_{1/3} \oplus \mathbf{7}_{2/3}}_V+\mathrm{Ind}_{\IE}(d',m_{G_2}+\eq\alpha^\vee,\eq,\et-\eq)\\&+\mathrm{Ind}_{\IE}(d'',m_{G_2}+\et\alpha^\vee,\eq-\et,\et)
=\mathrm{Ind}_{\Lambda}+\mathrm{Ind}_{\IE}(d,m_{G_2},\eq,\et).
\ea
\ee

\subsection{Recursion formula for twisted elliptic genera}\label{sec:recursion}
In general, for $\fn\ge 3$ and paired matter content, there exist a sufficient number of unity blowup equations which together give a recursion formula for the elliptic genera with respect to the string number $d$. For the untwisted situation, this has been discussed extensively in Section 4.1 of \cite{Gu:2020fem}. For the current twisted situation, the recursion formula is very similar. We omit the derivation and only state the results here. It should be noted that for $\fn=1,2$, although there is no explicit recursion formula, the elliptic genera still can be solved for many theories, see Section 4.2 of \cite{Gu:2020fem}. For simplicity, in this subsection, we first discuss the four pure gauge twisted theories, then move to the theories with matters.

For the pure gauge twisted theories $G^{(n)}=\su(3)^{(2)},\so(8)^{(2)},\so(8)^{(3)},E_6^{(2)}$,
let us use the following shorthand notation for the classical term in blowup equations:
\begin{equation}\label{eq:classicalthetapure}
  \theta_{i,\{d_0,d_1,d_2\}}^{[a]}=
  \theta^{[a]}_i\(\fn\tau, -\fn m_{\alpha^{\vee}}
  +(\fn-2)(\eq+\et)-\fn((d_0+d_1)\eq+(d_0+d_2)\et)\) \,,
\end{equation}
where $m_{\alpha^\vee} = {m}\cdot \alpha^\vee$ and $m$ is the gauge fugacity for $\cirG$. Furthermore, we define
\begin{equation}\label{eq:Dd}
  D_d=\det\left(
    \begin{array}{ccc}
      \theta_{i,\{0,d,0\}}^{[a_1]}
      &\ \theta_{i,\{0,0,d\}}^{[a_1]} &\ \theta_{i,\{0,0,0\}}^{[a_1]}\\
      \theta_{i,\{0,d,0\}}^{[a_2]}
      &\ \theta_{i,\{0,0,d\}}^{[a_2]} &\ \theta_{i,\{0,0,0\}}^{[a_2]}\\
      \theta_{i,\{0,d,0\}}^{[a_3]}
      &\ \theta_{i,\{0,0,d\}}^{[a_3]} &\ \theta_{i,\{0,0,0\}}^{[a_3]}\\
    \end{array}
  \right),
\end{equation}
as well as
\begin{equation}\label{eq:Dda}
  D_{\{d_0,d_1,d_2\}}^{\alpha^{\vee}}=\det\left(
    \begin{array}{ccc}
      \theta_{i,\{0,d,0\}}^{[a_1]}
      &\ \theta_{i,\{0,0,d\}}^{[a_1]} &\ \theta_{i,\{d_0,d_1,d_2\}}^{[a_1]}\\
      \theta_{i,\{0,d,0\}}^{[a_2]}
      &\ \theta_{i,\{0,0,d\}}^{[a_2]} &\ \theta_{i,\{d_0,d_1,d_2\}}^{[a_2]}\\
      \theta_{i,\{0,d,0\}}^{[a_3]}
      &\ \theta_{i,\{0,0,d\}}^{[a_3]} &\ \theta_{i,\{d_0,d_1,d_2\}}^{[a_3]}\\
    \end{array}
  \right).
\end{equation}
Note that $D_d=D_{\{0,0,0\}}^{\alpha^{\vee}}$ does not depend on
$\alpha^{\vee}$.
Then the recursion formula of $d$-string twisted elliptic genera $\IE_d(m,\eq,\et)$ read
\begin{equation}\label{recursionZd}
\ba
  \IE_d=
  \sum_{d_0=\tfrac{1}{2}||\alpha^{\vee}||^2,\,d_{1,2}<d}^{d_0+d_1+d_2=d}
&  (-1)^{|\alpha^{\vee}|}
  \frac{D^{\alpha^{\vee}}_{\{d_0,d_1,d_2\}}}{D_d}
  A_{V}^{\cirG}( m,\alpha^{\vee})A_{V}^{frac}( m,\alpha^{\vee})\\[-3mm]
&\ \times  \IE_{d_1}(m+\eq {\alpha^{\vee}},\eq,\et-\eq)
  \IE_{d_2}( m+\et {\alpha^{\vee}},\eq-\et,\et).
  \ea
\end{equation}
Specializing to $d=1$, we obtain the following elegant formula for the twisted  one-string elliptic genera of all four pure gauge twisted theories:
\begin{equation}\label{Z1}
  \IE_1=\sum_{\alpha\in\Delta^\vee_l}
  % (-1)^{|\alpha^{\vee}|}
  \frac{D^{\alpha^{\vee}}_{\{1,0,0\}}}{D_1}
  \frac{\eta^4}
  {\theta_1(m_{\alpha})
    \theta_1(m_{\alpha}-\e_{1,2})
    \theta_1(m_{\alpha}-2\e_+)}
  \prod_{\displaystyle \substack{\beta\in \Delta\\
      \alpha \cdot \beta= 1}}
  \frac{\eta}{\theta_1(m_{\beta})}\prod^k_{\displaystyle \substack{\gamma\in \Delta_s\\
      \alpha \cdot \gamma= 1}}
  \frac{\eta}{\theta_1^{[k]}(m_{\gamma})}.
\end{equation}
Recall $\Delta_s$ denotes the set of short roots, while $k$ denotes all fractional KK-charges of $\bf F$ appearing in the twist of adjoint representation. For example, for $\so(8)^{(3)}$, $k$ takes values $1/3,2/3$.

For twisted theories with paired matters, we just need to add the hypermultiplet contribution to the above formulas. The classical term in \eqref{eq:classicalthetapure} is replaced as 
\be
\theta_{i}^{[a]}(\fn\tau,-\fn m_{\alpha^{\vee}}^{\cirG} +
    k_{\cirF} m_{\lambda}^{\cirF}+(y-\fn d_0)(\eq+\et)
    -\fn d'\eq-\fn d''\et),
\ee
with a fixed unity $\lambda=\lambda_{\cirF}$. Then $D_d$ and $D_{\{d_0,d_1,d_2\}}^{\alpha^{\vee}}$ are defined the same with the pure gauge cases \eqref{eq:Dd}, \eqref{eq:Dda}. The universal recursion formula for twisted $d$-string  elliptic genera $\IE_d(\mG,\mF,\eq,\et)$ read
\begin{align}\label{recursionZdmatter}\nonumber
  \IE_d=
  \sum_{d_0=\tfrac{1}{2}||\alpha^{\vee}||^2,\,d_{1,2}<d}^{d_0+d_1+d_2=d}
&  (-1)^{|\alpha^{\vee}|}
  \frac{D^{\alpha^{\vee}}_{\{d_0,d_1,d_2\}}}{D_d}
  A_{V}^{\cirG}( \mG,\alpha^{\vee})A_{V}^{frac}( \mG,\alpha^{\vee})A_{{H}}^{\mathring{R}}(\mG,\mF,\alpha^{\vee},\lambda)\\[-3mm]\nonumber
&\ \times  \IE_{d_1}(\mG+\eq {\alpha^{\vee}},\mF+\eq\lambda ,\eq,\et-\eq)\\
&\ \times  \IE_{d_2}( \mG+\et {\alpha^{\vee}},\mF+\et\lambda,\eq-\et,\et).
\end{align}
Specializing to $d=1$, we obtain the following universal formula for the twisted one-string elliptic genera:
\begin{equation}\label{Z1m}
\ba
  \IE_1=\sum_{\alpha\in\Delta^\vee_l}&
  % (-1)^{|\alpha^{\vee}|}
  \frac{D^{\alpha^{\vee}}_{\{1,0,0\}}}{D_1}
  \frac{\eta^4}
  {\theta_1(m_{\alpha})
    \theta_1(m_{\alpha}-\e_{1,2})
    \theta_1(m_{\alpha}-2\e_+)}\\
    &\times
  \prod_{\displaystyle \substack{\beta\in \Delta\\
      \alpha \cdot \beta= 1}}
  \frac{\eta}{\theta_1(m_{\beta})}\prod_{\displaystyle \substack{\gamma\in \Delta_s\\
      \alpha\cdot \gamma= 1}}^k
  \frac{\eta}{\theta_1^{[k]}(m_{\gamma})}
  \prod_{\displaystyle \substack{w \in R\\
      \alpha \cdot w= 1/2}}
  \frac{\theta_1^{[k']}(m_{w}+m_f+\epsilon_+)}{\eta}.
  \ea
\end{equation}
Here $k'$ are the KK-charges of the twisted matter content, and the $w$ are the weights of $\cirG$ and $m_f$ are the flavor fugacities. We use this formula to compute the twisted one-string elliptic genera of lots of 6d twisted theories. In most cases, our results are new.

Though by the recursion formula, one can compute the elliptic genera to arbitrary number of strings, we are particular interested in the \emph{reduced twisted one-string elliptic genera} defined by
\be\label{eq:defred}
\IE_1^{\rm red}(\tau,\epsilon_+)=\frac{\theta_1(\tau,\eq)\theta_1(\tau,\et)}{\eta(\tau)^2}\IE_1(\tau,\eq,\et).
\ee
The prefactor here comes from the well-known center of motion contribution in $\mathbb{C}_{\eq,\et}^2$, see e.g. \cite{DelZotto:2016pvm,DelZotto:2018tcj}. 
The simplicity of $\IE_1^{\rm red}$ lies in that it only depends on $\epsilon_+$ but not on $\epsilon_-$. Since there is no room for confusion, in the following we often omit the term ``reduced", and just say twisted one-string elliptic genera. The reduced twisted one-string elliptic genera exhibit many fascinating properties such as $\Gamma_1(N)$ modularity and spectral flow symmetry, which will be discussed later.

\subsection{Vanishing equations and vanishing theta identities}\label{sec:vanish}
The vanishing elliptic blowup equations, although less constraining for elliptic genera than the unity ones, are still rather non-trivial functional equations. In particular, the leading degree of vanishing elliptic blowup equations does not involve contribution from elliptic genera, but only the classical and one-loop parts. Therefore, the leading degree equations usually give some interesting vanishing theta identities depending on the root and weight lattices of the 6d gauge algebras $G$, see many examples in 
\cite{Gu:2019dan,Gu:2020fem}. In the twisted cases, we find many similar identities depending on the root and weight lattices of the truncated algebras $\cirG$. In this subsection, we will explicitly show some leading degree identities for the vanishing twisted elliptic blowup equations given in Table \ref{tb:v}. We have checked these identities to sufficient high orders in the $q$ expansions.

Let us first study the $\so(8)^{(2)}$ cases, where $\cirG=\so(7)$ and the vanishing $\lambda_{\so(7)}$ can take value in the dominant orbit $\cO_{1,6}$ of vector representation $\bf 7$. For pure $\so(8)^{(2)}$ theory with $\fn=4$, we derive the following leading degree vanishing theta identity:
\be
\sum_{w\in\cO_{1,6}}(-1)^{|w|}\frac{\theta_3^{[a]}(4\tau,4m_w)}{\theta_4(\tau,m_w)}{\prod_{\alpha\in\Delta(B_3)}^{w\cdot\alpha=1}\theta_1(\tau,m_{\alpha})^{-1}}=0.
\ee
Here $a=\pm 1/8,\pm 3/8$. It is easy to check for each $w\in\cO_{1,6}$, the summand has index $\frac42m_w^2-\frac12m_w^2-\frac52m_w^2=-m_w^2$ which is invariant for the whole Weyl orbit. We checked this identity to high $q$ orders. For $\fn=3$, we have one vector hypermultiplet with mass $m_v$ and two spinor hypermultiplets with mass $m_s$. Note $\forall w\in\cO_{1,6}$ and $w'\in \mathbf{8_s}$, $w\cdot w'=\pm1/2$. Thus the spinor hypermultiplets do not contribute to the leading degree vanishing theta identity. We then have
\be
\sum_{w\in\cO_{1,6}}(-1)^{|w|}\frac{\theta_4^{[a]}(3\tau,-3m_w+y)\theta_1(\tau,m_w+y)}{\theta_4(\tau,m_w)}{\prod_{\alpha\in\Delta(B_3)}^{w\cdot\alpha=1}\theta_1(\tau,m_{\alpha})^{-1}}=0.
\ee
Here $y=m_f+\e_+$ and $a=0,\pm 1/3$. It is easy to check for each $w$, the summand has index $\frac32(m_w-\frac13y)^2+\frac12(m_w+y)^2-\frac12m_w^2-\frac52m_w^2=-m_w^2+\frac23y^2$ which is invariant for the whole Weyl orbit. We also checked this identity for general $y$ to high $q$ orders. In fact, for 6d $(1,0)$ $\so(8)^{(2)}$ theories  with $\fn=1,2,3,4$, we find the following unified formula for the leading degree vanishing theta identities:
\be
\sum_{w\in\cO_{1,6}}(-1)^{|w|}\frac{\theta_i^{[a]}(\fn\tau,-\fn m_w+\sum y_j)\prod_{j=1}^{4-\fn}\theta_1(\tau,m_w+y_j)}{\theta_4(\tau,m_w)}{\prod_{\alpha\in\Delta(B_3)}^{w\cdot\alpha=1}\theta_1(\tau,m_{\alpha})^{-1}}=0.
\ee
Recall $i=3$ for even $n$ and $i=4$ for odd $n$. 

Consider the $\so(2r+8)^{(2)}$ theories  with $\fn=4$, where $\cirG=\so(2r+7)$ and the vanishing $\lambda_{B_{r+3}}$ can take value in the dominant orbit $\cO_{1,2(r+3)}$ of the vector representation. In these cases, the vector hypermultiplets can contribute to the leading degree of vanishing blowup equations. To be precise, we find the following leading degree vanishing theta identity for general $r$:
\be
\sum_{w\in\cO_{1,2(r+3)}}(-1)^{|w|}\frac{\theta_4^{[a]}(4\tau,-4m_w+\sum y_j)}{\theta_4(\tau,m_w)}{\prod_{\alpha\in\Delta(B_{r+3})}^{w\cdot\alpha=1}\theta_1(\tau,m_{\alpha})^{-1}}\prod_{j=1}^{2r}\theta_1(\tau,m_w+y_j)=0.
\ee
Here $a=\pm 1/8,\pm 3/8$ and $y_i=m_{f_i}+\e_+$. 
We checked this identity for various $r$ up to high $q$ orders.
More $\so$ type examples with $\IZ_2$ twist include the $\so(10)^{(2)}$ theory  with $\fn=2$, for which we find the following leading degree vanishing theta identities:
\be
\sum_{w\in\cO_{1,8}}(-1)^{|w|}\frac{\theta_3^{[a]}(2\tau,-2m_w+\sum y_i)\prod_{i=1}^{4}\theta_1(\tau,m_w+y_i)}{\theta_4(\tau,m_w)}{\prod_{\alpha\in\Delta(B_4)}^{w\cdot\alpha=1}\theta_1(\tau,m_{\alpha})^{-1}}=0.
\ee
Here $a=\pm1/4$. Besides, for $\so(12)^{(2)}$ theories with $\fn=2$, we find the following leading degree vanishing theta identities:
\be
\sum_{w\in\cO_{1,10}}(-1)^{|w|}\frac{\theta_3^{[a]}(2\tau,-2m_w+\sum y_i)\prod_{i=1}^{6}\theta_1(\tau,m_w+y_i)}{\theta_4(\tau,m_w)}{\prod_{\alpha\in\Delta(B_5)}^{w\cdot\alpha=1}\theta_1(\tau,m_{\alpha})^{-1}}=0.
\ee
We checked all these identities for general $y_i$ up to high $q$ orders.

Now consider the $\su({2r})^{(2)}$ theories with $\fn=2$ where $\cirG=\Sp(r)$. First, for $\su(2)^{(2)}$ theory, the leading vanishing identity is trivial. For $\su({4})^{(2)}$ theory, the leading degree vanishing identity is for $C_2$:
\be
\sum_{w\in\cO_{1,4}}(-1)^{|w|}\frac{\theta_3^{[a]}(2\tau,-2m_w)}{\theta_4(\tau,m_w)}{\prod_{\alpha\in\Delta(C_2)}^{w\cdot\alpha=1}\theta_1(\tau,m_{\alpha})^{-1}}=0,
\ee
where $a=\pm1/4$ and $\cO_{1,4}$ is the dominant Weyl orbit in $\bf 5$. We checked this identity to high $q$ orders. For arbitrary $r$, it is easy to find the leading degree vanishing theta  identities for $C_r$ can be uniformly written as
\be\label{n2sunvanish}
\sum_{w\in\cO_{1,2r}}(-1)^{|w|}{\theta_3^{[a]}(2\tau,-2m_w)}{\prod_{\alpha\in\Delta(C_r)}^{w\cdot\alpha=1}\theta_1(\tau,m_{\alpha})^{-1}}{\prod_{\beta\in \mathbf{\Lambda}^2}^{w\cdot\beta=1}\theta_4(\tau,m_{\alpha})^{-1}}=0.
\ee
Again $a=\pm1/4$ and the $ \cO_{1,2r}$ is the dominant Weyl orbit in the $\mathbf{\Lambda}^r$ representation of $C_r$. Note each $w\in\cO_{1,2r}$ intersect with all weights of $\mathbf{\Lambda}^2$ by $r(r-1)/2$ number of $1$, $r(r-1)/2$ number of $-1$, while the rest intersection numbers are zero. We have checked identity \eqref{n2sunvanish} for $r=2,3,4,5$ to high $q$ orders.

\section{Structure of twisted elliptic genera}\label{sec:spectralflow}
We use the twisted elliptic blowup equations and the recursion formula \eqref{Z1} introduced in the last section to compute the twisted one-string elliptic genera of all twisted 6d $(1,0)$ SCFTs. These data enable us to study some universal features of the twisted elliptic genera. Many of these features are the generalization of those of the ordinary $(0,4)$ elliptic genera observed by del Zotto and Lockhart in \cite{DelZotto:2016pvm,DelZotto:2018tcj}. According to the universal formula \eqref{eq:dc} and \eqref{eq:defred}, the reduced twisted one-string  R-R elliptic genera have the following form
\be\label{eq:dc1}
\IE_1(q,v,x,m_{\cirG},m_{\cirF})= q^{\frac16-c}\Big(Z_1^{5d}(v,x,m_{\cirG},m_{\cirF})+ \mathcal{O}(q^{1/n_G}) \Big),
\ee
where $Z_1^{5d}$ is the reduced one-instanton partition function of the 5d circle reduction theory.

\subsection{Spectral flow symmetry}\label{subsec:sfs}
Spectral flow is a characteristic feature of 2d $\mathcal{N}=(2,2)$ SCFTs, see e.g. \cite[Chapter 5.4]{Blumenhagen:2009zz}. The different sectors of 2d SCFTs are interpolated by a spectral flow parameter $\eta$.  When $\eta\in \IZ+\frac12$, the spectral flow interpolates the NS sector with half-integer modes and the R sector with integer modes. For chiral primary $|h,q\rangle$ of $(2,2)$ SCFT, the conformal weight $h$ and charge $q$ transform under spectral flow with general $\eta$ as
\be
h_{\eta}=h-\eta q+\frac{\eta^2}{6}c',\qquad q_\eta=q-\frac{c'}{3}\eta.
\ee
Here the $c'$ is the central charge of the 2d SCFT. 
We are interested in the  $\eta=\frac12$ case. Using $h=q/2$, the spectral flow induces transformation
\be
\Big|\frac{q}{2},q\Big\rangle_{\rm NS}\rightarrow \Big|\frac{c'}{24},q-\frac{c'}{6}\Big\rangle_{\rm R}.
\ee
Note the weight $\frac{c'}{24}$ indicates the ground state of the R sector. This phenomenon suggests an interesting transformation between NS-R elliptic genera and R-R elliptic genera.

Similar phenomena appear in 2d $(0,4)$ SCFTs as well \cite{DelZotto:2016pvm,DelZotto:2018tcj}.  
Let us focus on the 2d $(0,4)$ SCFTs of \emph{one} BPS string in 6d $(1,0)$ SCFTs, possibly with twisted compactification.  
In general, the spectral flow from R-R sector to NS-R sector for the \emph{same} 6d $(1,0)$ theory possibly with a twist is induced by the following transformation:
\be\boxed{
\IE_{\mathrm{NS-R}}^{\mathring{R}_{\mathrm{KK}}}\Big(q,v\Big)=\pm\Big(\frac{q^{1/4}}{v}\Big)^{\fn-\dualCox}\IE_{\mathrm{R-R}}^{\mathring{R}_{\mathrm{KK}}}\Big(q,\frac{q^{1/2}}{v}\Big).}
\ee
On the other hand, from the viewpoint of $(0,4)$ NLSM, the NS-R elliptic genera are obtained by imposing anti-periodicity on the 2d chiral fermions. 
This suggests the following elegant phenomenon for the twisted one-string elliptic genera: the spectral flow transformation shifts the KK charges of all hypermultiplets by half. To be precise, we find that for twisted one-string elliptic genera of different sectors:
\be
\boxed{
\IE_{\mathrm{NS-R}}^{\mathring{R}_{\mathrm{KK}\phantom{+1}}}(q,v)=\IE_{\mathrm{R-R}}^{\mathring{R}_{\mathrm{KK}+1/2}}(q,v).}
\ee
Note this is even true for the untwisted cases, which explains the NS-R elliptic genera in \cite{DelZotto:2018tcj} have simple low energy behaviors as 5d pure gauge theories. With this understanding, we know that the NS-R elliptic genera have similar expansion as \eqref{eq:dc1}, as long as both  the $c$ and $Z_1^{5d}$ are derived from the twisted matter content with all KK-charges shifted by half. 
When the twisted matter content is invariant under KK-charge shift by half,\footnote{For pure gauge theories, this is automatically satisfied.} the above two properties induce a nontrivial symmetry for the R-R elliptic genus itself, i.e., 
\be\label{sfgeneral}
\boxed{\IE_1\Big(q,\frac{q^{1/2}}{v}\Big)=\Big(-\frac{q^{1/4}}{v}\Big)^{\fn-\dualCox}\,\IE_1(q,v).}
\ee
We have checked this symmetry for all admissible twisted theories. The prefactor on the right hand side does not involve any information on the twist. 
This means that \emph{the spectral flow symmetry is preserved upon twisted compactification}. 

\subsection{The pure gauge cases}
We first discuss the four pure gauge cases $\su(3)^{(2)}$, $\so(8)^{(2)}$, $\so(8)^{(3)}$ and $E_6^{(2)}$. 
Denote $\theta$ and $f$ as the weight generating the adjoint and fundamental representation respectively. 
We find that for 6d $(1,0)$ pure gauge $G^{(n)}$ twisted theories with twist coefficient $N=2,3,4$, the reduced twisted one-string elliptic genera have the following universal expansion
\be
\ba
\IE_1^{G^{(n)}}=q^{\frac16-c}v^{h^\vee_{\mathring{G}}-1}\bigg(\sum_{n=0}^{\infty}\chi^{\mathring{G}}_{k\theta}v^{2n}
+q^{{\frac{1}{N}}}(1+v^2)\sum_{n=0}^{\infty}\chi^{\mathring{G}}_{k\theta+f}v^{2n}+O(q^{\frac{2}{N}})\bigg).
\ea
\ee
For the $N=2$ cases, we further find that the $q^1$ term in the above expansion is $\chi^{\mathring{G}}_{2f}+\chi^{\mathring{G}}_{\theta}+2+\mathcal{O}(v^2)$. 
We also find the following interesting \emph{spectral flow symmetry}:
\be\label{sfpure}
\IE_1^{G^{(n)}}\Big(q,\frac{q^{1/2}}{v}\Big)=\Big(-\frac{q^{1/2}}{v^2}\Big)^{\fn-3}\,\IE_1^{G^{(n)}}(q,v),
\ee
as a special case of \eqref{sfgeneral}. This resembles the spectral flow symmetry for the reduced one-string elliptic genera associated to untwisted 6d $(1,0)$ SCFTs studied in \cite{DelZotto:2016pvm}.

\subsubsection{$\su(3)^{(2)}$}\label{sec:su3z2}
The 6d $(1,0)$ pure $\su(3) $ SCFT enjoys a series of fascinating 2d quiver constructions for the worldsheet theories of the BPS strings such that the elliptic genera are exactly computable to arbitrary number of strings \cite{Kim:2018gjo}. The $\IZ_2$ twisted circle compactification gives the 5d $\su(3) $ gauge theory with Chern-Simons level 9 \cite{Jefferson:2017ahm}. The blowup equations for this twisted theory have been studied before in \cite[Section 3.2.3]{Kim:2020hhh}, which are consistent with our universal elliptic form. As we have reviewed, this twisted theory can be Higgsed from 6d $(1,0)$ $G_2+\bf F$ theory \cite{Kim:2021cua}, thus the twisted elliptic genera are also exactly computable to  arbitrary number of strings. Here we provide a new observation that is the spectral flow symmetry of its twisted one-string elliptic genus.

For $\su(3)^{(2)}$, the low energy gauge algebra is $\cirG=\su(2)$. 
Let us denote the reduced twisted one-string elliptic genus as
\be
\IE_1=q^{-\frac{7}{48}}\sum_{i=n}^{\infty}q^n g_n\left(v,m_{\su(2)}\right)=q^{-\frac{7}{48}}\sum_{i=0}^{\infty}b_{ij}q^i v^{-2i+j},\qquad 4n\in\IZ.
\ee
We calculate $g_n$ up to $n=4$ from the unity twisted elliptic blowup equations with characteristics $a=0,\pm1/3$. The results are in perfect agreement with the computations in \cite{Kim:2020hhh,Kim:2021cua,Hayashi:2020hhb}. For example, when turning off the $\su(2)$ fugacity, we have
\begin{align}
g_0&=\frac{v(v^2+1)}{\left(v^2-1\right)^2},\qquad g_{\frac14}=\frac{2  v (v^2+1)}{\left(v^2-1\right)^2},\qquad g_{\frac12}=\frac{ (v^6+2 v^4+2 v^2+1)}{v \left(v^2-1\right)^2}, \\
g_{\frac34}&=\frac{2  (v^6+2 v^4+2 v^2+1)}{v \left(v^2-1\right)^2},\qquad g_1=\frac{ \left(4 v^6+9 v^4+9 v^2+4\right)}{v \left(v^2-1\right)^2}. 
\end{align}
With the $\su(2)$ fugacity, we find
\begin{align}
g_0&=v\sum_{n=0}^{\infty}\chi^{A_1}_{n\theta}v^{2n}=v+\mathbf{3}v^3+\mathbf{5}v^5+\mathbf{7}v^7+\dots,\\
g_{\frac14}&=v(1+v^2)\sum_{n=0}^{\infty}\chi^{A_1}_{f+n\theta}v^{2n}=\mathbf{2}v+(\mathbf{2}+\mathbf{4})v^3+(\mathbf{4}+\mathbf{6})v^5+(\mathbf{6}+\mathbf{8})v^7+\dots,\\
g_{\frac12}&=v^{-1}(1+v^2+v^4)\sum_{n=0}^{\infty}\chi^{A_1}_{n\theta}v^{2n}=v^{-1}+(\mathbf{1}+\mathbf{3})v+(\mathbf{1}+\mathbf{3}+\mathbf{5})v^3
+\dots,\\
g_{\frac34}&=v^{-1}(1+2v^2+2v^4+v^6)\sum_{n=0}^{\infty}\chi^{A_1}_{f+n\theta}v^{2n}=\mathbf{2}v^{-1}+(2\times\mathbf{2}+\mathbf{4})v
+\dots.
\end{align}
The $g_0$ result shows that the leading $q$ order of the twisted one-string elliptic genus gives the one $\su(2)$ instanton Hilbert series \cite{Benvenuti:2010pq}, which is expected since the circle reduction of the 6d $\su(3)^{(2)}$ theory should be a pure $\su(2)$ theory with $\theta$ angle $0$. We also have the following Table \ref{tab:bijsu3z2} for the coefficients $b_{ij}$. The manifest symmetry of the coefficient matrix suggests the following spectral flow symmetry of the twisted one-string elliptic genus
\be\label{spectralflowA2}
\IE_1^{\su(3)^{(2)}}(q,{q^{1/2}}/{v})= \IE_1^{\su(3)^{(2)}}(q,v).
\ee
This resembles the spectral flow symmetry of the one-string elliptic genus of untwisted 6d  $\su(3)$ theory observed in \cite{DelZotto:2016pvm}. It is actually easy to prove this identity from the exact formula of $\IE_1$ in \eqref{su3z2Ed}.

\begin{table}[h]
\begin{center}
$\begin{tabu}{c|ccccccccccccccccccc}
q\backslash v & 0 & & 1 & & 2 & & 3 & & 4 & & 5 & & 6 & & 7 & & 8 \\ \hline
0 & \phantom{.}0\phantom{.} & 0 & \color{orange}1 & 0 & 0 & 0 & \color{orange}3 & 0 & 0 & 0 & \color{orange}5 & 0 & 0 & 0 & \color{orange}7 & 0 & 0  \\
1/4& 0 & 0 & 0 & 2 & 0 & 0 & 0 & 6 & 0 & 0 & 0 & 10 & 0 & 0 & 0 & 14 & 0  \\
1/2& \color{orange}1 & 0 & 0 & 0 & 4 & 0 & 0 & 0 & 9 & 0 & 0 & 0 & 15 & 0 & 0 & 0 & 21 \\
3/4& 0 & 2 & 0 & 0 & 0 & 8 & 0 & 0 & 0 & 18 & 0 & 0 & 0 & 30 & 0 & 0 & 0  \\
1& 0 & 0 & 4 & 0 & 0 & 0 & 17 & 0 & 0 & 0 & 39 & 0 & 0 & 0 & 65 & 0 & 0  \\
5/4& 0 & 0 & 0 & 8 & 0 & 0 & 0 & 30 & 0 & 0 & 0 & 66 & 0 & 0 & 0 & 110 & 0  \\
3/2& \color{orange}3 & 0 & 0 & 0 & 17 & 0 & 0 & 0 & 51 & 0 & 0 & 0 & 105 & 0 & 0 & 0 & 170  \\
7/4& 0 & 6 & 0 & 0 & 0 & 30 & 0 & 0 & 0 & 86 & 0 & 0 & 0 & 174 & 0 & 0 & 0  \\
2& 0 & 0 & 9 & 0 & 0 & 0 & 51 & 0 & 0 & 0 & 147 & 0 & 0 & 0 & 297 & 0 & 0  \\
9/4& 0 & 0 & 0 & 18 & 0 & 0 & 0 & 86 & 0 & 0 & 0 & 236 & 0 & 0 & 0 & 468 & 0  \\
5/2& \color{orange}5 & 0 & 0 & 0 & 39 & 0 & 0 & 0 & 147 & 0 & 0 & 0 & 370 & 0 & 0 & 0 & 708  \\
11/4& 0 & 10 & 0 & 0 & 0 & 66 & 0 & 0 & 0 & 236 & 0 & 0 & 0 & 576 & 0 & 0 & 0  \\
3& 0 & 0 & 15 & 0 & 0 & 0 & 105 & 0 & 0 & 0 & 370 & 0 & 0 & 0 & 896 & 0 & 0  \\
\end{tabu}$
\caption{The coefficient matrix $b_{ij}$ for the  $\IE_1$ of 6d pure $\su(3)^{(2)}$ theory.}
\label{tab:bijsu3z2}
\end{center}
\end{table}

Interestingly, we notice that the twisted elliptic blowup equations for $\su(3)^{(2)}$ allow another solution for the twisted one-string elliptic genera if we set the characteristic $a=1/2,\pm 1/6$. We call this as the $\Sp(1)_{\pi}$ case, as the low energy limit gives the one-instanton partition function of 5d $\mathcal{N}=1$ pure $\Sp(1)_{\pi}$ theory. In this case, combining together the twisted elliptic blowup equations and modular ansatz which will be discussed later, we solve the twisted one-string elliptic genus with the $\Sp(1)$ fugacity turned off. Let us denote the reduced twisted one-string elliptic genus as
\be
\IE_1^{\pi}(q,v)=q^{-\frac{7}{48}}\sum_{i=0}^{\infty}q^n g_n^{\pi}\left(v,m_{\su(2)}\right)=q^{-\frac{7}{48}}\sum_{i=0}^{\infty}b_{ij}^{\pi}q^i v^{-2i+j},\qquad 4n\in\IZ.
\ee
We calculate $g_n^{\pi}$ up to $n=4$. For example,
\be
\ba
g_0^{\pi}&=\frac{2v^2}{\left(v^2-1\right)^2},\qquad g_{\frac14}^{\pi}=\frac{4  v^2 }{\left(v^2-1\right)^2},\qquad g_{\frac12}^{\pi}=\frac{ 6v^2}{ \left(v^2-1\right)^2}, \\
g_{\frac34}^{\pi}&=\frac{12v^2}{ \left(v^2-1\right)^2},\qquad g_1^{\pi}=\frac{ 2(v^8+11 v^4+1)}{v^2 \left(v^2-1\right)^2}. 
\ea
\ee
With the gauge $\su(2)$ fugacity, we find
\be
\ba
g_0^{\pi}=v^2\sum_{n=0}^{\infty}\chi^{\su(2)}_{n\theta+f}v^{2n}=\mathbf{2}v^2+\mathbf{4}v^4+\mathbf{6}v^6+\dots.
\ea
\ee
This shows the leading $q$ order of twisted elliptic genus $\IE_1^{\pi}$ is indeed the 5d one $\Sp(1)_{\pi}$ instanton partition function. We also have the Table \ref{tab:bijsu3z2pi} for the coefficients $b_{ij}^\pi$. The manifest symmetry of the coefficient matrix shows that the solution $\IE_1^{\pi}$ also satisfies the spectral flow symmetry \eqref{spectralflowA2}. We regard this as further evidence that this serves as a good elliptic completion of 5d one $\Sp(1)_{\pi}$ instanton partition function. 
However, to push forward to two-strings, we meet some inconsistency when solving the blowup equations. It is not clear to us whether there would exist a $\IZ_2$ twist of the two-string elliptic genus of 6d $(1,0)$ pure $\SU(3)$ theory that has low energy limit as 5d two $\Sp(1)_{\pi}$ instantons. 

\begin{table}[h]
\begin{center}
$\begin{tabu}{c|ccccccccccccccccccc}
q\backslash v & 0 & & 1 & & 2 & & 3 & & 4 & & 5 & & 6 & & 7 & & 8 \\ \hline
0& \phantom{.}0\phantom{.} & 0 & 0 & 0 & \color{orange}2 & 0 & 0 & 0 & \color{orange}4 & 0 & 0 & 0 & \color{orange}6 & 0 & 0 & 0 & \color{orange}8 \\
1/4& 0 & 0 & 0 & 0 & 0 & 4 & 0 & 0 & 0 & 8 & 0 & 0 & 0 & 12 & 0 & 0 & 0 \\
1/2& 0 & 0 & 0 & 0 & 0 & 0 & 6 & 0 & 0 & 0 & 12 & 0 & 0 & 0 & 18 & 0 & 0 \\
3/4& 0 & 0 & 0 & 0 & 0 & 0 & 0 & 12 & 0 & 0 & 0 & 24 & 0 & 0 & 0 & 36 & 0 \\
1& \color{orange}2 & 0 & 0 & 0 & 4 & 0 & 0 & 0 & 28 & 0 & 0 & 0 & 52 & 0 & 0 & 0 & 78 \\
5/4& 0 & 4 & 0 & 0 & 0 & 8 & 0 & 0 & 0 & 48 & 0 & 0 & 0 & 88 & 0 & 0 & 0 \\
3/2& 0 & 0 & 6 & 0 & 0 & 0 & 12 & 0 & 0 & 0 & 74 & 0 & 0 & 0 & 136 & 0 & 0 \\
7/4& 0 & 0 & 0 & 12 & 0 & 0 & 0 & 24 & 0 & 0 & 0 & 124 & 0 & 0 & 0 & 224 & 0 \\
2& \color{orange}4 & 0 & 0 & 0 & 28 & 0 & 0 & 0 & 52 & 0 & 0 & 0 & 220 & 0 & 0 & 0 & 388 \\
9/4& 0 & 8 & 0 & 0 & 0 & 48 & 0 & 0 & 0 & 88 & 0 & 0 & 0 & 348 & 0 & 0 & 0 \\
5/2& 0 & 0 & 12 & 0 & 0 & 0 & 74 & 0 & 0 & 0 & 136 & 0 & 0 & 0 & 520 & 0 & 0 \\
11/4& 0 & 0 & 0 & 24 & 0 & 0 & 0 & 124 & 0 & 0 & 0 & 224 & 0 & 0 & 0 & 804 & 0 \\
3& \color{orange}6 & 0 & 0 & 0 & 52 & 0 & 0 & 0 & 220 & 0 & 0 & 0 & 388 & 0 & 0 & 0 & 1272 \\
\end{tabu}$
\caption{The coefficient matrix $b_{ij}$ for the  $\IE_1^{\pi}$ of 6d pure $\su(3)^{(2)}$ theory.}
\label{tab:bijsu3z2pi}
\end{center}
\end{table}

\subsubsection{$\so(8)^{(2)}$}\label{sec:so8z2}
The $\IZ_2$ twisted circle compactification of 6d $(1,0)$ $\so(8)$ SCFT gives a non-Lagrangian 5d KK theory. The twisted elliptic genera can be exactly computed to an arbitrary number of strings by 2d localization \cite{Kim:2021cua}. The unity blowup equations for this twisted theory have been studied before in \cite[Section 6.2]{Kim:2020hhh}, which are consistent with our elliptic form. 
Using our universal $\IE_1$ formula (\ref{Z1}), we calculate the twisted one-string elliptic genus for this $\so(8)^{(2)}$ theory and find it in complete agreement with the 2d localization results. Let us denote the reduced twisted one-string elliptic genus of $\so(8)^{(2)}$ as
\be
\IE_1=q^{-\frac{7}{12}}\sum_{n=0}^{\infty}q^n g_n(v,m_{\so(7)})=q^{-\frac{7}{12}}\sum_{i,j=0}^{\infty}b_{ij}q^{{i}} v^{-2i+j},\qquad 2n\in \IZ.
\ee
When turning off the $\so(7)$ fugacities, we obtain
\begin{align}
g_0=&\,\frac{v^4 \left(v^8+13 v^6+28 v^4+13 v^2+1\right)}{\left(1-v^2\right)^8},\\
g_{\frac12}=&\,\frac{7 v^4 \left(v^2+1\right)^2 \left(v^4+6 v^2+1\right)}{\left(1-v^2\right)^8},\\
g_1=&\,\frac{2 v^4 \left(25 v^8+189 v^6+300 v^4+189 v^2+25\right)}{\left(1-v^2\right)^8}.
\end{align}
With the $\so(7)$ gauge fugacities, we have
\be
g_0=v^4\sum_{n=0}^{\infty}\chi_{n\theta}^{\so(7)}v^{2n}=v^4+\mathbf{21}v^6+ \mathbf{168} v^8 + \mathbf{825} v^{10} + \mathbf{3003} v^{12} + \dots.
\ee
This is exactly the one $\so(7)$ instanton Hilbert series \cite{Benvenuti:2010pq}, thus agrees with the expectation that the circle reduction of the twisted theory should be a 5d pure $\SO(7)$ gauge theory. Besides, we have
\be\nonumber
g_{\frac12}=v^4(1+v^2)\sum_{n=0}^{\infty}\chi^{\so(7)}_{f+n\theta}v^{2n}=\mathbf{7} v^4+({\mathbf{105}+\mathbf{7}}) v^6+(\mathbf{798}+\mathbf{105}) v^8+(\mathbf{3696}+\mathbf{798}) v^{10}+\dots
\ee
and $g_{1}=(\mathbf{27}+\mathbf{21}+2) v^4+\dots$. We also have the Table \ref{tab:bijso8z2} for the Fourier coefficients $b_{ij}$. The representations $\chi_{n\theta}^{\so(7)}$ are colored in orange. 
The manifest anti-symmetric coefficient matrix implies the following {spectral flow symmetry}
\be
\IE_1^{\so(8)^{(2)}}(q,{q^{1/2}}/{v})=-q^{1/2}v^{-2}\,\IE_1^{\so(8)^{(2)}}(q,v).
\ee
\begin{table}[h]
\begin{center}
$\begin{tabu}{c|cccccccccccc}
q\backslash v&  1 & 2 & 3 & 4 & 5 & 6 & 7 & 8 & 9 & 10& 11\\ \hline
0& 0 & 0 & 0 & \color{orange}1 & 0 & \color{orange}21 & 0 & \color{orange}168 & 0 & \color{orange}825 & 0 \\
1/2 & 0 & 0 & 0 & 0 & 7 & 0 & 112 & 0 & 798 & 0 & 3696  \\
1& 0 & 0 & 0 & 0 & 0 & 50 & 0 & 778 & 0 & 5424 & 0 \\
3/2& \color{orange}-1 & 0 & 0 & 0 & 1 & 0 & 238 & 0 & 3206 & 0 & 20860  \\
2& 0 & -7 & 0 & -1 & 0 & 7 & 0 & 1101 & 0 & 14225 & 0 \\
5/2& \color{orange}-21 & 0 & -50 & 0 & -7 & 0 & 50 & 0 & 4242 & 0 & 49756 \\
3& 0 & -112 & 0 & -238 & 0 & -50 & 0 & 238 & 0 & 15802 & 0  \\
7/2& \color{orange}-168 & 0 & -778 & 0 & -1101 & 0 & -238 & 0 & 1101 & 0 & 52900  \\
4& 0 & -798 & 0 & -3206 & 0 & -4242 & 0 & -1101 & 0 & 4221 & 0  \\
9/2& \color{orange}-825 & 0 & -5424 & 0 & -14225 & 0 & -15802 & 0 & -4221 & 0 & 15690 \\
5& 0 & -3696 & 0 & -20860 & 0 & -49756 & 0 & -52900 & 0 & -15690 & 0  \\
\end{tabu}$
\caption{The coefficient matrix $b_{ij}$ for the  $\IE_1$ of 6d pure $\so(8)^{(2)}$ theory.}
\label{tab:bijso8z2}
\end{center}
\end{table}

\subsubsection{$\so(8)^{(3)}$}\label{sec:so8z3}
The $\IZ_3$ twisted circle compactification of 6d $(1,0)$ $\so(8)$ SCFT gives the 5d $\su(4) $ gauge theory with Chern-Simons level 8 \cite{Jefferson:2017ahm}. The circle reduction on the other hand gives a 5d pure $G_2$ gauge theory. The worldsheet theories of the BPS strings for 6d $\so(8)^{(3)}$ theory do not have 2d quiver descriptions, thus the computation of elliptic genera is quite nontrivial. The blowup equations for this twisted theory have been studied before in \cite[Section 6.1]{Kim:2020hhh}, which are consistent with our elliptic form. There are in total four unity blowup equations and no vanishing one. Let us denote the reduced twisted one-string elliptic genus as
\be
\IE_1=q^{-\frac{7}{18}}\sum_{n=0}^{\infty}q^n g_n(v,m_{G_2})=q^{-\frac{7}{18}}\sum_{i,j=0}^{\infty}b_{ij}q^{{i}} v^{-2i+j},\qquad 3n\in\IZ.
\ee
From our universal $\IE_1$ formula (\ref{Z1}), we calculate $g_n$ up to $n=4$. For example, turning off the $G_2$ fugacities, we have
\be
\ba
g_0=&\,\frac{v^9+8 v^7+8 v^5+v^3}{\left(1-v^2\right)^6}=v^3+14 v^5+77 v^7+273 v^9+\dots,\\
g_{\frac13}=&\,\frac{v^3 \left(v^2+1\right) \left(7 v^4+22 v^2+7\right)}{\left(1-v^2\right)^6}=7 v^3+71 v^5+350 v^7+\dots,\\
g_{\frac23}=&\,\frac{7  v^3 \left(v^2+1\right) \left(5 v^4+8 v^2+5\right)}{\left(1-v^2\right)^6}=35 v^3+301 v^5+1372 v^7+\dots.
\ea
\ee
Keeping the $G_2$ gauge fugacities, we find the following exact formulas by the infinite summations of $G_2$ characters
\be\nonumber
\ba
g_{0}&= v^3\sum_{n=0}^{\infty}\chi_{n\theta}^{G_2}v^{2n}, \qquad\quad  
g_{\frac13}= v^3(1+v^2)\sum_{n=0}^{\infty}\chi^{G_2}_{f+n\theta}v^{2n} ,\\
g_{\frac23}&=\sum_{n=0}^{\infty}(\chi^{G_2}_{2f+n\theta}+\chi^{G_2}_{f+n\theta}+\chi^{G_2}_{n\theta})v^{2n+3}+\sum_{n=0}^{\infty}(\chi^{G_2}_{2f+n\theta}+\chi^{G_2}_{f+n\theta})v^{2n+5}+\sum_{n=0}^{\infty}\chi^{G_2}_{2f+n\theta}v^{2n+7}.
\ea
\ee
The $g_0$ formula shows that the circle reduction of 6d $\so(8)^{(3)}$ theory is indeed a 5d pure $G_2$ theory. 
We also have the Table \ref{tab:bijso8z3} for the coefficients $b_{ij}$, $i,j=0,1,2,\dots$, where the coefficients $\chi_{n\theta}^{G_2}$ in $g_0$ in colored orange. From the obvious anti-symmetry coefficient matrix, we find the following {spectral flow symmetry}
\be
\IE_1^{\so(8)^{(3)}}(q,{q^{1/2}}/{v})=-q^{1/2}v^{-2}\,\IE_1^{\so(8)^{(3)}}(q,v).
\ee

\begin{table}[h]
\begin{center}
$\begin{tabu}{c|ccccccccccccc}
q\backslash v & 1 & & & 3 & & & 5 & & & 7 & & \\ \hline
0& 0 & 0 & 0 & \color{orange}1 & 0 & 0 & \color{orange}14 & 0 & 0 & \color{orange}77 & 0 & 0  \\
1/3& 0 & 0 & 0 & 0 & 7 & 0 & 0 & 71 & 0 & 0 & 350 & 0  \\
2/3& 0 & 0 & 0 & 0 & 0 & 35 & 0 & 0 & 301 & 0 & 0 & 1372  \\
1& \color{orange}-1 & 0 & 0 & 0 & 0 & 0 & 141 & 0 & 0 & 1127 & 0 & 0 \\
4/3& 0 & -7 & 0 & 0 & 0 & 0 & 0 & 497 & 0 & 0 & 3648 & 0  \\
5/3& 0 & 0 & -35 & 0 & 0 & 0 & 0 & 0 & 1582 & 0 & 0 & 10836  \\
2& \color{orange}-14 & 0 & 0 & -141 & 0 & 0 & 0 & 0 & 0 & 4650 & 0 & 0  \\
7/3& 0 & -71 & 0 & 0 & -497 & 0 & 0 & 0 & 0 & 0 & 12838 & 0  \\
8/3& 0 & 0 & -301 & 0 & 0 & -1582 & 0 & 0 & 0 & 0 & 0 & 33621  \\
3& \color{orange}-77 & 0 & 0 & -1127 & 0 & 0 & -4650 & 0 & 0 & 0 & 0 & 0  \\
10/3& 0 & -350 & 0 & 0 & -3648 & 0 & 0 & -12838 & 0 & 0 & 0 & 0 \\
11/3& 0 & 0 & -1372 & 0 & 0 & -10836 & 0 & 0 & -33621 & 0 & 0 & 0 \\
\end{tabu}$
\caption{The coefficient matrix $b_{ij}$ for the  $\IE_1$ of 6d pure $\so(8)^{(3)}$ theory.}
\label{tab:bijso8z3}
\end{center}
\end{table}

\subsubsection{$E_6^{(2)}$}\label{sec:e6z2}

The $\IZ_2$ twisted circle compactification of 6d $(1,0)$ $E_6$ SCFT gives a non-Lagrangian 5d KK theory, while the circle reduction gives a 5d pure $F_4$ gauge theory. This twisted theory has six unity blowup equations and no vanishing one.  
The twisted elliptic genera of this theory have not been computed before. 
We use the universal $\IE_1$ formula (\ref{Z1}) to compute the twisted one-string elliptic genus. Denote the reduced twisted elliptic genus as
\be
\IE_1=q^{-\frac{13}{12}}\sum_{n=0}^{\infty}q^n g_n(v,m_{F_4})=q^{-\frac{13}{12}} \sum_{i,j=0}^{\infty}b_{ij}q^{{i} } v^{-2i+j}, \qquad 2n\in\IZ.
\ee
Turning off the $F_4$ fugacities, we obtain 
\be
\ba
g_0=&\,\frac{v^8 (v^{16}+36 v^{14}+341 v^{12}+1208 v^{10}+1820 v^8+1208 v^6+341 v^4+36 v^2+1)}{(1-v^2)^{16}},\\
g_{\frac12}=&\,\frac{13 v^8 \left(v^2+1\right)^2 \left(2 v^{12}+47 v^{10}+274 v^8+506 v^6+274 v^4+47 v^2+2\right)}{\left(1-v^2\right)^{16}},\\
g_1=&\,\frac{2 v^8 }{\left(1-v^2\right)^{16}}\big(189 v^{16}+3931 v^{14}+24233 v^{12}+64761 v^{10}+88332 v^8+64761 v^6\\
&\phantom{-----}+24233 v^4+3931 v^2+189\big).
\ea
\ee
When keeping the $F_4$ fugacities, we have
\be
g_0=v^8\sum_{n=0}^{\infty}\chi_{n\theta}^{F_4}v^{2n}=v^8+\mathbf{52} v^{10}+\mathbf{1053} v^{12}+\mathbf{12376} v^{14}+\dots.
\ee
This is exactly the one $F_4$ instanton Hilbert series \cite{Benvenuti:2010pq}, thus agrees with the expectation that the circle dimension reduction  should be a 5d pure $F_4$ gauge theory. Besides, we find
\be
g_{\frac12}=v^8(1+v^2)\sum_{n=0}^{\infty}\chi^{F_4}_{n\theta+f}v^{2n}=\mathbf{26} v^8+({\mathbf{1053}+\mathbf{26}}) v^{10}+(\mathbf{17901}+\mathbf{1053}) v^{12}
+\dots
\ee
and $g_{1}=(\mathbf{324}+\mathbf{52}+2) v^8+\dots$. 
We also have the Table \ref{tab:bijE6z2} for the coefficients $b_{ij}$. From the anti-symmetric coefficient matrix, we observe the following {spectral flow symmetry}
\be
\IE_1^{E_6^{(2)}}(q,{q^{1/2}}/{v})=-q^{3/2}v^{-6}\,\IE_1^{E_6^{(2)}}(q,v).
\ee

\begin{table}[h]
\begin{center}
$\begin{tabu}{c|ccccccccccc}
q\backslash v&  3 & 4 & 5 & 6 & 7 & 8 & 9 & 10 & 11 & 12 & 13\\ \hline
0& 0 & 0 & 0 & 0 & 0 & \color{orange}1 & 0 & \color{orange}52 & 0 & \color{orange}1053 & 0 \\
1/2& 0 & 0 & 0 & 0 & 0 & 0 & 26 & 0 & 1079 & 0 & 18954 \\
1& 0 & 0 & 0 & 0 & 0 & 0 & 0 & 378 & 0 & 13910 & 0 \\
3/2& 0 & 0 & 0 & 0 & 0 & 0 & 0 & 0 & 4056 & 0 & 134342 \\
2& 0 & 0 & 0 & 0 & 0 & -1 & 0 & 0 & 0 & 35362 & 0 \\
5/2& \color{orange}-1 & 0 & 0 & 0 & \phantom{-}1\phantom{-}  & 0 & -26 & 0 & 0 & 0 & 264576 \\
3& 0 & -26 & 0 & 0 & 0 & 26 & 0 & -378 & 0 & 0 & 0 \\
7/2& \color{orange}-52 & 0 & -378 & 0 & 0 & 0 & 378 & 0 & -4004 & 0 & 0 \\
4& 0 & -1079 & 0 & -4056 & 0 & 0 & 0 & 4004 & 0 & -34283 & 0 \\
5/2& \color{orange}-1053& 0& -13910& 0& -35362& 0& 0& 0& 34283& 0& \!-250666\!\\
5& 0& -18954& 0& \!-134342\!& 0& \!-264576\!& 0& 0& 0& 250666& 0 \\
\end{tabu}$
\caption{The coefficient matrix $b_{ij}$ for the  $\IE_1$ of 6d pure $E_6^{(2)}$ theory.}
\label{tab:bijE6z2}
\end{center}
\end{table}

\subsection{Universal features with matters}
In this section, we discuss the universal features of the reduced twisted one-string elliptic genera for twisted 6d theories with matters. Most of our observations are the twisted generalization of those in \cite{DelZotto:2018tcj}. These include the spectral flow symmetry and the precise spectrum of the twisted R-R vacuum, NS-R vacuum and possibly the excited states with fractional KK-charges above the  NS-R vacuum. For all reduced twisted one-string elliptic genera we solved from blowup equations, we analyze these features in the $(q,v)$ expansion:
\be
\IE_1(q,v)=q^{\frac{1}{6}-c}\sum_{i,j=0}^{\infty}c_{ij}q^{{i}} v^{j}=q^{\frac{1}{6}-c}\sum_{i,j=0}^{\infty}b_{ij}q^{{i}} v^{-2i+j},\qquad Ni,j\in\IZ,
\ee
where $N$ is twist coefficient. All coefficients $c_{ij}$ should be integers as they represent the states in the 2d worldsheet theories, in particular form representations of $\cirG$ and $\cirF$ when turning on the fugacities. The $b_{ij}$ coefficients are slightly shifted from the $c_{ij}$ coefficients. 
As we have seen in the pure gauge cases, the $b_{ij}$ coefficients are very useful to show explicitly the spectral flow symmetry. Therefore, we will show the $b_{ij}$ coefficient matrix for all examples in the following.

\subsubsection{$\fn=4,\so{(2r+8)}^{(2)}$}
The $\IZ_2$ twist of 6d $(1,0)$ $\so{(2r+8)}+2r\bf V$ theory has low energy gauge algebra $\cirG=\so{(2r+7)}$ and twisted matter content $2r\mathbf{V}_0$. The circle reduction gives a 5d $\mathcal{N}=1$ $\so{(2r+7)}+2r\bf F$ theory whose 5d Nekrasov partition function is exactly computable by localization. On the other hand, the circle reduction of the NS-R sector gives a 5d $\mathcal{N}=1$ pure $\so{(2r+7)}$ theory. For all $\fn=4,\so{(2r+8)}^{(2)}$ theories with $r\ge 0$, we use the recursion formula \eqref{Z1m} to compute the one-string elliptic genera and find complete agreement with the 2d localization formulas. 
Let us study the following expansion for the reduced twisted one-string elliptic genera
\be
\IE_1^{\so{(2r+8)}^{(2)}}(q,v)=q^{-\frac{7}{12}}\sum_{i,j=0}^{\infty}c_{ij}q^{{i}} v^{j}=q^{-\frac{7}{12}}\sum_{i,j=0}^{\infty}b_{ij}q^{{i}} v^{-2i+j}.
\ee
We observe the following patterns for the Fourier coefficients $c_{ij}$:
\begin{itemize} 
    \item $c_{0,j}=0$ for $j<2r+4$. This is consistent with the fact that the reduced one-instanton partition function of 5d $ \so{(2r+7)}+2r\bf F$ theory has $v$ expansion starting from $v^{2r+4}$.  The $v$ expansion coefficients are colored orange in the Tables \ref{tab:bijn4so10z2}, \ref{tab:bijn4so12z2} and \ref{tab:bijn4so14z2}.
    \item $c_{\frac32+n,-2-2n}=\chi^{\so{(2r+7)}}_{n\theta}$ and $c_{2+n,-3-2n}=0$. This shows that the leading $q$ order of the NS-R twisted elliptic genera indeed gives the 5d one $\so{(2r+7)}$ instanton Hilbert series \cite{Benvenuti:2010pq}. The coefficients are colored red in the Tables \ref{tab:bijn4so10z2}, \ref{tab:bijn4so12z2} and \ref{tab:bijn4so14z2}.
    \item $c_{\frac52+n,-1-2n}=-\chi^{\so{(2r+7)}}_{f+n\theta}\chi^{\Sp{(2r)}}_f$. These are the first excited states with integer modes and a half KK-charge above the NS-R vacuum. The coefficients are colored blue in the Tables \ref{tab:bijn4so10z2}, \ref{tab:bijn4so12z2} and \ref{tab:bijn4so14z2}.
    \item $c_{2+n,-2-2n}=\chi^{\so{(2r+7)}}_{f+n\theta}+\chi^{\so{(2r+7)}}_{f+(n-1)\theta}$. These are the first excited states with half-integer modes and a half KK-charge above the NS-R vacuum. The coefficients are colored cyan in the Tables \ref{tab:bijn4so10z2}, \ref{tab:bijn4so12z2} and \ref{tab:bijn4so14z2}. 
   \item $c_{2+n,-1-2n}=-\chi^{\Sp{(2r)}}_{f}$. These are the second excited states with half-integer modes and KK-charge 1 above the NS-R vacuum. 
    \item $c_{\frac52,-2}=\chi^{\so{(2r+7)}}_{\theta}+\chi^{\so{(2r+7)}}_{2f}+\chi^{\Sp{(2r)}}_{2f}+2$. These are the second excited states with integer modes and KK-charge 1 above the NS-R vacuum. The coefficients are colored purple in the Tables \ref{tab:bijn4so10z2}, \ref{tab:bijn4so12z2} and \ref{tab:bijn4so14z2}.
\end{itemize}

\paragraph{\underline{$\fn=4$, $\so(10)^{(2)}$}}
From the recursion formula \eqref{Z1m}, we compute the twisted one-string elliptic genus to $q$ order $4$ and find complete agreement with the 2d quiver formula. In particular, we obtain the coefficients $b_{ij}$ in Table \ref{tab:bijn4so10z2}. We checked that the coefficients satisfy all universal behaviors summarized earlier with $r=1$. 

\begin{table}[h]
\begin{center}
$\begin{tabu}{c|cccccccccc}
q\backslash v & 1 &2 &3  &4 &5 &6 &7 & 8 & 9 & 10  \\ \hline
0 & 0 & 0& 0& 0& 0 & \color{orange}-5 & \color{orange}36 & \color{orange}-224 & \color{orange}924 & \color{orange}-3385 \\
1/2& 0 & 0 & 0 & 0 & 0 & 4 & -54 & 324 & -1596 & 5940 \\
1& 0 & 0 & 0 & 0 & 0 & 0 & 36 & -506 & 3132 & -15568 \\
3/2& \red1 & 0 & 0 & 0 & -1 & -4 & 0 & 348 & -3498 & 19368 \\
2& 0 & \cyan9 & -4 & 0 & 4 & -4 & -36 & 0 & 2208 & -21438 \\
5/2& \red{36} & \blue{-36} & \color{purple}92 & -36 & 0 & 36 & -38 & -312 & -36 & 13624 \\
3& 0 & \cyan{240} & -324 & 642 & -348 & 0 & 312 & -136 & -1884 & -240 \\
7/2& \red{495} & \blue{-924} & 2776 & -3132 & 4236 & -2208 & 0 & 1884 & -738 & -10492 \\
4& 0 & \cyan{2805} & -5940 &15744 & -19368 & 23838 & -13624 & 224 & 10492 & -2400 \\
9/2& \red{4004} & \blue{-10296} & 35240 & -59004 & 106468 & -118080 & 124392 & -69372 & 1596 & 50004 \\
\end{tabu}$
\caption{The coefficient matrix $b_{ij}$ for the $\IE_1$ of 6d $\so(10)^{(2)}$ theory with $\fn=4$.}
\label{tab:bijn4so10z2}
\end{center}
\end{table}

\paragraph{\underline{$\fn=4,\so(12)^{(2)}$}}
From the recursion formula \eqref{Z1m}, we compute the twisted one-string elliptic genus to $q$ order $4$. 
We obtain the following Fourier coefficients $b_{ij}$ in the $(q,v)$ expansion in Table \ref{tab:bijn4so12z2}. We checked that the coefficients satisfy all universal behaviors summarized earlier with $r=2$. 

\begin{table}[h]
\begin{center}
$\begin{tabu}{c|cccccccccc}
q\backslash v & 1 &2 &3  &4 &5 &6 &7 & 8 & 9 & 10  \\ \hline
0 & 0 & 0 & 0 & 0 & 0 & 0 & 0 & \color{orange}-42 & \color{orange}528 & \color{orange}-4065 \\
1/2& 0 & 0 & 0 & 0 & 0 & 0 & 0 & 48 & -759 & 6328 \\
1& 0 & 0 & 0 & 0 & 0 & 0 & 0 & 0 & 528 & -8748 \\
3/2& \color{red}1 & 0 & 0 & 0 & -1 & 0 & 0 & -48 & 0 & 6792 \\
2& 0 & \color{cyan}11 & -8 & 0 & 8 & -11 & 0 & 42 & -528 & 0 \\
5/2& \color{red}55 & \color{blue}-88 & \color{purple}158 & -88 & 0 & 88 & -158 & 88 & 704 & -6792 \\
3& 0 & \color{cyan}440 & -968 & 1606 & -1144 & 0 & 1144 & -1606 & 440 & 8308 \\
7/2& \color{red}1144 & \color{blue}-3432 & 8630 & -12584 & 14501 & -9152 & 0 & 9152 & -14501 & 6256 \\
4& 0 & \color{cyan}7579 & -26312 & 65230 & -101152 & 115434 & -76192 & 0 & 76192 & -111369 \\
9/2& \color{red}13650 & \color{blue}-57200 & 180372 & -378136 & 640565 & -814352 & 800642 & -493416 & 0 & 493416 \\
\end{tabu}$
\caption{The coefficient matrix $b_{ij}$ for the $\IE_1$ of 6d  $\so(12)^{(2)}$ theory with $\fn=4$.}
\label{tab:bijn4so12z2}
\end{center}
\end{table}

\paragraph{\underline{$\fn=4,\so(14)^{(2)}$}}
From the recursion formula \eqref{Z1m}, we compute the twisted one-string elliptic genus to $q$ order $4$. 
We obtain the following coefficients $b_{ij}$ in the $(q,v)$ expansion  in Table \ref{tab:bijn4so14z2}. We checked that the coefficients satisfy all universal behaviors summarized earlier with $r=3$. 

\begin{table}[h]
\begin{center}
$\begin{tabu}{c|cccccccccc}
q\backslash v & 1 &2 &3  &4 &5 &6 &7 & 8 & 9 & 10  \\ \hline
0& 0 & 0 & 0 & 0 & 0 & 0 & 0 & 0 & 0 & \color{orange}-429 \\
1/2& 0 & 0 & 0 & 0 & 0 & 0 & 0 & 0 & 0 & 572 \\
1& 0 & 0 & 0 & 0 & 0 & 0 & 0 & 0 & 0 & 0 \\
3/2& \color{red}1 & 0 & 0 & 0 & -1 & 0 & 0 & 0 & 0 & -572 \\
2& 0 & \color{cyan}13 & -12 & 0 & 12 & -13 & 0 & 0 & 0 & 429 \\
5/2& \color{red}78 & \color{blue}-156 & \color{purple}248 & -156 & 0 & 156 & -248 & 156 & -78 & 0 \\
3& 0 & \color{cyan}728 & -2028 & 3354 & -2612 & 0 & 2612 & -3354 & 2028 & -728 \\
7/2& \color{red}2275 & \color{blue}-8580 & 21476 & -33956 & 38456 & -25376 & 0 & 25376 & -38456 & 33956 \\
4& 0 & \color{cyan}17290 & -76440 & 200824 & -340028 & 396643 & -271984 & 0 & 271984 & -396643 \\
\end{tabu}$
\caption{The coefficient matrix $b_{ij}$ for the $\IE_1$ of 6d $\so(14)^{(2)}$ theory with $\fn=4$.}
\label{tab:bijn4so14z2}
\end{center}
\end{table}

\subsubsection{$\fn=2,\su(2r)^{(2)}$}
The $\IZ_2$ twist of 6d $(1,0)$ $\su(2r)+4r\bf F$ theory has low energy gauge algebra $\cirG=\Sp(r)$ and twisted matter content $2r(\mathbf{F}_0+\mathbf{F}_{1/2})$ which is invariant under the KK-charge shift by $1/2$.  The circle reduction gives a 5d $\mathcal{N}=1$ $\Sp(r)+2r\bf F$ theory whose 5d Nekrasov partition function is exactly computable by localization. However, the twisted elliptic genera, albeit the simple look, do not to our knowledge have a 2d quiver gauge theory description. We utilize the blowup equations to compute the twisted one-string elliptic genera. For arbitrary $r\ge1$, the twisted elliptic blowup equations can have two possibilities for the characteristics $a=\pm1/4$. 
From the unity twisted elliptic blowup equations, we solve the one-string twisted elliptic genera for $r=1,2,3,4$. We find the reduced twisted one-string elliptic genus has the following spectral flow symmetry
\be
\IE^{\SU(2r)^{(2)}}_1\Big(q,\frac{\sqrt{q}}{v}\Big)=-\Big(\frac{\sqrt{q}}{v^2}\Big)^{r-1}\IE^{\SU(2r)^{(2)}}_1(q,{v}).
\ee
We also find the following compact formula for the twisted one-string elliptic genus when turning off all gauge and flavor fugacities:
\be
\IE^{\SU(2r)^{(2)}}_1(\tau,\epsilon_+)=\frac{2^{2r-1}\eta(\tau)^{8r-3}}{\eta(\frac{\tau}{2})^{4r-2}}\frac{\theta_4(\tau,2\epsilon_+)}{(\theta_2(\tau,\epsilon_+)\theta_3(\tau,\epsilon_+))^{2r}}.
\ee
It is easy to prove the spectral flow symmetry from this expression. In the following, we explicitly show the results for $r=1,2,3$. 

\paragraph{\underline{$\fn=2,\su(2)^{(2)}$}}
From the unity twisted elliptic blowup equations, we calculate the twisted one-string elliptic genus to $q$ order $5$. We can also use the 2d localization to compute the twisted elliptic genus in this case as the $\IZ_2$ twist for the gauge algebra $\su(2)$ is trivial. We only need to fold the four fundamentals. For one-string, we have
    \be\label{localizationn2su2z2}
\IE_1=\frac{\prod_{i=1}^2\theta_1(a+\epsilon_+-m_i)\theta_4(a+\epsilon_++m_i)}{\eta^2\theta_1(2a)\theta_1(2a+2\epsilon_+)}+(a\rightarrow -a),
\ee
We checked this agrees with our results solved from blowup equations to high $q$ orders. We also obtain the coefficients $b_{ij}$ in Table \ref{tab:bijn2su2z2}. The spectral flow symmetry manifests as expected. We also checked the spectral flow symmetry with all gauge and flavor parameters turned on. Notice the leading orders of both R-R and NS-R elliptic genera give the one-instanton partition function of 5d $\SU(2)+2\bf F$. With gauge and flavor fugacities turned on, we have the following exact formula for the $v$ expansion:
\be
Z_1^{5d}=\sum_{n=0}^{\infty}\Big(-\chi^b_{(1)}\chi^{\SU(2)}_{n\theta}v^{1+2n}+\chi^a_{(1)}\chi^{\SU(2)}_{f+n\theta}v^{2+2n}\Big).
\ee
Here the truncated flavor symmetry $\SO(4)\sim \SU(2)_a\times \SU(2)_b$.
\begin{table}[h]
\begin{center}
$\begin{tabu}{c|cccccccc}
q\backslash v &  0 & 1 & 2 & 3 & 4 & 5 & 6 &7 \\ \hline
0& 0 & \color{orange}-2 & \color{orange}4 & \color{orange}-6 & \color{orange}8 & \color{orange}-10 & \color{orange}12 & \color{orange}-14 \\
1/2& \color{orange}2 & 0 & -6 & 16 & -24 & 32 & -40 & 48 \\
1& \color{orange}-4 & 6 & 0 & -22 & 52 & -84 & 112 & -140 \\
3/2& \color{orange}6 & -16 & 22 & 0 & -58 & 144 & -234 & 320 \\
2& \color{orange}-8 & 24 & -52 & 58 & 0 & -152 & 368 & -606 \\
5/2& \color{orange}10 & -32 & 84 & -144 & 152 & 0 & -352 & 864 \\
3& \color{orange}-12 & 40 & -112 & 234 & -368 & 352 & 0 & -796 \\
7/2& \color{orange}14 & -48 & 140 & -320 & 606 & -864 & 796 & 0 \\
\end{tabu}$
\caption{The coefficient matrix $b_{ij}$ for the  $\IE_1$ of 6d $\su(2)^{(2)}$ theory with $\fn=2$.}
\label{tab:bijn2su2z2}
\end{center}
\end{table}

\paragraph{\underline{$\fn=2,\su(4)^{(2)}$}}
From the unity twisted elliptic blowup equations, we compute the twisted one-string elliptic genus to $q$ order $5$. We obtain the coefficients $b_{ij}$ in Table \ref{tab:bijn2su4z2}. The obvious anti-symmetric coefficient matrix shows the self-duality under spectral flow as expected. The leading orders of both R-R and NS-R elliptic genera should give the one-instanton partition function of 5d $\Sp(2)+4\bf F$. From ADHM construction, we obtain the following exact formula for the $v$ expansion:
\be
Z_{1}^{ \Sp(2)+4\bf F}=\sum_{n=0}^{\infty}\Big(\chi^{\so(8)}_{S}\chi^{\Sp(2)}_{n\theta}v^{2+2n}-\chi^{\so(8)}_{C}\chi^{\Sp(2)}_{f+n\theta}v^{3+2n}\Big) .
\ee
where $\so(8)$ is the pertubative 5d flavor symmetry. The Fourier coefficients completely agree with the orange numbers in Table \ref{tab:bijn2su4z2}.

\begin{table}[h]
\begin{center}
$\begin{tabu}{c|cccccccc}
q\backslash v &   1 & 2 & 3 & 4 & 5 & 6 &7 & 8 \\ \hline
0& 0 & \color{orange}-8 & \color{orange}32 & \color{orange}-80 & \color{orange}160 & \color{orange}-280 & \color{orange}448 & \color{orange}-672 \\
1/2& \color{orange}8 & 0 & -96 & 384 & -960 & 1920 & -3360 & 5376 \\
1& \color{orange}-32 & 96 & 0 & -752 & 2848 & -7008 & 13920 & -24272 \\
3/2& \color{orange}80 & -384 & 752 & 0 & -4464 & 16256 & -39296 & 77440 \\
2& \color{orange}-160 & 960 & -2848 & 4464 & 0 & -22152 & 78048 & -185552 \\
5/2& \color{orange}280 & -1920 & 7008 & -16256 & 22152 & 0 & -96320 & 330112 \\
3& \color{orange}-448 & 3360 & -13920 & 39296 & -78048 & 96320 & 0 & -378096 \\
7/2& \color{orange}672 & -5376 & 24272 & -77440 & 185552 & -330112 & 378096 & 0 \\
\end{tabu}$
\caption{The coefficient matrix $b_{ij}$ for the  $\IE_1$ of 6d $\su(4)^{(2)}$ theory with $\fn=2$.}
\label{tab:bijn2su4z2}
\end{center}
\end{table}

\paragraph{\underline{$\fn=2,\su(6)^{(2)}$}}
From the unity twisted elliptic blowup equations, we calculate the twisted one-string elliptic genus to $q$ order $5$. We obtain the coefficients $b_{ij}$ in Table \ref{tab:bijn2su6z2}. The obvious anti-symmetric coefficient matrix shows the self-duality under spectral flow. The leading orders of both R-R and NS-R elliptic genera give the one-instanton partition function of 5d $\Sp(3)+6\bf F$. From ADHM construction, we find the following exact formula for the $v$ expansion:
\be
Z_{1}^{ \Sp(3)+6 \bf F}=\sum_{n=0}^{\infty}\Big(\chi^{\so(12)}_{S}\chi^{\Sp(3)}_{n\theta}v^{1+2n}-\chi^{\so(12)}_{C}\chi^{\Sp(3)}_{f+n\theta}v^{2+2n}\Big) .
\ee
where $\so(12)$ is the pertubative 5d flavor symmetry. We checked that these coefficients completely agree with the orange numbers in Table \ref{tab:bijn2su6z2}.

\begin{table}[h]
\begin{center}
$\begin{tabu}{c|cccccccc}
q\backslash v &    2 & 3 & 4 & 5 & 6 &7 & 8  \\ \hline
0& 0 & \color{orange}-32 & \color{orange}192 & \color{orange}-672 & \color{orange}1792 & \color{orange}-4032 & \color{orange}8064  \\
1/2& \color{orange}32 & 0 & -800 & 4352 & -14560 & 37888 & -84000  \\
1& \color{orange}-192 & 800 & 0 & -11104 & 55424 & -177184 & 449024  \\
3/2& \color{orange}672 & -4352 & 11104 & 0 & -109760 & 512512 & -1574720 \\
2& \color{orange}-1792 & 14560 & -55424 & 109760 & 0 & -862848 & 3819200  \\
5/2& \color{orange}4032 & -37888 & 177184 & -512512 & 862848 & 0 & -5727648  \\
3& \color{orange}-8064 & 84000 & -449024 & 1574720 & -3819200 & 5727648 & 0  \\
\end{tabu}$
\caption{The coefficient matrix $b_{ij}$ for the  $\IE_1$ of 6d $\su(6)^{(2)}$ theory with $\fn=2$.}
\label{tab:bijn2su6z2}
\end{center}
\end{table}

\subsubsection{$\fn=2,\su(2r+1)^{(2)}$}\label{sec:n2su2r1}
The $\IZ_2$ twist of 6d $(1,0)$ $\su(2r+1)+2(2r+1)\bf F$ theory has low energy gauge algebra $\cirG=\Sp(r)$ and twisted matter content $(2r+1)(\mathbf{F}_0+\mathbf{F}_{1/2})$ which is invariant under the KK-charge shift by $1/2$.  The circle reduction gives  5d $\mathcal{N}=1$ $\Sp(r)+(2r+1)\bf F$ 
gauge theories whose 5d Nekrasov partition function is exactly computable by localization. The twisted circle compactification gives 5d $\mathcal{N}=1$ $\so(2r+3)+(2r+1)\bf V$ KK theories \cite{Hayashi:2015vhy}. To our knowledge, there is no 2d quiver gauge theory description in the current situation. Therefore, it is important to use blowup equations to solve the twisted elliptic genera and study their properties. Indeed, we successfully solve the twisted one-string elliptic genera for $r=1,2,3,4$ to high $q$ orders. We find the reduced twisted one-string elliptic genus has the following spectral flow symmetry
\be
\IE^{\SU(2r+1)^{(2)}}_1\Big(q,\frac{\sqrt{q}}{v}\Big)=-\Big(\frac{{q}^{1/4}}{v}\Big)^{2r-1}\IE^{\SU(2r+1)^{(2)}}_1(q,{v}).
\ee
We also manage to find the following compact formula for the twisted one-string elliptic genus for arbitrary $r$:
\be
\IE_1^{\SU(2r+1)^{(2)}}(\tau,\epsilon_+)=\frac{\eta(\frac{\tau}{4})^2\theta_4(\frac{\tau}{2},\epsilon_+)}{\eta(\frac{\tau}{2})^2\theta_4(\frac{\tau}{2})}\Big(\frac{2\theta_1(\tau,\epsilon_+)\theta_4(\tau,\epsilon_+)}{\theta_1(\tau,2\epsilon_+)\theta_4(\tau)}\Big)^{2r}.
\ee
We checked this formula for $r=1,2,3,4$ against the twisted elliptic genera solved from blowup equations and found perfect agreement. 
It is easy to prove the spectral flow symmetry from this expression. 

It is interesting to remark that the twisted elliptic blowup equations also allow nice solutions for the twisted matter content $(2r+1)(\mathbf{F}_{1/4}+\mathbf{F}_{3/4})$. This choice of the KK-charges of the hypermultiplets $\bf F$ has its merit in that they coincide with those of the $\bf F$ in the twist of vector multiplet, i.e. $\mathbf{Adj}_0+\mathbf{F}_{1/4}+\mathbf{F}_{3/4}+\mathbf{\Lambda}^2_{1/2}$. This brings in some simplification that the $B$ field of the tensor parameter becomes integral. In fact, there are two possibilities for such $B$ field -- even and odd, corresponding to two non-equivalent circle reductions to 5d $\mathcal{N}=1$ pure $\Sp(r)_0$ or $\Sp(r)_\pi$ theory. We leave the discussion for these situations in Appendix \ref{app:C}.

\paragraph{\underline{$\fn=2,\su(3)^{(2)}$}}
The $\IZ_2$ twisted circle compactification of 6d $(1,0)$ $\su(3)+6\bf F$ theory gives a 5d rank-3 KK theory which has gauge theory description as 5d $\Sp(2)+3\bf \Lambda^2$ or 5d $\SO(5)+3\bf V$ theory \cite{Hayashi:2015vhy}. 
From the unity twisted elliptic blowup equations, we compute the twisted one-string elliptic genus to $q$ order $5$. The coefficients $b_{ij}$ are listed in Table \ref{tab:bijn2su3z2F0}. The obvious anti-symmetric coefficient matrix shows the self-duality under spectral flow. For the low energy limit, the 5d one-instanton partition function of $\su(2)+3\bf F$ theory can be computed by localization as
\be
Z_{1}^{ \su(2)+3\bf F}=\sum_{n=0}^{\infty}\Big(\chi^{\so(6)}_{S}\chi^{\SU(2)}_{n\theta}v^{1+2n}-\chi^{\so(6)}_{C}\chi^{\SU(2)}_{f+n\theta}v^{2+2n}\Big).
\ee
where $\so(6)$ is the pertubative 5d flavor symmetry. It is easy to check that the $v$ expansion coefficients are consistent with the orange numbers in Table \ref{tab:bijn2su3z2F0}.

\begin{table}[h]
\begin{center}
$\begin{tabu}{c|ccccccccc}
q\backslash v &  &  1 &  & 2 &   & 3 & & 4  \\ \hline
0& 0 & \color{orange}4 & 0 & \color{orange}-8 & 0 & \color{orange}12 & 0 & \color{orange}-16 & 0 \\
1/4& \color{orange}-4 & 0 & 8 & 0 & -16 & 0 & 24 & 0 & -32 \\
1/2& 0 & -8 & 0 & 36 & 0 & -72 & 0 & 108 & 0 \\
3/4& \color{orange}8 & 0 & -36 & 0 & 80 & 0 & -144 & 0 & 216 \\
1& 0 & 16 & 0 & -80 & 0 & 224 & 0 & -416 & 0 \\
5/4& \color{orange}-12 & 0 & 72 & 0 & -224 & 0 & 472 & 0 & -812 \\
3/2& 0 & -24 & 0 & 144 & 0 & -472 & 0 & 1092 & 0 \\
7/4& \color{orange}16 & 0 & -108 & 0 & 416 & 0 & -1092 & 0 & 2168 \\
2& 0 & 32 & 0 & -216 & 0 & 812 & 0 & -2168 & 0 \\
\end{tabu}$
\caption{The coefficient matrix $b_{ij}$ for the  $\IE_1$ of 6d $\su(3)^{(2)}$ theory with $\fn=2$.}
\label{tab:bijn2su3z2F0}
\end{center}
\end{table}

\paragraph{\underline{$\fn=2,\su(5)^{(2)}$}}
The $\IZ_2$ twisted circle compactification of 6d $(1,0)$ $\su(5)+10\bf F$ theory gives a 5d rank-4 KK theory which has gauge theory description as 5d  $\SO(7)+5\bf V$ 
 theory \cite{Hayashi:2015vhy}. 
From the unity twisted elliptic blowup equations, we compute the twisted one-string elliptic genus to $q$ order $5$. The coefficients $b_{ij}$ are listed in Table \ref{tab:bijn2su5z2F0}. The obvious anti-symmetric coefficient matrix  shows the self-duality under spectral flow. The 5d one-instanton partition function of $\Sp(2)+5\bf F$ theory can be computed by localization as
\be
Z_{1}^{ \Sp(2)+5\bf F}=\sum_{n=0}^{\infty}\Big(\chi^{\so(12)}_{S}\chi^{\Sp(2)}_{n\theta}v^{1+2n}-\chi^{\so(12)}_{C}\chi^{\Sp(2)}_{f+n\theta}v^{2+2n}\Big) .
\ee
where $\so(10)$ is the pertubative 5d flavor symmetry. It is easy to check that the $v$ expansion coefficients are consistent with the orange numbers in Table \ref{tab:bijn2su5z2F0}.

\begin{table}[h]
\begin{center}
$\begin{tabu}{c|ccccccccc}
q\backslash v &  &  1 &  & 2 &   & 3 & & 4  \\ \hline
0& 0 & \color{orange}16 & 0 & \color{orange}-64 & 0 &\color{orange} 160 & 0 & \color{orange}-320 & 0 \\
1/4& \color{orange}-16 & 0 & 64 & 0 & -176 & 0 & 384 & 0 & -720 \\
1/2& 0 & -64 & 0 & 400 & 0 & -1280 & 0 & 2976 & 0 \\
3/4& \color{orange}64 & 0 & -400 & 0 & 1344 & 0 & -3376 & 0 & 7040 \\
1& 0 & 176 & 0 & -1344 & 0 & 5312 & 0 & -14528 & 0 \\
5/4& \color{orange}-160 & 0 & 1280 & 0 & -5312 & 0 & 15488 & 0 & -36192 \\
3/2& 0 & -384 & 0 & 3376 & 0 & -15488 & 0 & 48976 & 0 \\
7/4& \color{orange}320 & 0 & -2976 & 0 & 14528 & 0 & -48976 & 0 & 128768 \\
2& 0 & 720 & 0 & -7040 & 0 & 36192 & 0 & -128768 & 0 \\
\end{tabu}$
\caption{The coefficient matrix $b_{ij}$ for the  $\IE_1$ of 6d $\su(5)^{(2)}$ theory with $\fn=2$.}
\label{tab:bijn2su5z2F0}
\end{center}
\end{table}

\subsubsection{Other examples}

\paragraph{\underline{$\fn=4,E_6^{(2)}$}}
The $\IZ_2$ twist of 6d $(1,0)$ $E_6+2\bf F$ theory has low energy gauge algebra $\cirG=F_4$ and twisted matter content $\mathbf{26}_0+\mathbf{26}_{1/2}$ which is invariant under the KK-charge shift by $1/2$.  The twisted partition function has not been computed before. 
From the recursion formula \eqref{Z1m}, we compute the twisted one-string elliptic genus to $q$ order $5$. We summarize the coefficients $b_{ij}$ in Table \ref{tab:bijn4E6z2}. Notice the obvious anti-symmetric coefficient matrix, this shows that the twisted theory has self-dual spectral flow symmetry, albeit the nontrivial matter content. The leading $q$ order of $\IE_1$ should give the one-instanton Nekrasov partition function of 5d $F_4+\bf F$ theory, whose exact $v$ expansion formula can be found in \cite[Equation (H.36)]{DelZotto:2018tcj}. For example, when turning off all fugacities,
\be
\ba
Z_1^{F_4+\bf F}(v)=\,&\frac{v^7}{(v-1)^{10} (v+1)^{16}} (v^{12}+10 v^{11}-49 v^{10}+266 v^9-549 v^8+1068 v^7\\
&-1110 v^6+1068 v^5-549 v^4+266 v^3-49 v^2+10 v+1)\\
=\,& v^7+4 v^8-78 v^9+754 v^{10}-4433 v^{11}+21060 v^{12}+\dots
\ea
\ee
We have checked they are in complete agreement. The $v$ expansion coefficients of 5d $Z_1^{F_4+\bf F}$ are colored in orange in Table \ref{tab:bijn4E6z2}. 
\begin{table}[h]
\begin{center}
$\begin{tabu}{c|ccccccccc}
q\backslash v & 4 & 5&6 &7 &8 &9 & 10 & 11 & 12  \\ \hline
0& 0 & 0 & 0 & \color{orange}-1 & \color{orange}-4 & \color{orange}78 & \color{orange}-754 & \color{orange}4433 & \color{orange}-21060 \\
1/2& 0 & 0 & 0 & 2 & -27 & -108 & 2509 & -21654 & 124956 \\
1& 0 & 0 & 0 & 3 & 54 & -512 & -1736 & 43680 & -363866 \\
3/2& \color{orange}1 & -2 & -3 & 0 & 86 & 874 & -7436 & -19680 & 552578 \\
2& \color{orange}4 & 27 & -54 & -86 & 0 & 1437 & 10756 & -88505 & -178544 \\
5/2& \color{orange}-78 & 108 & 512 & -874 & -1437 & 0 & 16930 & 113530 & -895387 \\
3& \color{orange}754 & -2509 & 1736 & 7436 & -10756 & -16930 & 0 & 154075 & 1072946 \\
7/2& \color{orange}-4433 & 21654 & -43680 & 19680 & 88505 & -113530 & -154075 & 0 & 1117564 \\
4& \color{orange}21060 & -124956 & 363866 & -552578 & 178544 & 895387 & -1072946 & -1117564 & 0 \\
\end{tabu}$
\caption{The coefficient matrix $b_{ij}$ for the  $\IE_1$ of 6d $E_6^{(2)}$ theory with $\fn=4$.}
\label{tab:bijn4E6z2}
\end{center}
\end{table}

\paragraph{\underline{$\fn=3,\so(8)^{(2)}$}}
The $\IZ_2$ twist of 6d $(1,0)$ $\so(8)+\bf V+S+C$ theory has low energy gauge algebra $\cirG=\so(7)$ and twisted matter content $\mathbf{7}_0+\mathbf{8}_{0}+\mathbf{8}_{1/2}$.  
From the recursion formula \eqref{Z1m}, we computed the twisted one-string elliptic genus to $q$ order $5$. 
We obtain the Fourier coefficients $b_{ij}$ in the $(q,v)$ expansion in Table \ref{tab:bijn3so8z2}.
\begin{table}[h]
\begin{center}
$\begin{tabu}{c|ccccccccc}
q\backslash v  & 1& 2& 3 & 4 & 5 & 6 & 7 & 8 & 9  \\ \hline
0& 0 & 0 & 0 & \color{orange}4 & \color{orange}-30 & \color{orange}132 & \color{orange}-434 & \color{orange}1184 & \color{orange}-2826 \\
1/2& 0 & 0 & 0 & 0 & 40 & -272 & 1152 & -3696 & 9936 \\
1& \color{red}2 & 0 & -2 & -4 & 0 & 324 & -2168 & 8868 & -27860 \\
3/2& \color{red}-8 & 16 & 0 & -16 & -32 & 0 & 1936 & -12416 & 49464 \\
2& \color{red}42 & -108 & 142 & 0 & -112 & -216 & -42 & 10392 & -64496 \\
5/2& \color{red}-112 & 432 & -848 & 848 & 0 & -576 & -1088 & -432 & 48448 \\
3& \color{red}336 & -1428 & 3606 & -5528 & 4814 & -132 & -2646 & -4864 & -3606 \\
7/2& \color{red}-720 & 3712 & -10976 & 21744 & -29312 & 23056 & -1152 & -10640 & -19024 \\
4& \color{red}1650 & -8964 & 29700 & -68096 & 115992 & -140732 & 103936 & -8868 & -39006 \\
9/2& \color{red}-3080 & 18800 & -68256 & 176608 & -349888 & 534816 & -603264 & 424864 & -49464 \\
\end{tabu}$
\caption{The coefficient matrix $b_{ij}$ for the  $\IE_1$ of 6d $\SO(8)^{(2)}$ theory with $\fn=3$.}
\label{tab:bijn3so8z2}
\end{center}
\end{table}

We find the leading $q$ order of the R-R sector, i.e., the orange coefficients gives the one-instanton Nekrasov partition function of 5d $\SO(7)+\bf V+S$ theory, while the leading $q$ order of the NS-R sector, i.e., the red coefficients gives the one of 5d $\SO(7)+\bf S$ theory. For 5d $\SO(7)+\bf S$ theory, the ADHM description has been studied in \cite{Kim:2018gjo}. For example, the reduced one-instanton partition function is given by 
\begin{equation}
  Z_1^{\SO(7)+\bf S}(u_i,m)=
  \sum_{i=1}^4\frac{\sinh(4\epsilon_+-2u_i)
  \sinh(m\pm(u_i-\epsilon_+))}
  {\prod_{j(\neq i)}\sinh(u_{ij})\sinh(2\epsilon_+-u_{ij})
  \sinh(2\epsilon_+-u_i-u_j)}.
\end{equation}
Here $u_i$ satisfying $u_1+u_2+u_3+u_4=0$ are the fugacities of $\SU(4)$ embedded in $\SO(7)$. 
From this formula, we find
\be
Z_1^{\SO(7)+\bf S}(v)=\frac{2 v^4 \left(v^6-2 v^5+8 v^4-6 v^3+8 v^2-2 v+1\right)}{(v-1)^6 (v+1)^8}.
\ee
\be
Z_1^{\SO(7)+\bf S}(v,m_{\SO(7)},m_{\Sp(1)})=\sum_{n=0}^{\infty}\chi^{\SO(7)}_{n\theta}\chi^{\Sp(1)}_{(1)}v^{2n+4}-\chi^{\SO(7)}_{n\theta+s}v^{2n+5}.
\ee
This perfectly agrees with the leading $q$ order of the NS-R twisted elliptic genus. For $\SO(7)+\bf V+ S$ theory, we can compare with the reduced one-instanton partition function of 5d $\SO(7)+{\bf V}+4\bf S$ theory given in (H.15) of \cite{DelZotto:2018tcj}. After decoupling three $\bf S$, we find the resulting 5d partition function is
\be
\ba
Z_1^{\SO(7)+\bf V+ S}(v,m_{\SO(7)},m_{v},m_s)=\sum_{n=0}^{\infty}&\chi^{\SO(7)}_{n\theta}\chi^{\Sp(1)}_{(1)_v}\chi^{\Sp(1)}_{(1)_s}v^{2n+4}+\chi^{\SO(7)}_{n\theta+v+s}v^{2n+6}\\
& -\left(\chi^{\SO(7)}_{n\theta+v}\chi^{\Sp(1)}_{(1)_s}+\chi^{\SO(7)}_{n\theta+s}\chi^{\Sp(1)}_{(1)_v}\right)v^{2n+5}.
\ea
\ee
This perfectly agrees with the leading $q$ order of the R-R twisted elliptic genus that is when turning off all gauge and flavor fugacities:
\be
Z_1^{\SO(7)+\bf V+ S}(v)=\frac{2 v^4 \left(2 v^2-3 v+2\right)}{(v-1)^2 (v+1)^8}.
\ee

\paragraph{\underline{$\fn=3,\so(8)^{(3)}$}}
The $\IZ_3$ twist of 6d $(1,0)$ $\so(8)+\bf V+S+C$ theory has low energy gauge algebra $\cirG=G_2$ and twisted matter content $\mathbf{7}_0+\mathbf{7}_{1/3}+\mathbf{7}_{2/3}$. 
From the recursion formula \eqref{Z1}, we compute the twisted one-string elliptic genus to $q$ order $11/3$. In particular, we obtain the  coefficients $b_{ij}$ in Table \ref{tab:bijn3so8z3}. We observe the  leading $q$ order of R-R elliptic genus gives the one-instanton partition function of 5d $G_2+\bf F$ theory \cite{DelZotto:2018tcj}:
    \be
Z_1^{G_2+\bf F}=    \sum_{n=0}\chi^{G_2}_{n\theta}\chi^{\Sp(1)}_{(1)}v^{2n+3}- \sum_{n=0}\chi^{G_2}_{f+n\theta}v^{2n+4}.
    \ee
Besides, the leading order of NS-R elliptic genus colored red has coefficients $\chi^{G_2}_{n\theta}$ as the one $G_2$ instanton Hilbert series. The subleading order of NS-R elliptic genus colored blue has coefficients $-\chi^{G_2}_{f+n\theta}\chi^{\Sp(1)}_{(1)}$, which are the excited states over NS-R vacuum with KK-charge $1/3$. The subsubleading order of NS-R elliptic genus colored cyan has coefficients $(\chi^{G_2}_f+\chi^{\Sp(1)}_{(2)})\chi^{G_2}_{n\theta}$, which are the excited states over NS-R vacuum with KK-charge $2/3$.
\begin{table}[h]
\begin{center}
$\begin{tabu}{c|cccccccccc}
q\backslash v &\ 2 & 3 & 4 & 5 & 6 & 7 & 8 & 9 & 10 & 11 \\ \hline
0&\ 0 & \color{orange}2 & \color{orange}-7 & \color{orange}28 &\color{orange}-64 & \color{orange}154 & \color{orange}-286 & \color{orange}546 & \color{orange}-896 & \color{orange}1496 \\
1/3& 0 & -1 & 16 & -70 & 224 & -563 & 1232 & -2412 & 4368 & -7413 \\
2/3& \ \red{1}\  & -2 & -10 & 104 & -420 & 1260 & -3070 & 6566 & -12680 & 22722 \\
1& 0 & \textcolor{cyan}{10} & -16 & -60 & 528 & -2016 & 5824 & -13812 & 29120 & -55488 \\
4/3& 0 & \blue{-14} & 67 & -90 & -288 & 2270 & -8344 & 23198 & -54112 & 112056 \\
5/3& 0 & \red{14} & -112 & 358 & -416 & -1192 & 8720 & -30820 & 83216 & -190402 \\
2& 0 & 0 & \cyan{140} & -630 & 1598 & -1668 & -4412 & 30566 & -104388 & 275170 \\
7/3& 0 & 0 & \blue{-128} & 840 & -2912 & 6288 & -6032 & -15004 & 99296 & -329752 \\
8/3& 0 & 0 & \red{77} & -870 & 4032 & -11676 & 22508 & -20150 & -47592 & 303454 \\
3 &0 & 0 & 0 & \cyan{770} & -4528 & 16611 & -42096 & 74444 & -62896 & -142378 \\
10/3& 0 & 0 & 0 & \blue{-572} & 4438 & -19694 & 60998 & -140026 & 230718 & -185756 \\
11/3& 0 & 0 & 0 & \red{273} & -3680 & 20356 & -74720 & 205436 & -435408 & 677190 \\
\end{tabu}$
\caption{The coefficient matrix $b_{ij}$ for the  $\IE_1$ of 6d $\su(8)^{(3)}$ theory with $\fn=3$.}
\label{tab:bijn3so8z3}
\end{center}
\end{table}

\paragraph{\underline{$\fn=2,\so(8)^{(2)}$}}
The $\IZ_2$ twist of 6d $(1,0)$ $\so(8)+2(\bf V+S+C)$ theory has low energy gauge algebra $\cirG=\so(7)$ and twisted matter content $2(\mathbf{7}_0+\mathbf{8}_{0}+\mathbf{8}_{1/2})$. From the unity twisted elliptic blowup equations, we compute the twisted one-string elliptic genus $\IE_1$ to $q^7$ order. In particular, we obtain the following coefficients $b_{ij}$ in Table \ref{tab:bijn2so8z2}.
\begin{table}[h]
\begin{center}
$\begin{tabu}{c|cccccccc}
q\backslash v  & 1 & 2 & 3 & 4 & 5 & 6 & 7 & 8 \\ \hline
0& 0 & \color{orange}-1 & \color{orange}-4 & \color{orange}-18 & \color{orange}300 & \color{orange}-1603 & \color{orange}5768 & \color{orange}-16548 \\
1/2& \color{red}5 & 4 & -12 & -68 & -152 & 3460 & -18980 & 68796 \\
1& \color{red}-32 & 50 & 100 & -110 & -780 & -1340 & 31516 & -169674 \\
3/2& \color{red}140 & -524 & 426 & 1004 & -620 & -6840 & -6896 & 212152 \\
2& \color{red}-448 & 2413 & -5060 & 2332 & 8440 & -3452 & -46800 & -35810 \\
5/2& \color{red}1218 & -8232 & 25172 & -38908 & 12255 & 54640 & -14240 & -279572 \\
3& \color{red}-2880 & 22724 & -86492 & 195414 & -245816 & 50690 & 315700 & -61000 \\
7/2& \color{red}6204 & -54888 & 242564 & -681100 & 1265024 & -1377048 & 206878 & 1592400 \\
\end{tabu}$
\caption{The coefficient matrix $b_{ij}$ for the  $\IE_1$ of 6d $\SO(8)^{(2)}$ theory with $\fn=2$.}
\label{tab:bijn2so8z2}
\end{center}
\end{table}

Our results are consistent with the 5d limits in both R-R and NS-R sectors. The leading $q$ order of the NS-R sector gives the 5d $\so(7)+2\bf S$ theory whose 5d partition function can be computed  by localization \cite{Kim:2018gjo} or  by 5d blowup equations \cite{Kim:2019uqw}. We use the localization formula in \cite{Kim:2018gjo} to compute the 5d one-instanton partition function as 
\be
Z_{1}^{ \so(7)+2\bf S}=\frac{v^4 \left(5 v^4-12 v^3+22 v^2-12 v+5\right)}{(v-1)^4 (v+1)^8}=5 v^4-32 v^5+140 v^6-448 v^7+\dots.
\ee
One can see the coefficients are in agreement with the red numbers in Table \ref{tab:bijn2so8z2}. One can also turn on the gauge and flavor fugacities, in which cases the $Z_1$ formula as $v$ expansion can be found in \cite[Equation (H.27)]{DelZotto:2018tcj}. 
The leading $q$ order of the R-R sector should give the 5d $\so(7)+2\mathbf{V}+2\bf S$ theory whose partition function to our knowledge has not been computed before. Luckily we can compare with the reduced one-instanton partition function of 5d $\SO(7)+2{\bf V}+6\bf S$ theory given in  \cite[Equation (E.2)]{Gu:2020fem}. After decoupling four $\bf S$, we find the resulting 5d partition function is
\be
\ba
Z_1^{\SO(7)+2\mathbf{V}+2\mathbf{S}}=&\,v^2+\chi^{\Sp(2)}_{(10)_v}v^3-\chi^{\SO(7)}_{(100)}v^4+\sum_{n=0}^{\infty}\chi^{\SO(7)}_{n\theta}\chi^{\Sp(2)}_{(01)_v}\chi^{\Sp(2)}_{(01)_s}v^{2n+4}\\
&-\left(\chi^{\SO(7)}_{(1n0)}\chi^{\Sp(2)}_{(10)_v}\chi^{\Sp(2)}_{(01)_s}+\chi^{\SO(7)}_{(0n1)}\chi^{\Sp(2)}_{(01)_v}\chi^{\Sp(2)}_{(10)_s}\right)v^{2n+5}\\
&+\left(\chi^{\SO(7)}_{(2n0)}\chi^{\Sp(2)}_{(01)_s}+
\chi^{\SO(7)}_{(1n1)}\chi^{\Sp(2)}_{(10)_v}\chi^{\Sp(2)}_{(10)_s}+\chi^{\SO(7)}_{(0n2)}\chi^{\Sp(2)}_{(01)_v}\right)v^{2n+6}\\
&-\left(\chi^{\SO(7)}_{(2n1)}\chi^{\Sp(2)}_{(10)_s}+\chi^{\SO(7)}_{(1n2)}\chi^{\Sp(2)}_{(10)_v}\right)v^{2n+7}+  \chi^{\SO(7)}_{(2n2)}v^{2n+8}.
\ea
\ee
We checked that this completely agrees with leading $q$ order of the R-R elliptic genus up to an overall minus sign:
\be\nonumber
Z_1^{\SO(7)+2\mathbf{V}+2\mathbf{S}}(v)= \frac{v^2 \left(v^4+12 v^3+78 v^2+12 v+1\right)}{(v+1)^8}=v^2+4 v^3+18 v^4-300 v^5+\dots
\ee
\paragraph{\underline{$\fn=2,\so(8)^{(3)}$}}
The $\IZ_3$ twist of 6d $(1,0)$ $\so(8)+2(\bf V+S+C)$ theory has low energy gauge algebra $\cirG=G_2$ and twisted matter content $2(\mathbf{7}_0+\mathbf{7}_{1/3}+\mathbf{7}_{2/3})$. From the unity twisted elliptic blowup equations, we compute the twisted one-string elliptic genera $\IE_1$ to the order $q^{16/3}$. We summarize the coefficients $b_{ij}$ in Table \ref{tab:bijn2so8z3}. The leading $q$ order of the R-R elliptic genus should give the 5d partition function of $G_2+2\bf F$ theory. We compute the 5d one-instanton partition function of $G_2+2\bf F$ theory from the localization formula in \cite{Kim:2018gjo} as
\be
Z_{1}^{ G_2+2\bf F}=-\frac{v^3 \left(5 v^6-28 v^5+67 v^4-88 v^3+67 v^2-28 v+5\right)}{\left(1-v^2\right)^6}.
\ee
We checked that the Fourier coefficients here are in agreement with  the orange numbers in Table \ref{tab:bijn2so8z3}. 
The leading order of NS-R elliptic genus colored red has coefficients $\chi^{G_2}_{n\theta}$ as expected, since the circle reduction of the NS-R sector should be a 5d pure $G_2$ theory. We find the subleading order of NS-R elliptic genus colored blue has coefficients $-\chi^{G_2}_{f+n\theta}\chi^{\Sp(2)}_{(10)}$, which represent the first excited states with KK-charge $1/3$ above the NS-R vacuum. 
Interestingly, we notice the diagonal elements in Table \ref{tab:bijn2so8z3} are always zero, but we could not find an explanation for it. 
\begin{table}[h]
\begin{center}
$\begin{tabu}{c|ccccccccc}
q\backslash v  & \,2& 3 & 4 & 5 & 6 & 7 & 8 &9&10\\ \hline
0& \,0 & \color{orange}-5 & \color{orange}28 & \color{orange}-97 & \color{orange}256 & \color{orange}-574 & \color{orange}1144 & \color{orange}-2094 & \color{orange}3584 \\
1/3& \,\red{1} & 0 & -57 & 336 & -1159 & 3072 & -6881 & 13728 & -25119 \\
2/3& \,0 & 17 & 0 & -455 & 2488 & -8349 & 21760 & -48293 & 95728 \\
1& \,0 & \color{blue}-28 & 170 & 0 & -2802 & 14400 & -46814 & 119808 & -262906 \\
4/3& \,0 & \red{14} & -336 & 1157 & 0 & -14386 & 70388 & -222392 & 559616 \\
5/3& \,0 & 0 & 373 & -2488 & 6466 & 0 & -64869 & 303888 & -935643 \\
2& \,0 & 0 & \color{blue}-256 & 3240 & -14400 & 31001 & 0 & -263943 & 1191224 \\
7/3& \,0 & 0 & \red{77} & -3072 & 20251 & -70388 & 132353 & 0 & -987985 \\
8/3& \,0 & 0 & 0 & 2254 & -21760 & 103822 & -303888 & 515468 & 0 \\
3& \,0 & 0 & 0 & \color{blue}-1144 & 18935 & -119808 & 462279 & -1191224 & 1861127 \\
10/3& \,0 & 0 & 0 & \red{273} & -13728 & 115465 & -559616 & 1850023 & -4316352 \\
\end{tabu}$
\caption{The coefficient matrix $b_{ij}$ for the  $\IE_1$ of 6d $\so(8)^{(3)}$ theory with $\fn=2$.}
\label{tab:bijn2so8z3}
\end{center}
\end{table}

\paragraph{\underline{$\fn=2,\so(10)^{(2)}$}} 
The $\IZ_2$ twist of 6d $(1,0)$ $\so(10)+4{\bf V}+2\bf S$ theory has low energy gauge algebra $\cirG=\so(9)$ and twisted matter content $4\mathbf{V}_0+\mathbf{S}_{0}+\mathbf{S}_{1/2}$. From the unity twisted elliptic blowup equations, we compute the twisted one-string elliptic genus $\IE_1$ to $q^7$ order. In particular, we obtain the  Fourier coefficients $b_{ij}$ in the $(q,v)$ expansion in Table \ref{tab:bijn2so10z2}. The leading $q$ order of the NS-R sector should give the 5d $\so(9)+\bf S$ theory. To our knowledge, 5d $\so(9)+\bf S$ theory does not have an ADHM construction. Luckily, the exact $v$ expansion formula of the  one-instanton partition function of  5d $\so(9)+2\mathbf{V}+\bf S$ theory has been found in \cite[Equation (H.29)]{DelZotto:2018tcj}. By decoupling the two $\bf V$, we obtain the following exact $v$ expansion formula
\be
\ba
Z_1^{\SO(9)+\mathbf{S}}=\sum_{n=0}^{\infty}\chi^{\SO(9)}_{n\theta}\chi^{\Sp(1)}_{(2)_s}v^{2n+6}-\chi^{\SO(7)}_{(0n01)}\chi^{\Sp(1)}_{(1)_s}v^{2n+7}+\chi^{\SO(7)}_{(0n10)}v^{2n+8}.
\ea
\ee
This is in complete agreement with our leading $q$ order of the NS-R elliptic genus whose coefficients are colored red in Table \ref{tab:bijn2so10z2}. Utilizing the Weyl dimension formula, we further find the following rational expression of $v$ with all gauge and flavor fugacities turned off:
\be
Z_1^{\SO(9)+\mathbf{S}}(v)=\frac{v^6 \left(3 v^8-20 v^7+58 v^6-116 v^5+134 v^4-116 v^3+58 v^2-20 v+3\right)}{(v-1)^8 (v+1)^{12}}.
\ee
On the other hand, the leading $q$ order of the R-R sector should give the 5d $\so(9)+4\bf V+ S$ theory. The exact $v$ expansion formula of the  one-instanton partition function of  5d $\so(9)+4\mathbf{V}+3\bf S$ theory has been found in \cite[Equation (E.6)]{Gu:2020fem}. By decoupling two $\bf S$, we find the resulting 5d partition function is in agreement with the leading $q$ order of the one-string R-R elliptic genus:
\be\nonumber
\ba
Z_1^{\SO(9)+4\mathbf{V}+\mathbf{S}}(v)=-\frac{v^2 \left(v^8+12 v^7+66 v^6+268 v^5+954 v^4+268 v^3+66 v^2+12 v+1\right)}{(v+1)^{12}},
\ea
\ee
up to an overall sign. The Fourier coefficients are colored orange in Table \ref{tab:bijn2so10z2}.

\begin{table}[h]
\begin{center}
$\begin{tabu}{c|cccccccc}
q\backslash v  & 1 & 2 & 3 & 4 & 5 & 6 & 7 & 8 \\ \hline
0&  0 & \color{orange}-1 & 0 & 0 & \color{orange}-48 & \color{orange}117 & \color{orange}2288 & \color{orange}-23760 \\
1/2& \color{red}3 & 8 & -10 & -24 & -12 & -264 & 66 & 31448 \\
1& \color{red}-32 & 30 & 112 & -82 & -96 & -184 & -5344 & 4314 \\
3/2& \color{red}192 & -600 & 150 & 1352 & -500 & -1344 & -472 & -42496 \\
2& \color{red}-864 & 4346 & -7664 & 1260 & 11520 & -6583 & -1056 & 16448 \\
5/2& \color{red}3135 & -20664 & 57492 & -72752 & 2499 & 96792 & -60738 & -11760 \\
3& \color{red}-9856 & 78646 & -286704 & 586292 & -616336 & 27138 & 701296 & -625918 \\
7/2& \color{red}27456 & -252120 & 1101390 & -2917296 & 4860776 & -4412640 & 15894 & 4855392 \\
\end{tabu}$
\caption{The coefficient matrix $b_{ij}$ for the  $\IE_1$ of 6d $\SO(10)^{(2)}$ theory with $\fn=2$.}
\label{tab:bijn2so10z2}
\end{center}
\end{table}

\paragraph{\underline{$\fn=2,E_6^{(2)}$}}
The $\IZ_2$ twist of 6d $(1,0)$ $E_6+4\bf F$ theory has low energy gauge algebra $\cirG=F_4$ and twisted matter content $2(\mathbf{26}_0+\mathbf{26}_{1/2})$ which is invariant under the KK-charge shift by $1/2$.  From the unity twisted elliptic blowup equations, we compute the twisted one-string elliptic genera $\IE_1$ to the $q^{3}$ order. We summarize the coefficients $b_{ij}$ in the $(q,v)$ expansion in Table \ref{tab:bijn2e6z2}. From the obvious anti-symmetric coefficient matrix, we recognize  that the $\IE_1$ is anti self-dual upon the spectral flow. Besides, the leading $q$ order of $\IE_1$ should give the one-instanton Nekrasov partition function of 5d $F_4+2\bf F$ theory, which as $v$ expansion can be found in \cite[Equation (H.34)]{DelZotto:2018tcj}. We have checked they are in complete agreement. The $v$ expansion coefficients in 5d are colored in orange in Table \ref{tab:bijn2e6z2}. 

\begin{table}[h]
\begin{center}
$\begin{tabu}{c|ccccccc}
q\backslash v  & 5&6 &7 &8 &9 & 10 & 11   \\ \hline
0& 0 & \color{orange}-5 & \color{orange}-20 & \color{orange}282 & \color{orange}-52 & \color{orange}-14377 & \color{orange}133136 \\
1/2& \color{orange}5 & 0 & -160 & -848 & 11354 & -8144 & -491872 \\
1& \color{orange}20 & 160 & 0 & -4490 & -18476 & 263092 & -322556 \\
3/2& \color{orange}-282 & 848 & 4490 & 0 & -94702 & -273520 & 4456372 \\
2& \color{orange}52 & -11354 & 18476 & 94702 & 0 & -1598277 & -3093340 \\
5/2& \color{orange}14377 & 8144 & -263092 & 273520 & 1598277 & 0 & -21992316 \\
3& \color{orange}-133136 & 491872 & 322556 & -4456372 & 3093340 & 21992316 & 0 \\
\end{tabu}$
\caption{The coefficient matrix $b_{ij}$ for the $\IE_1$ of 6d $E_6^{(2)}$ theory with $\fn=2$.}
\label{tab:bijn2e6z2}
\end{center}
\end{table}

\paragraph{\underline{$\fn=1,\SU(3)^{(2)}$}} 
The $\IZ_2$ twist of 6d $(1,0)$  $\SU(3)+12\bf F$ theory has twisted matter content $6\mathbf{F}_0\oplus 6\mathbf{F}_{1/2}$ which is invariant under KK-charge shift by $1/2$.  The circle reduction should give a 5d $\su(2)+6\bf F$ theory. The twisted circle compactification gives a highly nontrivial 5d KK theory with several 5d gauge theory descriptions which are 5d $\SU(3)_4 +6\bf F$, $G_2 +6\bf F$ and $\Sp(2)+2\mathbf{\Lambda}^2 +4\bf F$ theories \cite{Jefferson:2018irk}. To our knowledge, there is no 2d quiver gauge theory description for the twisted elliptic genera in this case. From the unity twisted elliptic blowup equations, we compute its twisted one-string elliptic genus $\IE_1$ to the order $q^{4}$. We obtain the coefficients $b_{ij}$ in Table \ref{tab:bijn1su3z2}. The symmetry coefficient matrix indicates the self-dual spectral flow symmetry as expected. The 5d one-instanton partition function of $\su(2)+6\bf F$ theory can be computed as
\be
Z_{1}^{ \su(2)+6\bf F}=\sum_{n=0}^{\infty}\Big(\chi^{\so(12)}_{S}\chi^{\SU(2)}_{n\theta}v^{1+2n}-\chi^{\so(12)}_{C}\chi^{\SU(2)}_{f+n\theta}v^{2+2n}\Big).
\ee
where $\so(12)$ is the pertubative 5d flavor symmetry and $S$ and $C$ are the spinor and conjugate spinor representations which both have dimension 32. 
One can see these coefficients are in total agreement with  the orange numbers in Table \ref{tab:bijn1su3z2}.

\begin{table}[h]
\begin{center}
$\begin{tabu}{c|ccccccccc}
q\backslash v  & 1 & & 2 & & 3 & & 4 & \\ \hline
0& \color{orange}32 & 0 & \color{orange}-64 & 0 & \color{orange}96 & 0 & \color{orange}-128 \\
1/4& 0 & 64 & 0 & -128 & 0 & 192 & 0 \\
1/2& \color{orange}-64 & 0 & 480 & 0 & -960 & 0 & 1440 \\
3/4& 0 & -128 & 0 & 960 & 0 & -1920 & 0 \\
1& \color{orange}96 & 0 & -960 & 0 & 4160 & 0 & -8128 \\
5/4& 0 & 192 & 0 & -1920 & 0 & 8192 & 0 \\
3/2& \color{orange}-128 & 0 & 1440 & 0 & -8128 & 0 & 27040 \\
\end{tabu}$
\caption{The coefficient matrix $b_{ij}$ for the  $\IE_1$ of 6d $\su(3)^{(2)}$ theory with $\fn=1$.}
\label{tab:bijn1su3z2}
\end{center}
\end{table}

\paragraph{\underline{$\fn=1,\so(8)^{(2)}$}}
The $\IZ_2$ twist of 6d $(1,0)$ $\so(8)+3(\bf V+S+C)$ theory has low energy gauge algebra $\cirG=\so(7)$ and twisted matter content $3(\mathbf{V}_0+\mathbf{S}_{0}+\mathbf{S}_{1/2})$. From the unity twisted elliptic blowup equations, we compute its twisted one-string elliptic genus $\IE_1$ to the order $q^{7}$. We summarize the Fourier coefficients $b_{ij}$ in Table \ref{tab:bijn1so8z2}. 
\begin{table}[h]
\begin{center}
$\begin{tabu}{c|ccccccccc}
q\backslash v  & 1&2& 3 & 4 & 5 & 6  \\ \hline
0& \color{red}14 & \color{orange}44 & \color{orange}42 &\color{orange} -168 & \color{orange}-2490 & \color{orange}18564 \\
1/2& \color{red}-112 & 128 & 896 & 896 & -5376 & -29568 \\
1& \color{red}504 & -2212 & -224 & 10768 & 12152 & -66528 \\
3/2& \color{red}-1680 & 11648 & -24416 & -16000 & 98672 & 130048 \\
2& \color{red}4620 & -41496 & 148526 & -198688 & -212310 & 751104 \\
5/2&\color{red} -11088 & 118272 & -557088 & 1380736 & -1323168 & -1918080 \\
3& \color{red}24024 & -290136 & 1629600 & -5386024 & 10423952 & -7603608 \\
\end{tabu}$
\caption{The coefficient matrix $b_{ij}$ for the  $\IE_1$ of 6d $\so(8)^{(2)}$ theory with $\fn=1$.}
\label{tab:bijn1so8z2}
\end{center}
\end{table}

The leading $q$ order of the NS-R sector should give the 5d $\so(7)+3\bf S$ theory whose 5d partition function can be computed  by localization \cite{Kim:2018gjo} or 5d blowup equations \cite{Kim:2019uqw}. We use the localization formula in \cite{Kim:2018gjo} to compute the 5d one-instanton partition function as 
\be
Z_{1}^{ \so(7)+3\bf S}=\frac{14 v^4}{(v+1)^8}=14v^4-112 v^5+504 v^6-1680 v^7+4620 v^8 -\dots
\ee
One can see these coefficients are in agreement with  the red numbers in Table \ref{tab:bijn1so8z2}. 
The leading $q$ order of the R-R elliptic genus should give the on-instanton partition function of 5d $\so(7)+3\mathbf{V}+3\bf S$ theory whose $v$ expansion coefficients are colored orange. We are not aware that any other method can compute the partition function of 5d $\so(7)+3\mathbf{V}+3\bf S$ theory.

\paragraph{\underline{$\fn=1,\so(8)^{(3)}$}}
The $\IZ_3$ twist of 6d $(1,0)$ $\so(8)+3(\bf V+S+C)$ theory has low energy gauge algebra $\cirG=G_2$ and twisted matter content $3(\mathbf{7}_0+\mathbf{7}_{1/3}+\mathbf{7}_{2/3})$. The circle reduction gives the 5d $G_2+3\bf F$ theory. From the unity twisted elliptic blowup equations, we compute its twisted one-string elliptic genus $\IE_1$ to the order $q^{16/3}$. We summarize the Fourier coefficients $b_{ij}$ in $(q,v)$ expansion in Table \ref{tab:bijn1so8z3}. The leading $q$ order of the R-R elliptic genus should give the 5d partition function of $G_2+3\bf F$ theory. We compute the 5d one-instanton partition function of $G_2+3\bf F$ theory from the localization formula in \cite{Kim:2018gjo} as
\be\nonumber
\ba
Z_{1}^{ G_2+3\bf F}&=\frac{-v^{12}+7 v^{10}+14 v^9-119 v^8+274 v^7-350 v^6+274 v^5-119 v^4+14 v^3+7 v^2-1}{\left(1-v^2\right)^6}\\
&=-1+v^2+14 v^3-98 v^4+358 v^5-973 v^6+2212 v^7-4452 v^8+8196 v^9-\dots
\ea
\ee
One can see these Fourier coefficients are in agreement with  the orange numbers in Table \ref{tab:bijn1so8z3} up to the first gauge singlet $1$. 
On the other hand, the leading $q$ order of the NS-R elliptic genus gives the 5d one $G_2$ instanton Hilbert series, as shown from the red numbers which are the $\chi_{n\theta}^{G_2}$ characters. The subleading $q$ order of the NS-R elliptic genus has coefficients $-\chi^{G_2}_{f+n\theta}\chi^{\Sp(3)}_{(100)}$ colored in blue in Table \ref{tab:bijn1so8z3}. These are the first excited states with KK charge $1/3$ above the NS-R vacuum.

\begin{table}[h]
\begin{center}
$\begin{tabu}{c|ccccccccc}
q\backslash v  & \phantom{,}2& 3 & 4 & 5 & 6 & 7 & 8 &9 \\ \hline
0& \phantom{,}\color{red}1 & \color{orange}14 & \color{orange}-98 & \color{orange}358 & \color{orange}-973 & \color{orange}2212 & \color{orange}-4452 & \color{orange}8196 \\
1/3& \phantom{,}0 & 28 & 176 & -1448 & 5488 & -15148 & 34720 & -70224 \\
2/3& \phantom{,}0 & \color{blue}-42 & 308 & 1498 & -12712 & 48230 & -133028 & 304682 \\
1& \phantom{,}0 & \color{red}14 & -784 & 2472 & 10304 & -85456 & 319712 & -874888 \\
4/3& \phantom{,}0 & 0 & 770 & -7630 & 16226 & 59970 & -480164 & 1763734 \\
5/3& \phantom{,}0 & 0 & \color{blue}-384 & 9793 & -54080 & 91168 & 306432 & -2363200 \\
2& \phantom{,}0 & 0 & \color{red}77 & -8246 & 79058 & -316218 & 452881 & 1409772 \\
7/3& \phantom{,}0 & 0 & 0 & 4802 & -79856 & 497280 & -1603616 & 2035490 \\
8/3& \phantom{,}0 & 0 & 0 & \color{blue}-1716 & 62083 & -557458 & 2643312 & -7287124 \\
3& \phantom{,}0 & 0 & 0 & \color{red}273 & -38304 & 498064 & -3169600 & 12400754 \\
\end{tabu}$
\caption{The coefficient matrix $b_{ij}$ for the  $\IE_1$ of 6d $\so(8)^{(3)}$ theory with $\fn=1$.}
\label{tab:bijn1so8z3}
\end{center}
\end{table}

\section{Modular bootstrap of the twisted elliptic genera}\label{sec:MA}
In this section, we utilize the modularity of elliptic genera to obtain some all order results beyond the $q$ expansion.  This compensates the results of twisted elliptic genera for $\fn=1,2$ twisted theories where from elliptic blowup equations we can only solve $\IE_1$ as $q$ expansion but cannot have a compact formula. The modular bootstrap approach views the elliptic genera as Jacobi forms on certain modular groups, and 
utilizes the finite generation property of modular forms and Jacobi forms to obtain a compact modular expression of elliptic genera which works for all $q$ orders.  

\subsection{The modular ansatz for twisted elliptic genera}
For a 6d $(1,0)$ SCFT with gauge group $G$ and tensor coefficient $\fn$, the modular ansatz for its reduced one-string elliptic genus was proposed by Del Zotto and Lockhart in \cite{DelZotto:2018tcj} as
\be\label{MADL}
\IE_1(\tau,\epsilon_+)=\frac{ \mathcal{N}(\tau,\epsilon_+)}{\eta^{12(\fn-2)-4+24\delta_{\fn,1}}\phi_{-2,1}(\tau,2\epsilon_+)^{h^\vee_{G}-1}}.
\ee
Here all gauge and flavor fugacities are turned off, and only $\epsilon_+$ serves as the elliptic parameter of the Jacobi form.  
The $\delta_{\fn,1}$ comes from the special phenomenon of $\fn=1$ theories, where the leading $q$ order of elliptic genus is always trivial, and the 5d theory information only emerges from the subleading $q$ order. Utilizing several other constraints, mostly from the universal features of the leading order of R-R and NS-R elliptic genera, \cite{DelZotto:2018tcj} successfully determined the modular ansatz for most rank-one theories. The modular ansatz for several remaining theories was determined in \cite{Gu:2020fem} with the help of elliptic blowup equations.

We would like to find a generalization of \eqref{MADL} for the reduced twisted one-string elliptic genera. 
An inspiration comes from the recent study on the modularity of topological strings on $N$-section Calabi-Yau threefolds \cite{Cota:2019cjx,Knapp:2021vkm,Schimannek:2021pau}. The modular ansatz for the topological string partition function on such geometries was first proposed in \cite{Cota:2019cjx}. Combining \eqref{MADL} and the results in \cite{Cota:2019cjx} together, we propose the following modular ansatz for the reduced one-string elliptic genus for twisted 6d (1,0) theories with twist coefficient $N=2,3,4$:
\be\label{MA}
\boxed{
\IE_1(\tau,\epsilon_+)=\frac{ \mathcal{N}(\tau,\epsilon_+)}{\eta^{12(\fn-2)-4+24\delta_{\fn,1}}\Delta_{2N}(\frac{\tau}{N})^{s}\phi_{-2,1}(\tau,2\epsilon_+)^{h^\vee_{\cirG}-1}},}
\ee
with
\be\label{eq:s}
\boxed{s=\frac{N}{N-1}\Big(c-\frac{\fn-2}{2}-\delta_{\fn,1}\Big).}
\ee
Recall the $c$ is the $c$ constant of twisted theories defined in  \eqref{goldall1} that  depends only on the tensor coefficient $\fn$ and twist coefficient $N$. The $\Delta_{2N}$ are some $\Gamma_0(N)$ cusp forms defined in Appendix \ref{app:B}. 
The numerator $\mathcal{N}(\tau,\epsilon_+)$ is of weight $6(\fn-2)+2Ns+12\delta_{\fn,1}-2h^\vee_{\cirG}$ and index $4({h^\vee_{\cirG}-1})+\fn-\dualCox $. The  $\mathcal{N}(\tau,\epsilon_+)$ usually has fractional $q^{1/N}$ orders thus it is convenient to scale $q ^{1/N}$ to $q$. Then we expect $\mathcal{N}(N\tau,\epsilon_+)$ to be a $\Gamma_1(N)$ weak Jacobi form with positive integer index. By the finite generation property of weak Jacobi forms \cite{eichler1985theory}:, we have
\be
\mathcal{N}(N\tau,\epsilon_+)\in M_{\star}(N)[\phi_{-2,1}(N\tau,\epsilon_+),\phi_{0,1}(N\tau,\epsilon_+)].
\ee
Here $M_{\star}(N)$ denotes the ring of modular forms on $\Gamma_1(N)$. The $\phi_{-2,1}(\tau,z)$ and $\phi_{0,1}(\tau,z)$ are the well-known Eichler-Zagier generators for weak Jacobi forms whose expressions are collected in Appendix \ref{app:B}. The $s$ here measure the difference between the twisted $c$ constant and the untwisted one and  is related to a geometric quantity $r_{\beta}=N s$ which is always integral \cite{Cota:2019cjx}. Interestingly, we find that except for the $\fn=1,\so(8)^{(3)}$ twisted theory, the modular ansatz of all twisted one-string elliptic genera encountered in the current paper satisfy a bigger congruence subgroup that is $\Gamma_0(N)$. We  collect the definition and modular generators for $\Gamma_0(N)$ and $\Gamma_1(N)$ in Appendix \ref{app:B}.

Some more explanation for the $N=4$ cases are needed. We notice that for some theories with twist coefficients $4$, it is possible that besides the $\Gamma_1(4)$ elements from $\Delta_8$, there can be extra $\Gamma_1(2)$ elements. In such case, the $\Delta_8(\frac{\tau}{4})^{s}$ in \eqref{MA} need to be extended to  $\Delta_8(\frac{\tau}{4})^{s_1}\Delta_4(\frac{\tau}{4})^{s_2}$. This suggests that the associated Calabi-Yau geometries contain both 4-section and 2-section components. 

We collect in Table \ref{tb:mainMA1} the relevant data for the modular ansatz for all rank-one twisted theories. The $\#$ means the number of independent parameters to be fixed in the numerator $\mathcal{N}$. The $\fn=2,\so(12)^{(2)}$ theory is marked with $*$ because we do not have enough data to fix the modular ansatz, as it involves half-hyper. Except for this theory, we successfully fix its modular ansatz for all other theories in Table \ref{tb:mainMA1} and check the ansatz against the results from twisted elliptic blowup equations to high $q$ orders. 

\begin{table}[h]
\centering
%\resizebox{\linewidth}{!}{
\begin{tabular}{|c|l|l|c|c|c|c|c|c|}\hline
  $\fn$
  &$G$ & $\mathring{G}$ & $N$
  & $c$ & $s$ & weight & index & \#
  \\ \hline
  6&$E_6^{(2)}$& $F_4$ & 2 & 5/4 & $-3/2$ & 0 & 26 & 196
   \\
\hline
5 &$E_6^{(2)}$& $F_4$ &2&   3/2 & 0 & 0 & 25 & 182
   \\
& &  & 2 & 3/4 & $-3/2$ & $-6$ & 25 & 144
   \\
\hline
  4&$D_4^{(3)}$& $G_2$ & 3&  5/9 & $-2/3$ & 0 & 10 & 44
 \\
  4&$D_{4}^{(2)}$& $B_{3}$ & 2 & 3/4 & $-1/2$ & 0 & 14 & 64
 \\
   4&$D_{5}^{(2)}$& $B_{4}$ & 2 & 3/4 & $-1/2$ & $-4$ & 20 &  100
 \\
   4&$D_{6}^{(2)}$& $B_{5}$& 2  & 3/4 & $-1/2$ & $-8$ & 26 &   144
 \\
    4&$D_{7}^{(2)}$& $B_{6}$ & 2 & 3/4 & $-1/2$ & $-12$ & 32 &   196
 \\
   4&$D_{r+4}^{(2)}$& $\!\!B_{r+3}\!\!$ & 2 & 3/4 & $-1/2$ & $-4r$ & $14+6r$ & $\cdots$\\
  4 &$E_6^{(2)}$& $F_4$ & 2 & 1 & 0 & $-6$ & 24 & 132 
   \\
\hline
  3&$A_2^{(2)}$  & $C_1$& 4  & 5/16 & $-1/4$ & 0 & 4 & 15
   \\
  3&$D_4^{(2)}$& $B_3$ & 2&   1/2 & 0 & $-4$ & 13 & 42 
\\
  3& $D_4^{(3)}$& $G_2$ & 3 & 1/2 & 0 & $-2$ & 9 & 30
  \\
 \hline
       2&$A_{1}^{(2)}$&$C_1$& 2 & 1/4 & 1/2 & $-2$ & 4 & 6
    \\
       2& $A_{2}^{(2)}$&$C_{1}$ & 4 &  $3/16$ & $1/4$ & $-2$ & $3$ & $ 6$ 
     \\
 &     &$C_{1,\theta=0/\pi}$\!\! & 4 &  3/8 & 1/2 & 0 & $3$ &  10    \\
          2&$A_{3}^{(2)}$&$C_2$& 2 & 1/4 & 1/2 & $-4$ & 6 & 9
    \\
       2& $A_{4}^{(2)}$&$C_{2}$ & 4 &  $3/16$ & $1/4$ & $-4$ & $5$ & $ 10$ 
     \\

             &  &  $C_{2,\theta=0/\pi}$ & 4 &  1/2 & $(1/2,1/2)$ & 0 & $5$ & 21  \\
          2&$A_{5}^{(2)}$&$C_3$& 2 & 1/4 & 1/2 & $-6$ & $8$ & 12
    \\
    2&$A_{2r-1}^{(2)}$&$C_{r}$& 2 & 1/4  & 1/2 & $-2r$ & $2r+2$ & $\cdots$
    \\
         2& $A_{2r}^{(2)}$&$C_{r}$ & 4 &  $3/16$ & $1/4$ & $-2r$ & $2r+1$ & $\cdots$ 
     \\
 &     &$C_{r,\theta=0/\pi}$\!\! & 4 &  $\frac{r}{8}+\frac{1}{4} $ & $(\frac12,\frac{r-1}{2}) $ & 0 & $2r+1$ &  $\cdots$    \\
    2&$D_4^{(2)}$& $B_3$& 2  & 1/4 & 1/2 & $-8$ & 12 & 25
  \\
    2& $D_4^{(3)}$& $G_2$& 3  & 4/9 & 2/3 & $-4$ & 8 & 19
   \\
    2&$D_5^{(2)}$ & $B_4$& 2 & 1/4 & 1/2 & $-12$ & 18 & 49
    \\
       2&$D_6^{(2)}$& $B_5$ & 2  &  1/4 & 1/2 & $-16$ & 24 &  $81^*$ \\
    2 &$E_6^{(2)}$& $F_4$ & 2&   $3/4$ & 3/2 & $-12$ & 22 & 81
    \\
   \hline
   1& $A_0^{(2)}$ & $C_0$ &2 & 0 & $-1$ & 0 & 0 & 1  \\
    1& $A_2^{(2)}$ & $C_1$ &  4 & 1/16 & $(-5/4,2)$ & 0& 2& 6 \\
    1&$D_4^{(2)}$& $B_3$ &  2&   0 & $-1$ & $-8$ & 11 & 20
 \\
    1&$D_4^{(3)}$& $G_2$ &3 & 7/18 & $-1/6$ & $-3$ & 7 & 15\\ 
   \hline
\end{tabular}
%}
\caption{Data of $\Gamma_1(N)$ modular ansatz for the twisted one-string elliptic genera of twisted 6d $(1,0)$ rank-one theories. For $N=4$, the $(s_1,s_2)$ means that the denominator contains $\Delta_8(\frac{\tau}{4})^{s_1}\Delta_4(\frac{\tau}{4})^{s_2}$ and only one $s$ means the $(s,0)$.}
\label{tb:mainMA1}
\end{table}

\subsection{Computational results}
We pick some interesting theories to explicitly show the modular ansatz results. The results for other theories in Table \ref{tb:mainMA1} can be shared to interested readers upon request.

\paragraph{\underline{$\fn=3,\SU(3)^{(2)}$}} 
For 6d pure $\su(3)^{(2)}$ theory studied in Section \ref{sec:su3z2}, let us first discuss the $\Sp(1)_{0}$ case. From the general formula \eqref{MA}, we have the following modular ansatz for its reduced twisted one-string elliptic genus
\be\label{MAsu3z2}
\IE_1(\tau,\epsilon_+)=\frac{\Delta_8(\tau/4)^{\frac{1}{4}}\mathcal{N}(\tau,\epsilon_+)}{\eta(\tau)^{8}\phi_{-2,1}(\tau,2\epsilon_+)}.
\ee
Here $\mathcal{N}(\tau,\epsilon_+)$ is of weight 0 and index 4. Interestingly, we notice that in this case the $\Delta_8(\tau/4)^{\frac{1}{4}}$ has taken care of all the $\Gamma_0(4)$ elements such that $\mathcal{N}(\tau,\epsilon_+)$  contains no $q^{1/4+n/2}$ orders, i.e., $\mathcal{N}(2\tau,\epsilon_+)$ is a $\Gamma_0(2)$ Jacobi form. Using the generators $E_2(\tau)^{(2)},E_4(2\tau)$ and $\phi_{-2,1}(2\tau,\epsilon_+),\phi_{0,1}(2\tau,\epsilon_+)$ given in  Appendix \ref{app:B}, there are 9 parameters to fix the ansatz of $\mathcal{N}(2\tau,\epsilon_+)$. 
We find the ansatz can be determined by the known Fourier coefficients of $q^{0,1,2}$ of $\mathcal{N}(2\tau,\epsilon_+)$ as
\begin{align}\nonumber
\mathcal{N}(2\tau,\epsilon_+)&=\frac{1}{10368}\Big( \big(-4 (E_2^{(2)})^4+18 E_4 (E_2^{(2)})^2-9 E_4^2\big)\phi_{-2}^4+2 E_2^{(2)} \big(2 (E_2^{(2)})^2-9 E_4\big)\phi_{-2}^3\phi_0\\
&\phantom{=\frac{1}{10368}\Big(} -6 \big((E_2^{(2)})^2-3 E_4\big)\phi_{-2}^2\phi_0^2-2 E_2^{(2)}\phi_{-2}\phi_0^3-\phi_0^4
\Big) .
\end{align}
We then check the modular ansatz results to $q^{8}$ order. Here $\phi_{-2},\phi_0$ are the short notation of $\phi_{-2,1}(2\tau,\epsilon_+),\phi_{0,1}(2\tau,\epsilon_+)$. On the other hand, if we regard $\mathcal{N}(4\tau,\epsilon_+)$ as a $\Gamma_0(4)$ Jacobi form, there will be 15 coefficients to fix $\mathcal{N}(4\tau,\epsilon_+)$ using $E_2^{(2)}$ and $E_2^{(4)}$ given in  Appendix \ref{app:B} as modular generators. For the $\Sp(1)_{\pi}$ case, the modular ansatz takes the same form as \eqref{MAsu3z2} and the $\mathcal{N}(\tau,\epsilon_+)$ is still of weight 0 and index 4. Amusingly, we find in this case the $\mathcal{N}(2\tau,\epsilon_+)$ is again a $\Gamma_0(2)$ Jacobi form which we determine to be
\be
\mathcal{N}(2\tau,\epsilon_+)=-\frac{1}{10368}\Big(\big(2(E_2^{(2)})^2-3 E_4\big)\phi_{-2}^2 +2E_2^{(2)}\phi_{-2}\phi_0  -\phi_{0}^2\Big)^2 .
\ee
This numerator is still fixed by the known Fourier coefficients of $q^{0,1,2}$ and then checked to $q^8$ order against the twisted elliptic blowup equations. 
\paragraph{\underline{$\fn=4,\SO(8)^{(2)}$}} 
For 6d pure $\so(8)^{(2)}$ theory, from the general formula \eqref{MA}, we have the following modular ansatz for its reduced twisted one-string elliptic genus
\be
\IE_1(\tau,\epsilon_+)=\frac{\Delta_4(\tau/2)^{\frac{1}{2}}\mathcal{N}(\tau,\epsilon_+)}{\eta(\tau)^{20}\phi_{-2,1}(\tau,2\epsilon_+)^4}.
\ee
Here $\mathcal{N}(\tau,\epsilon_+)$ is of weight 0 and index 14. Based on the relation between twists and modularity, the $\mathcal{N}(2\tau,\epsilon_+)$ should be a  $\Gamma_0(2)$ Jacobi form. Again using the generators $E_2(\tau)^{(2)},E_4(2\tau)$ and $\phi_{-2,1}(2\tau,\epsilon_+),\phi_{0,1}(2\tau,\epsilon_+)$, we find there are 64 coefficients to fix $\mathcal{N}(2\tau,\epsilon_+)$. They can be fixed from the known Fourier coefficients up to $q^7$. We have
\begin{align}\nonumber
    \mathcal{N}(2\tau,\epsilon_+)=&\,-\frac{1}{160489808068608}\Big(\big(2(E_2^{(2)})^2-3 E_4\big)\phi_{-2}^2 +2E_2^{(2)}\phi_{-2}\phi_0  -\phi_{0}^2\Big)\Big((64 (E_2^{(2)})^{12}\\ \nonumber
    &-672 E_4 (E_2^{(2)})^{10}+3024 E_4^2 (E_2^{(2)})^8-9072 E_4^3 (E_2^{(2)})^6+17982 E_4^4 (E_2^{(2)})^4\\ \nonumber
    &-15066 E_4^5 (E_2^{(2)})^2+5103 E_4^6)\phi_{-2}^{12}+6 E_2^{(2)} \big(64 (E_2^{(2)})^{10}-800 E_4 (E_2^{(2)})^8\\ \nonumber
    &+3456 E_4^2 (E_2^{(2)})^6-5616 E_4^3 (E_2^{(2)})^4+2268 E_4^4 (E_2^{(2)})^2-1053 E_4^5\big)\phi_{-2}^{11}\phi_0\\ \nonumber
    &-18 \big(64 (E_2^{(2)})^{10}-720 E_4 (E_2^{(2)})^8+2544 E_4^2 (E_2^{(2)})^6-2340 E_4^3 (E_2^{(2)})^4\\ \nonumber
    &-1836 E_4^4 (E_2^{(2)})^2+513 E_4^5\big)\phi_{-2}^{10}\phi_0^2+2 E_2^{(2)} \big(1696 (E_2^{(2)})^8-18000 E_4 (E_2^{(2)})^6\\ \nonumber
    &+64368 E_4^2 (E_2^{(2)})^4-77760 E_4^3 (E_2^{(2)})^2+1701 E_4^4\big)\phi_{-2}^{9}\phi_0^3-9 \big(80 (E_2^{(2)})^8\\ \nonumber
    &-1648 E_4 (E_2^{(2)})^6+8316 E_4^2 (E_2^{(2)})^4-12276 E_4^3 (E_2^{(2)})^2-945 E_4^4\big)\phi_{-2}^{8}\phi_0^4\\ \nonumber
    &    +108 E_2^{(2)} \big(8 (E_2^{(2)})^6-72 E_4 (E_2^{(2)})^4+332 E_4^2 (E_2^{(2)})^2-597 E_4^3\big)\phi_{-2}^{7}\phi_0^5\\ \nonumber
    & +12 \big(128 (E_2^{(2)})^6-690 E_4 (E_2^{(2)})^4+846 E_4^2 (E_2^{(2)})^2+675 E_4^3\big) \phi_{-2}^{6}\phi_0^6\\ \nonumber
    & +36 E_2^{(2)} \big(8 (E_2^{(2)})^4+8 E_4 (E_2^{(2)})^2-63 E_4^2\big) \phi_{-2}^{5}\phi_0^7+135 \big(2 (E_2^{(2)})^4-6 E_4 (E_2^{(2)})^2\\ \nonumber
    &+7 E_4^2\big)  \phi_{-2}^{4}\phi_0^8 +2 E_2^{(2)} \big(52 (E_2^{(2)})^2-123 E_4\big)\phi_{-2}^{3}\phi_0^9 +6 \big(8 (E_2^{(2)})^2-19 E_4\big) \phi_{-2}^{2}\phi_0^{10}\\
    &+18 E_2^{(2)} \phi_{-2}\phi_0^{11}
    +7\phi_{0}^{12}\Big).
\end{align}
We then check the modular ansatz results to $q^{16}$ against the 2d localization formula and find perfect agreement. 

\paragraph{\underline{$\fn=4,\SO(8)^{(3)}$}} 
For 6d pure $\so(8)^{(3)}$ theory, from the general formula \eqref{MA}, we have the following modular ansatz for its reduced twisted one-string elliptic genus
\be
\IE_1(\tau,\epsilon_+)=\frac{\Delta_6(\tau/3)^{\frac{2}{3}}\mathcal{N}(\tau,\epsilon_+)}{\eta(\tau)^{20}\phi_{-2,1}(\tau,2\epsilon_+)^3}.
\ee
Here $\mathcal{N}(\tau,\epsilon_+)$ is of weight 0 and index 10. The $\mathcal{N}(3\tau,\epsilon_+)$ should be a  $\Gamma_0(3)$ Jacobi form. Using the generators $E_2(\tau)^{(3)},E_4(3\tau),E_6(3\tau)$ and $\phi_{-2,1}(3\tau,\epsilon_+),\phi_{0,1}(3\tau,\epsilon_+)$ given in  Appendix \ref{app:B}, we find there are 67 coefficients to fix $\mathcal{N}(3\tau,\epsilon_+)$. We find they can be fixed from the data up to $q^7$ up to the vanishing relation \eqref{relationgen3}, which involve in total 44 independent parameters. We obtain
\begin{align}\nonumber
    \mathcal{N}(3\tau,\epsilon_+)=&\,\frac{1}{10319560704}\Big(E_2^{(3)}\big(2268 (E_2^{(3)})^9-14661 E_4 (E_2^{(3)})^7+20880 E_6 (E_2^{(3)})^6\\ \nonumber
    &-25560 E_4 E_6 (E_2^{(3)})^4+22176 E_6^2 (E_2^{(3)})^3-7488 E_4 E_6^2 E_2^{(3)}+2560 E_6^3\big)\phi_{-2}^{10}\\ \nonumber
    &+\! 12 \big(39 (E_2^{(3)})^9-252 E_4 (E_2^{(3)})^7+300 E_6 (E_2^{(3)})^6-54 E_4 E_6 (E_2^{(3)})^4\!-96 E_6^2 (E_2^{(3)})^3\\ \nonumber
    &+96 E_4 E_6^2 E_2^{(3)}-128 E_6^3\big)\phi_{-2}^9\phi_{0}
+    27 \big(39 (E_2^{(3)})^8-252 E_4 (E_2^{(3)})^6+344 E_6 (E_2^{(3)})^5\\ \nonumber
    &-336 E_4 E_6 (E_2^{(3)})^3+256 E_6^2 (E_2^{(3)})^2+64 E_4 E_6^2\big)\phi_{-2}^8\phi_{0}^2
    +24 E_2^{(3)} \big(18 (E_2^{(3)})^6\\ \nonumber
    &-117 E_4 (E_2^{(3)})^4+93 E_6 (E_2^{(3)})^3+234 E_4 E_6 E_2^{(3)}-416 E_6^2\big)\phi_{-2}^7\phi_{0}^3-126 \big(6 (E_2^{(3)})^6\\ \nonumber
    &-39 E_4 (E_2^{(3)})^4+48 E_6 (E_2^{(3)})^3-28 E_4 E_6 E_2^{(3)}-16 E_6^2\big)
    \phi_{-2}^6\phi_{0}^4
    -1512 E_4 E_6\phi_{-2}^5\phi_{0}^5\\ \nonumber
    &+42 E_2^{(3)} \big(3 (E_2^{(3)})^3-18 E_4 E_2^{(3)}+20 E_6\big)
    \phi_{-2}^4\phi_{0}^6+
    72 (E_2^{(3)} E_4-E_6)\phi_{-2}^3\phi_{0}^7\\ 
    &
   + 27 E_4\phi_{-2}^2\phi_{0}^8-4 E_2^{(3)}\phi_{-2}\phi_{0}^9
    -3 \phi_{0}^{10} \Big).
\end{align}
Here $\phi_{-2},\phi_0$ are short for $\phi_{-2,1}(3\tau,\epsilon_+),\phi_{0,1}(3\tau,\epsilon_+)$. 
We then check the modular ansatz results to $q^{16}$ against the twisted elliptic genera solved from the recursion formula.
\paragraph{\underline{$\fn=6, E_6^{(2)}$}} 
For 6d pure $E_6^{(2)}$ theory, the general formula \eqref{MA} gives the following modular ansatz for its reduced twisted one-string elliptic genus
\be
\IE_1(\tau,\epsilon_+)=\frac{\Delta_4(\tau/2)^{\frac{3}{2}}\mathcal{N}(\tau,\epsilon_+)}{\eta(\tau)^{44}\phi_{-2,1}(\tau,2\epsilon_+)^8}.
\ee
Here $\mathcal{N}(\tau,\epsilon_+)$ is of weight 0 and index 26. The $\mathcal{N}(2\tau,\epsilon_+)$ should be a $\Gamma_0(2)$ Jacobi form. Using the generators $E_2(\tau)^{(2)},E_4(2\tau)$ and $\phi_{-2,1}(2\tau,\epsilon_+),\phi_{0,1}(2\tau,\epsilon_+)$, we find there are 196 coefficients to fix $\mathcal{N}(2\tau,\epsilon_+)$. Indeed we find they can be fixed from the data up to $q^{13}$. We then check the modular ansatz results against the twisted elliptic genera solved from the recursion formula to $q^{14}$. The explicit formula for $\mathcal{N}(2\tau,\epsilon_+)$ is too long to show here.

\paragraph{\underline{E-string}}
The elliptic genera of E-strings have been studied in \cite{Haghighat:2014pva,Kim:2014dza,Cai:2014vka,Kim:2015fxa,Duan:2018sqe}. For a 6d theory with trivial gauge group in the tensor branch, there is no discrete symmetry from the automorphism of the gauge algebra. However, one can still consider discrete symmetries of the flavor algebra. A typical example is the E-string theory, where the flavor symmetry is the exceptional group $E_8$, for which we may find a rich class of discrete subgroups. The twisted circle compactification with such a discrete symmetry shall lead to a theory whose partition function is the same as the partition function of E-strings on a circle with a special choice of the Wilson line parameters. At first glance, it looks like these theories are boring, but as have been studied in \cite{Cota:2019cjx,Knapp:2021vkm}, their geometries, which are genus-one fibrations with $N$-sections, have very rich structures. Thus it is worthy to have a study on them. For example, let us do a $\IZ_2$ twist to E-string theory formally. The twisted one-string elliptic genus is just
\be
\IE_1(\tau,\epsilon_+,m_i)=\frac{1}{2}\sum_{(r,s)=(1,4),(4,1)}^{(2,3),(3,2)}\prod_{i=1}^4\prod_{j=5}^8\frac{\theta_r(m_i)\theta_s(m_j)}{\eta^2} .
\ee
Turning off the flavor parameters,  the reduced twisted one-string elliptic genus is independent from $\epsilon_+$ and has the following nice modular ansatz as a special case of our general formula \eqref{MA}:
\be
\IE_1(\tau,\epsilon_+)=\frac{\theta_2^4\theta_3^4}{\eta^8}=\frac{16\Delta_4(\frac{\tau}{2})}{\eta^8} .
\ee
It is easy to prove this identity. This has been also studied in the setting of Calabi-Yau threefolds with 2-sections in \cite{Cota:2019cjx}.

\paragraph{\underline{$\fn=2,\SU(2)^{(2)}$}} 
This case is similar to the E-string theory, where the $\IZ_2$ twist is only performed on the flavor parameters. 
From the general formula \eqref{MA}, we find the following modular ansatz
\be\label{MA:n2su2}
\IE_1(\tau,\epsilon_+)=\frac{ \mathcal{N}(\tau,\epsilon_+)}{\eta(\tau)^{-4}\Delta_4(\frac{\tau}{2})^{\frac{1}{2}}\phi_{-2,1}(\tau,2\epsilon_+)}.
\ee
Here $\mathcal{N}(\tau,\epsilon_+)$ is of weight $-2$ and index 4. In the case, the $\mathcal{N}(2\tau,\epsilon_+)$ is a $\Gamma_0(2)$ Jacobi form. 
We fix it to be 
\be
\mathcal{N}(2\tau,\epsilon_+)=\frac{\phi _{-2}}{864}  \big(E^{(2)}_2 \phi _{-2}-\phi _0\big) \left(2 (E^{(2)}_2)^2 \phi _{-2}^2-3 E_4 \phi _{-2}^2+2 E^{(2)}_2 \phi _0 \phi _{-2}-\phi _0^2\right).
\ee
Recall $\phi_{-2}$ and $\phi_0$ are short for $\phi_{-2,1}(2\tau,\epsilon_+),\phi_{0,1}(2\tau,\epsilon_+)$. 
We have checked this modular ansatz against the localization formula \eqref{localizationn2su2z2} to $\mathcal{O}(q^{20})$. In fact, we find an even simpler form for the numerator:
\be
\mathcal{N}(2\tau,\epsilon_+)=2\phi_{-2}(\epsilon_+)h(\epsilon_+)^2h(2\epsilon_+),\qquad h(\tau,z)=\frac{\theta_4(2\tau,z)}{\theta_4(2\tau)}.
\ee
Here we notice that
\be
E^{(2)}_2 \phi _{-2}-\phi _0=-12h(\epsilon_+)^2,\\
\ee
\be
2 (E^{(2)}_2)^2 \phi _{-2}^2-3 E_4 \phi _{-2}^2+2 E^{(2)}_2 \phi _0 \phi _{-2}-\phi _0^2=-144h(2\epsilon_+).
\ee
In summary, we find the following compact formula for the twisted one-string elliptic genus:
\be
\IE_1(\tau,\epsilon_+)=\frac{\theta_2(\tau)\theta_3(\tau)\eta(\tau)\theta_4(\tau,2\epsilon_+)}{\theta_2(\tau,\epsilon_+)^2\theta_3(\tau,\epsilon_+)^2}=\frac{2\eta(\tau)^5}{\eta(\frac{\tau}{2})^2}\frac{\theta_4(\tau,2\epsilon_+)}{\theta_2(\tau,\epsilon_+)^2\theta_3(\tau,\epsilon_+)^2}
\ee

\paragraph{\underline{$\fn=2,\SU(3)^{(2)}$}} 
From the general formula \eqref{MA}, we have the following modular ansatz for its reduced twisted one-string elliptic genus
\be\label{MAn2su3z2}
\IE_1(\tau,\epsilon_+)=\frac{\mathcal{N}(\tau,\epsilon_+)}{\eta(\tau)^{-4}\Delta_8(\tau/4)^{\frac{1}{4}}\phi_{-2,1}(\tau,2\epsilon_+)}.
\ee
Here $\mathcal{N}(\tau,\epsilon_+)$ is of weight $-2$ and index 3. We find that  unlike the pure $\SU(3)^{(2)}$ case, now the $\mathcal{N}(\tau,\epsilon_+)$ does contain quarter $q$ orders. Therefore, there is no longer simplification to $\Gamma_0(2)$, i.e., the $\mathcal{N}(4\tau,\epsilon_+)$ should be a genuine  $\Gamma_0(4)$ Jacobi form. Using the $\Gamma_0(4)$ modular generators $E_2(\tau)^{(2)},E_2^{(4)}(\tau)$ and $\phi_{-2,1}(4\tau,\epsilon_+),\phi_{0,1}(4\tau,\epsilon_+)$ given in Appendix \ref{app:B}, we find there are 6 parameters to fix the modular ansatz. By the known coefficients of $\mathcal{N}(4\tau,\epsilon_+)$ up to $q^6$, we determine it to be
\be
\mathcal{N}(4\tau,\epsilon_+)=\frac{1}{288}\phi_{-2} \big(  (E_2^{(2)}+3E_2^{(4)})\phi_{-2}-4\phi_{0} \big)\big(  (E_2^{(2)}-3E_2^{(4)})\phi_{-2}+2\phi_{0} \big)  .
\ee
Here $\phi_{-2}$ and $\phi_0$ are short for $\phi_{-2,1}(4\tau,\epsilon_+),\phi_{0,1}(4\tau,\epsilon_+)$. We then check the modular ansatz against the results from blowup equations to $q^{16}$. In fact, we observe an even simpler form for the numerator:
\be
\mathcal{N}(4\tau,\epsilon_+)=-4\phi_{-2,1}(4\tau,\epsilon_+)h(2\tau,\epsilon_+)^2h(\tau,\epsilon_+),\qquad h(\tau,z)=\frac{\theta_4(2\tau,z)}{\theta_4(2\tau)}.
\ee
Here we notice that
\be
(E_2^{(2)}-3E_2^{(4)})\phi_{-2}+2\phi_{0} =24h(2\tau,\epsilon_+)^2,\\
\ee
\be
(E_2^{(2)}+3E_2^{(4)})\phi_{-2}-4\phi_{0}=-48h(\tau,\epsilon_+).
\ee
In summary, we find the following compact formula for the twisted one-string elliptic genus:
\be
\IE_1(\tau,\epsilon_+)=\frac{4\eta(\frac{\tau}{4})^2\theta_4(\frac{\tau}{2},\epsilon_+)}{\eta(\frac{\tau}{2})^2\theta_4(\frac{\tau}{2})}\Big(\frac{\theta_1(\tau,\epsilon_+)\theta_4(\tau,\epsilon_+)}{\theta_1(\tau,2\epsilon_+)\theta_4(\tau)}\Big)^2.
\ee

\paragraph{\underline{$\fn=2,\SU(5)^{(2)}$}} 
From the general formula \eqref{MA}, we have the following modular ansatz for its reduced twisted one-string elliptic genus
\be\label{MAn2su5z2}
\IE_1(\tau,\epsilon_+)=\frac{\mathcal{N}(\tau,\epsilon_+)}{\eta(\tau)^{-4}\Delta_8(\tau/4)^{\frac{1}{4}}\phi_{-2,1}(\tau,2\epsilon_+)^2}.
\ee
Here $\mathcal{N}(\tau,\epsilon_+)$ is of weight $-4$ and index 5. We find the $\mathcal{N}(4\tau,\epsilon_+)$ should be a $\Gamma_0(4)$ Jacobi form. Using the $\Gamma_0(4)$ modular generators $E_2(\tau)^{(2)},E_2^{(4)}(\tau)$ and $\phi_{-2,1}(4\tau,\epsilon_+),\phi_{0,1}(4\tau,\epsilon_+)$, we find there are 10 parameters to fix the modular ansatz. By the known coefficients of $\mathcal{N}(4\tau,\epsilon_+)$ up to $q^4$, we determine it to be
\be
\mathcal{N}(4\tau,\epsilon_+)=-\frac{1}{1728}\phi_{-2}^2 \big(  (E_2^{(2)}+3E_2^{(4)})\phi_{-2}-4\phi_{0} \big)\big(  (E_2^{(2)}-3E_2^{(4)})\phi_{-2}+2\phi_{0} \big)^2  .
\ee
Here $\phi_{-2}$ and $\phi_0$ are short for $\phi_{-2,1}(4\tau,\epsilon_+),\phi_{0,1}(4\tau,\epsilon_+)$. We then check the modular ansatz against the results from blowup equations to $q^{12}$.

\paragraph{\underline{$\fn=1,\SU(3)^{(2)}$}} 
The twisted one-string elliptic genus has been solved from blowup equations in early sections and the Fourier coefficients have been shown in Table \ref{tab:bijn1su3z2}. 
From the general formula \eqref{MA}, we have the following modular ansatz 
\be\label{MA:n1su3}
\IE_1(\tau,\epsilon_+)=\frac{\Delta_8^{\frac{5}{4}}({\tau}/{4}) \mathcal{N}(\tau,\epsilon_+)}{\eta^{8}\Delta_4^2({\tau}/{4})\phi_{-2,1}(\tau,2\epsilon_+)}.
\ee
Here $\mathcal{N}(\tau,\epsilon_+)$ is of weight $0$ and index 2. Note there are both $\Delta_8$ and $\Delta_4$ which indicates that both $\Gamma_0(4)$ and $\Gamma_0(2)$ play a role. The $\mathcal{N}(4\tau,\epsilon_+)$ is a genuine  $\Gamma_0(4)$ Jacobi form. Using the $\Gamma_0(4)$ modular generators $E_2(\tau)^{(2)},E_2^{(4)}(\tau)$ and $\phi_{-2,1}(4\tau,\epsilon_+),\phi_{0,1}(4\tau,\epsilon_+)$ given in Appendix \ref{app:B}, we find there are 6 parameters to fix for the modular ansatz of $\mathcal{N}(4\tau,\epsilon_+)$. Indeed, they are fixed by the known $q^{0,1,2}$ coefficients. We have 
\be
\mathcal{N}(4\tau,\epsilon_+)=\frac43 E_2^{(4)}\phi_{-2}\big(  (E_2^{(2)}-3E_2^{(4)})\phi_{-2}+2\phi_{0} \big)   .
\ee
Here $\phi_{-2}$ and $\phi_0$ are short for $\phi_{-2,1}(4\tau,\epsilon_+),\phi_{0,1}(4\tau,\epsilon_+)$. 
We then check the modular ansatz against the twisted elliptic genera solved from blowup equations to $q^{16}$ order.

\paragraph{\underline{$\fn=1,\SO(8)^{(3)}$}} 
In the end we show one example of $\IZ_3$ twist which is the only case we encounter in the current work that the modular group must be $\Gamma_1(N)$ instead of $\Gamma_0(N)$. This is the $\so(8)^{(3)}$ theory with $\fn=1$. The general formula \eqref{MA} gives the following modular ansatz for its reduced twisted one-string elliptic genus
\be
\IE_1(\tau,\epsilon_+)=\frac{\Delta_6(\tau/3)^{\frac{1}{6}}\mathcal{N}(\tau,\epsilon_+)}{\eta(\tau)^{8}\phi_{-2,1}(\tau,2\epsilon_+)^3}.
\ee
Here $\mathcal{N}(\tau,\epsilon_+)$ is of weight $-3$ and index 7. Due to the odd weight, the $\mathcal{N}(3\tau,\epsilon_+)$ must be a  $\Gamma_1(3)$ Jacobi form instead of a $\Gamma_0(3)$ one. Using the generators $E_2(\tau)^{(3)}$, $E_4(3\tau)$, $E_3^{(a)}(\tau)$, $E_3^{(b)}(\tau)$ and $\phi_{-2,1}(3\tau,\epsilon_+),\phi_{0,1}(3\tau,\epsilon_+)$ given in Appendix \ref{app:B}, we find there are 26 coefficients to fix $\mathcal{N}(3\tau,\epsilon_+)$. Indeed they can be fixed from the data up to $q^3$ up to the vanishing relations \eqref{relationgen3a}. We have
\begin{align}\nonumber
    \mathcal{N}(3\tau,\epsilon_+)=&-\frac{\phi_{-2}^3}{186624}\Big( E_2^{(3)} \big(19 (E_2^{(3)})^3 E_3^{(a)}\!+450 (E_2^{(3)})^3 E_3^{(b)}\!-45 E_4 E_2^{(3)} E_3^{(a)}\!-64 (E_3^{(a)})^3\big)\phi_{-2}^4\\ \nonumber
&\ +6 \big(17 (E_2^{(3)})^3 E_3^{(a)}-342 (E_2^{(3)})^3 E_3^{(b)}+45 E_4 E_2^{(3)} E_3^{(a)}+16 (E_3^{(a)})^3\big)    \phi_{-2}^3\phi_0\\ \nonumber
&\ 
-27 \big(21 (E_2^{(3)})^2 E_3^{(a)}-72 (E_2^{(3)})^2 E_3^{(b)}+11 E_4 E_3^{(a)}\big) \phi_{-2}^2\phi_0^2+36 E_2^{(3)} (19 E_3^{(a)}\\
&\ +69 E_3^{(b)})\phi_{-2}\phi_0^3  -18 (11 E_3^{(a)}+189 E_3^{(b)})\phi_0^4
    \Big).
\end{align}
We then checked it to $q^{11}$ order against the twisted elliptic blowup equations and found perfect agreement. 

\subsection{Full modular ansatz with gauge fugacities}
It is also interesting to consider the full modular ansatz with gauge fugacities. For untwisted 6d $(1,0)$ SCFTs, this has been studied in \cite{DelZotto:2017mee} for pure $\su(3)$ and $\so(8)$ theories, see also \cite{Kim:2018gak,Duan:2020imo}. The denominator of the modular ansatz of reduced one-string elliptic genus involves the block
\be
\prod_{\alpha\in\Delta_l^+(G)}\theta_1(m_{\alpha}\pm2\epsilon_+).
\ee
This is natural as all the poles of elliptic genera are associated to the 5d states related to the long roots of $G$.  
Analogously for twisted theories, given the understanding on 5d circle reduction theory, it is reasonable to expect the the denominator of the modular ansatz for twisted one-string elliptic genera involves the block 
\be
\prod_{\alpha\in\Delta_l^+(\mathring{G})}\theta_1(m_{\alpha}\pm2\epsilon_+),
\ee
where $\Delta_l^+$ is the long positive roots of $\mathring{G}$. More precisely, we propose the following full modular ansatz for the reduced  one-string twisted elliptic genera:
\be\label{MAfull}
\IE_1(\tau,\epsilon_+,m_{\cirG},m_{\cirF})=\frac{ \mathcal{N}(\tau,\epsilon_+,m_{\cirG},m_{\cirF})}{\eta^{12(\fn-2)-4+24\delta_{\fn,1}}\Delta_{2N}(\frac{\tau}{N})^{s}\prod_{\alpha\in\Delta_l^+(\mathring{G})}\phi_{-1,1/2}(m_{\alpha}\pm2\epsilon_+)},
\ee
where the $s$ is defined the same as in \eqref{eq:s}. The $\phi_{-1,1/2}(\tau,z):=\phi_{-2,1}(\tau,z)^{1/2}$ is a Jacobi form with weight $-1$ and index $1/2$. The weight and index of $\mathcal{N}(\tau,\epsilon_+,m_{\cirG},m_{\cirF})$ should be  coordinated with the weight 0 and index \eqref{eq:twistedEd-index} of the $\IE_1(\tau,\epsilon_+,m_{\cirG},m_{\cirF})$. In general, for twist coefficient $N$, the $\mathcal{N}(N\tau,\epsilon_+,m_{\cirG},m_{\cirF})$ should be a $\Gamma_1(N)$ Jacobi form of multi elliptic variables. In principle, one can fix the numerator by the $M_*(N)$ modular generators, the $\phi_{-2,1}(N\tau,\epsilon_+),\phi_{0,1}(N\tau,\epsilon_+)$ generators, the generators of $\cirG$ Weyl invariant weak Jacobi forms and the generators of $\cirF$ Weyl invariant weak Jacobi forms all together.
 
We give a simple example here, which is the pure $\su(3)^{(2)}$ theory with $\fn=3$. The low energy algebra is $\cirG=\Sp(1)$. We have the following anstaz for the reduced one-string elliptic genus with $\Sp(1)$ fugacity $m$ as
\be\label{MAsu3z2f}
\IE_1(\tau,\epsilon_+,m)=\frac{\Delta_8(\tau/4)^{\frac{1}{4}}\mathcal{N}(\tau,\epsilon_+,m)}{\eta(\tau)^{2}\theta_1(2m\pm2\epsilon_+)}.
\ee
Here $\mathcal{N}(\tau,\epsilon_+,m)$ has weight 0 and index 4 for $\epsilon_+$ and index 1 for $m$. We find the $\mathcal{N}(4\tau,\epsilon_+,m)$ is a $\Gamma_0(4)$ Jacobi form with two elliptic variables. Using the $\Gamma_0(4)$ modular generators $E_2(\tau)^{(2)},E_2^{(4)}(\tau)$ and $\phi_{-2,1}(4\tau,\epsilon_+),\phi_{0,1}(4\tau,\epsilon_+)$ and $\phi_{-2,1}(4\tau,m),\phi_{0,1}(4\tau,m)$ given in Appendix \ref{app:B}, we find there are 35 parameters to fix for the modular ansatz of $\mathcal{N}(4\tau,\epsilon_+,m)$. Indeed, they are fixed by the known coefficients up to $q^5$. We have 
\begin{align}\nonumber
    \mathcal{N}(4\tau,\epsilon_+,m)= &\,\frac{1}{127401984} \Big((5E_2^{(2)}-9E_2^{(4)}) \varphi_{-2}+4\varphi_0  \Big)\Big(
   \big((E_2^{(2)})^4+636 E_2^{(4)} (E_2^{(2)})^3\\ \nonumber
   &-3834 (E_2^{(4)})^2 (E_2^{(2)})^2+8316 (E_2^{(4)})^3 E_2^{(2)}-6399 (E_2^{(4)})^4\big) \phi_{-2}^4    
-16 \big((E_2^{(2)})^3\\ \nonumber
   &-117 E_2^{(4)} (E_2^{(2)})^2+567 (E_2^{(4)})^2 E_2^{(2)}-675 (E_2^{(4)})^3\big)   \phi_{-2}^3\phi_{0} +96 \big((E_2^{(2)})^2\\ \nonumber
   &+30 E_2^{(4)} E_2^{(2)}-63 (E_2^{(4)})^2\big)
   \phi_{-2}^2\phi_{0}^2   
  -256 (E_2^{(2)}-3 E_2^{(4)}) \phi_{-2}\phi_{0}^3+256 \phi _{0}^4
    \Big).
\end{align}
Here $\varphi_{-2},\varphi_0$ are short for $\phi_{-2,1}(4\tau,m),\phi_{0,1}(4\tau,m)$, while $\phi_{-2},\phi_0$ are short for $\phi_{-2,1}(4\tau,\epsilon_+)$ and $\phi_{0,1}(4\tau,\epsilon_+)$. 
We then check the modular ansatz against the twisted elliptic genera from 2d localization formula to $q^{10}$ order. Note in $\mathcal{N}(4\tau,\epsilon_+,m)$ the $m$ and $\epsilon_+$ dependence naturally factorizes. This is expected as there is no $m\epsilon_+$ term in its index quadratic form. 

\section{Twisting from Higgsing}\label{sec:higgsing}
Many twisted 6d $(1,0)$ SCFTs can be Higgsed from untwisted 6d $(1,0)$ SCFTs \cite{Kim:2019dqn,Hayashi:2021pcj}. Most previously known examples are recognized from brane webs. In particular, many examples of twisting from Higgsing of $\mf{so}$ type with $\fn=2,3,4$ were discussed in \cite{Kim:2019dqn}. In this section, we systematically study all possible twisting from Higgsing among rank-one theories and propose a simple method to obtain the precise Higgsing condition that does not rely on brane webs, but only on Lie algebra representations and their decompositions. 

A primary condition to establish a Higgsing is that the two theories must have the same tensor coefficients $\fn$. 
Suppose we want to establish a Higgsing from an untwisted theory with gauge group $\mathcal{G}$ to a twisted theory $G^{(n)}$ with low energy gauge group $\cirG$. Our trick is to perform the decomposition
\be
\mathcal{G}\to \cirG\times H_{res}.
\ee
Here the residue group $H_{res}$ is either $\su(2)$ or $\SO(2)$ for $\IZ_2$ twist, or $\su(3)$ for $\IZ_3$ twist. The Higgsing to twisted theory can be achieved by discretizing $H_{res}$, i.e., to take its continuous fugacities as some fractions of $\tau$. In the meantime, some mass parameters of the hypermulitplets of the untwisted theory must be taken as some delicate linear combinations of $\epsilon_+$ and $\tau$ to produce the correct one-loop partition function of the twisted theory. We find all previously known examples of twisting from Higgsing fall into this paradigm. For example, the $\mathbb{Z}_2$ twisted pure $\mf{su}(3)$ can be obtained from Higgsing of $G_2+\bf F$  theory. Here we show this from a purely algebraic viewpoint. Consider the decomposition $G_2\to\su(2)\times \su(2)$, we have the following representation decompositions
\begin{align}
   & \mathbf{14}\rightarrow (\mathbf{2},\mathbf{4})+(\mathbf{3},\mathbf{1})+(\mathbf{1},\mathbf{3}),\qquad\mathbf{7}\rightarrow  (\mathbf{1},\mathbf{3})+(\mathbf{2},\mathbf{2}).
\end{align}
The Higgsing process can be realized by simply discretizing the second $\su(2)$, which we have presented in \eqref{eq:higgsG2A2}. To check the one-loop part, it is only necessary to study the function $f(\epsilon_+,m_{G},m_F,q)$ inside the vector multiplet partition function
\be
\mathrm{PE}\left(\frac{f(\epsilon_+,m_{G},m_F,q)}{\big(q_1^{1/2}-q_1^{-1/2}\big)\big(q_2^{1/2}-q_2^{-1/2}\big)(1-q)}\right).
\ee
Under analytic continuations, we have
\be\nonumber
\ba
&\phantom{=}\left(-v\chi_{\bf 14}^{G_2}+e^{m}\chi_{\bf 7}^{G_2}\right)\Big|_{Higgsing}\\
&\sim  -v\left(\chi_{\bf 3}^{A_1}+\chi_{\bf 2}^{A_1}(q^{3/4}+q^{1/4}+q^{-1/4}+q^{-3/4}) \right)+vq^{-1/2}\left(\chi_{\bf 2}^{A_1}(q^{1/4}+q^{-1/4}) \right) ,\\
&= -v\left(\chi_{\bf 3}^{A_1}+\chi_{\bf 2}^{A_1}q^{1/4}+\chi_{\bf 2}^{A_1}q^{3/4} \right)    .
\ea
\ee
Here we have dropped those gauge singlets by $\sim$. This computation shows that the Higgsing indeed gives the required one-loop part of the twisted $\mf{su}(3)^{(2)}$ theory. 

A known infinite series of twisting from Higgsing is from $\mf{so}(2k+9)+(2k+1)\mathbf{V} $ to the $\IZ_2$ twist of $\mf{so}(2k+8)+2k\mathbf{V}$ with $k=0,1,2,\dots$. This type of Higgsing has been discussed from the viewpoint of brane webs in Section 5 of \cite{Kim:2019dqn}. Since the low energy gauge algebra of $\mf{so}(2k+8)^{(2)}$ is $\mf{so}(2k+7)$, we can consider the decomposition $\mf{so}(2k+9)\to \mf{so}(2k+7)\times \so(2)$. Let us just regard the $\so(2)$ as $\mf{u}(1)$. It is useful to remark the following representation decompositions
\be
\bf Adj\to (Adj+1)_{0}+V_{\pm2} ,\qquad V\to V_{0}+1_{\pm2}.
\ee
To be precise, we find the Higgsing condition is
\be
m_{\mf{u}(1)}\to \frac{\tau}{4},\qquad m_{j}\to m_{j},\ m_{j+k}\to m_{j}+\frac{\tau}{2},\ j=1,2,\dots,k,\quad m_{{\mathbf V}_{2k+1}}\to \epsilon_+-\frac{\tau}{2}.
\ee
Here we have used the orthogonal bases of $\mf{so}(2n+1)$ algebras such that the character of vector represention is $\chi_{\bf V}=1+\sum_{i=1}^nQ_i^{\pm 1}$. During the decomposition $m_{\mf{u}(1)}=2\log(Q_{k+4})$. 
For the one-loop function, we have
\begin{align}\nonumber
&\phantom{=}\bigg(-v\chi_{\bf Adj}^{\mf{so}(2k+9)}+\chi_{\bf V}^{\mf{so}(2k+9)}\sum_{i=1}^{2k+1}e^{m_i}\bigg)\Big|_{Higgsing}\\ \nonumber
&\sim  -v\left(\chi_{\bf Adj}^{\mf{so}(2k+7)}+\chi_{\bf V}^{\mf{so}(2k+7)}(q^{-1/2}+q^{1/2}) \right)+vq^{-1/2}\chi_{\bf V}^{\mf{so}(2k+7)}+   \chi_{\bf V}^{\mf{so}(2k+7)}(1+q^{1/2})\sum_{i=1}^k e^{m_i}  ,\\
&= -v\left(\chi_{\bf Adj}^{\mf{so}(2k+7)}+\chi_{\bf V}^{\mf{so}(2k+7)}q^{1/2} \right)+   \chi_{\bf V}^{\mf{so}(2k+7)}(1+q^{1/2})\sum_{i=1}^k e^{m_i} .
\end{align}
This shows that the above Higgsing indeed produces the correct one-loop part of the $\mf{so}(2k+8)^{(2)}$ twisted theory. 
It is also easy to check the elliptic genera obtained from the above Higgsing is precisely the same with expression from 2d localization.

\begin{figure}[]
\centering
\hspace{-0.3cm}
\begin{tikzpicture}
\tikzset{block/.style={draw, align=center, inner sep=0.7mm}};

\node (SO11) [block] {
$\,\SO(11)+5\mathbf{V}+\bf S\!\!$
};

\node (SO12b) [block, above = 0.5cm of SO11] {
$\,\SO(12)+6\mathbf{V}+\frac12\mathbf{S}+\frac12\bf C\!\!$
};

\node (SO12bZ2) [block, right = 1.0cm of SO12b] {
$(\SO(12)+6\mathbf{V}+\frac12\mathbf{S}+\frac12{\bf C})/\IZ_2\!\!$
};

\node (SO13) [block, above = 0.5cm of SO12b] {
$\,\SO(13)+7\mathbf{V}+\frac12\mathbf{S}\!\!$
};

\node (SO12a) [block, left = 0.1cm of SO12b] {
$\,\SO(12)+6\mathbf{V}+\bf S\!\!$
};

\node (SO10) [block, below = 0.5cm of SO11] {
$\,\SO(10)+4\mathbf{V}+2\bf S\!\!$
};

\node (SO10Z2) [block, right = 3.4cm of SO10] {
$(\SO(10)+4\mathbf{V}+2{\bf S})/\IZ_2\!\!$
};

\node (E7) [block, left = 0.9cm of SO11] {
$E_7+3\bf F\!\!$
};

\node (SO9) [block, below = 0.5cm of SO10] {
$\,\SO(9)+3\mathbf{V}+2\bf S\!\!$
};

\node (E6) [block, left = 0.5cm of SO10] {
$E_6+4\bf F\!\!$
};

\node (F4) [block, left = 0.1cm of SO9] {
$F_4+3{\bf F}\!\!$
};

\node (SO8) [block, below = 0.5cm of F4] {
$\,\SO(8)+2\mathbf{V}+2{\bf S}+2{\bf C}\!\!$
};

\node (E6Z2) [block, right = 0.7cm of SO10] {
$(E_6+4{\bf F})/\IZ_2\!\!$
};

\node (SO8Z2) [block, right = 0.3cm of SO8] {
$(\SO(8)+2\mathbf{V}+2{\bf S}+2{\bf C})/\IZ_2\!\!$
};

\node (SO8Z3) [block, right = 0.3cm of SO8Z2] {
$(\SO(8)+2\mathbf{V}+2{\bf S}+2{\bf C})/\IZ_3\!\!$
};

\node (SO7) [block, below = 1.5cm of SO9] {
$\,\SO(7)+\mathbf{V}+4{\bf S}\!\!$
};

\node (G2) [block, below = 0.5cm of SO7] {
$G_2+4\mathbf{F}\!\!$
};

\node (SUNZ2) [block, right = 3.5cm of SO7] {
$(\su(N)+2N\mathbf{F})/\IZ_2\!\!$
};

\node (SU4) [block, left = 0.5cm of G2] {
$\,\SU(4)+8\mathbf{F}\!\!$
};

\node (SUN) [block, left = 0.7cm of SO7] {
$\su(N)+2N\mathbf{F}\!\!$
};

\node (SU3) [block, below = 0.5cm of G2] {
$\,\SU(3)+6\mathbf{F}\!\!$
};

\node (SU4Z2) [block, right = 3.0cm of G2] {
$(\SU(4)+8\mathbf{F})/\IZ_2\!\!$
};

\node (SU3Z2) [block, below = 0.5cm of SU4Z2] {
$(\SU(3)+6\mathbf{F})/\IZ_2\!\!$
};

\node (Sp1) [block, below = 0.5cm of SU3] {
$\,\SU(2)+4\mathbf{F}\!\!$
};

\node (SU2Z2) [block, below = 0.5cm of SU3Z2] {
$(\SU(2)+4\mathbf{F})/\IZ_2\!\!$
};

\node (E) [block, below = 0.5cm of Sp1] {
\ M-string
};

\draw[-latex](SO11) -- (SO10);
\draw[-latex](SO11) -- (SO10Z2);
\draw[-latex](SO12a) -- (SO11);
\draw[-latex](SO12b) -- (SO11);
\draw[-latex](SO10) -- (SO9);

\draw[-latex](E7) -- (E6);
\draw[-latex](E6) -- (F4);
\draw[-latex](F4) -- (SO8);
\draw[-latex](SO9) -- (SO8);
\draw[-latex](SO9) -- (SO8Z2);
\draw[-latex](SO8) -- (SO7);
\draw[-latex](SO7) -- (G2);
\draw[-latex](SO7) -- (SU4);
\draw[-latex](SO7) -- (SU4Z2);
\draw[-latex](G2) -- (SU3);
\draw[-latex](SU4) -- (SU3);
\draw[-latex](SUN) -- (SU4);
\draw[-latex](G2) -- (SU3Z2);
\draw[-latex](SU3) -- (Sp1);
\draw[-latex](Sp1) -- (E);
\draw[-latex](SO13) -- (SO12b);
\draw[-latex](SO13) -- (SO12bZ2);
\draw[-latex](SO12bZ2) -- (SO10Z2);
\draw[-latex](SO10Z2) -- (SO8Z2);
\draw[-latex](SO8Z2) -- (SU4Z2);
\draw[-latex](SU4Z2) -- (SU3Z2);
\draw[-latex](SUNZ2) -- (SU4Z2);
%\draw[-latex](SUNZ2) -- (SU5Z2);
\draw[-latex](SU3Z2) -- (SU2Z2);

\draw[-latex](E6Z2) -- (SO8Z2);
\draw[-latex](E7) -- (E6Z2);
\draw[-latex](E6) -- (SO8Z3);
\end{tikzpicture}
\caption{Higgsing tree for untwisted/twisted 6d $(1,0)$ SCFTs with $\fn=2$.
}
\label{higgsingtree2} 
\end{figure}
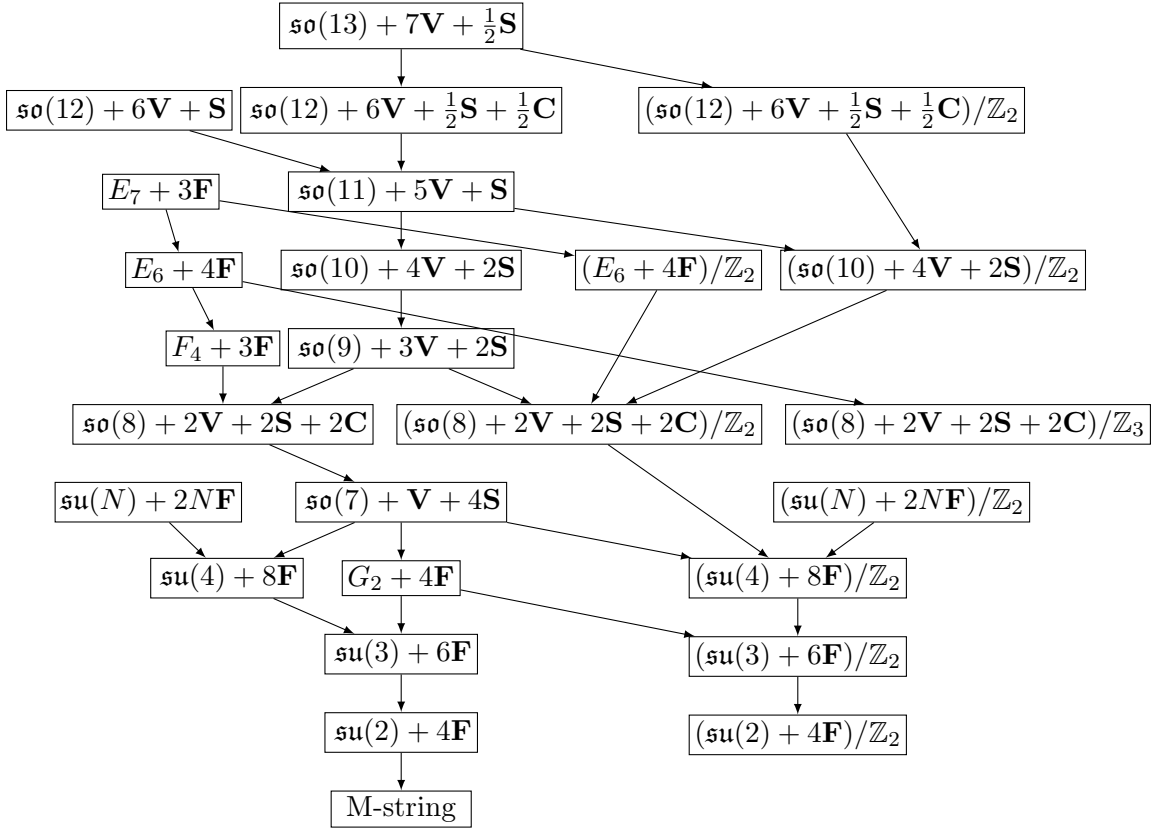

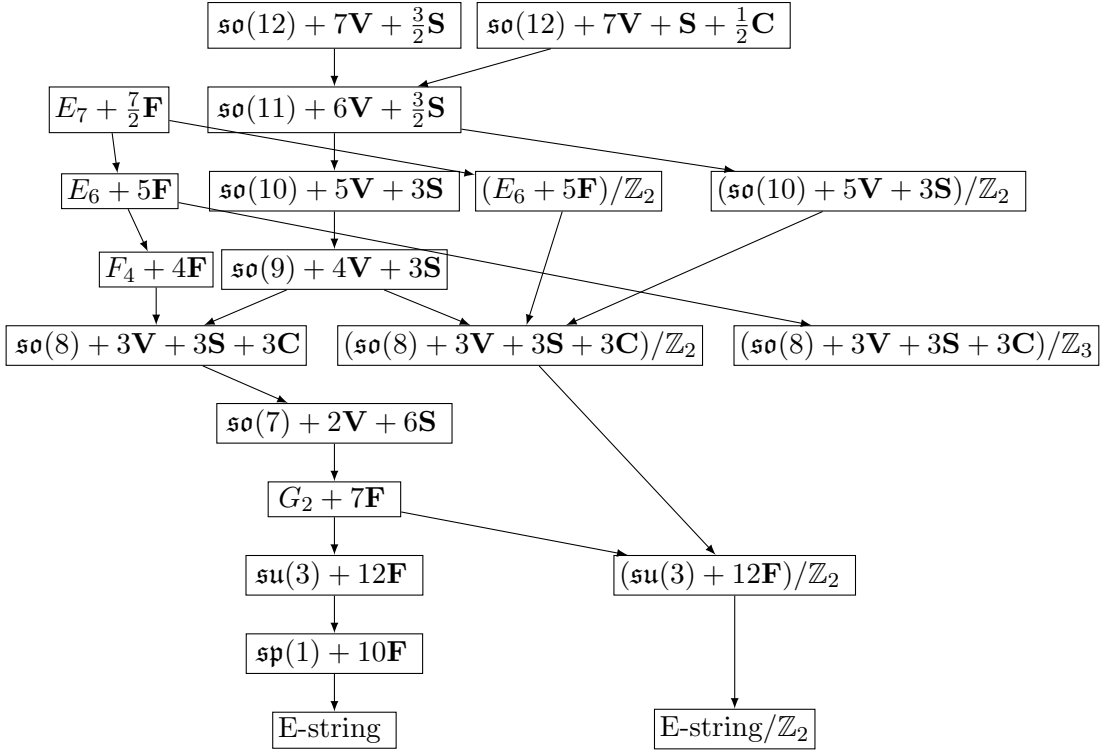
\begin{figure}[]
\centering
\hspace{-0.3cm}
\begin{tikzpicture}
\tikzset{block/.style={draw, align=center, inner sep=0.7mm}};
\node (SO11) [block] {
$\,\SO(11)+6\mathbf{V}+\frac{3}{2}\bf S$
};

\node (SO12a) [block, above = 0.5cm of SO11] {
$\,\SO(12)+7\mathbf{V}+\frac32\bf S$
};

\node (SO12b) [block, right = 0.2cm of SO12a] {
$\,\SO(12)+7\mathbf{V}+\mathbf{S}+\frac12\bf C$
};

\node (SO10) [block, below = 0.5cm of SO11] {
$\,\SO(10)+5\mathbf{V}+3\bf S$
};

\node (E6) [block, left = 0.4cm of SO10] {
$E_6+5\bf F\!\!$
};

\node (E6Z2) [block, right = 0.2cm of SO10] {
$(E_6+5{\bf F})/\IZ_2\!\!$
};

\node (E7) [block, left = 0.5cm of SO11] {
$E_7+\frac{7}{2}\bf F\!\!$
};

\node (SO10Z2) [block, right = 3.3cm of SO10] {
$(\SO(10)+5\mathbf{V}+3{\bf S})/\IZ_2$
};

\node (SO9) [block, below = 0.5cm of SO10] {
$\,\SO(9)+4\mathbf{V}+3\bf S\!\!$
};

\node (F4) [block, left = 0.1cm of SO9] {
$F_4+4{\bf F}\!\!$
};

\node (SO8) [block, below = 0.5cm of F4] {
$\,\SO(8)+3\mathbf{V}+3{\bf S}+3{\bf C}\!\!$
};

\node (SO8Z2) [block, right = 0.4cm of SO8] {
$(\SO(8)+3\mathbf{V}+3{\bf S}+3{\bf C})/\IZ_2\!\!$
};

\node (SO8Z3) [block, right = 0.4cm of SO8Z2] {
$(\SO(8)+3\mathbf{V}+3{\bf S}+3{\bf C})/\IZ_3\!\!$
};

\node (SO7) [block, below = 1.5cm of SO9] {
$\,\SO(7)+2\mathbf{V}+6{\bf S}$
};

\node (G2) [block, below = 0.5cm of SO7] {
$\,G_2+7\mathbf{F}$
};

\node (SU3) [block, below = 0.5cm of G2] {
$\,\SU(3)+12\mathbf{F}$
};

\node (SU3Z2) [block, right = 2.5cm of SU3] {
$(\SU(3)+12\mathbf{F})/\IZ_2$
};

\node (Sp1) [block, below = 0.5cm of SU3] {
$\,\mathfrak{sp}(1)+10\mathbf{F}$
};

\node (E) [block, below = 0.5cm of Sp1] {
E-string
};

\node (EZ2) [block, below = 1.5cm of SU3Z2] {
E-string$/\IZ_2\!\!$
};

\draw[-latex](SO11) -- (SO10);
\draw[-latex](SO11) -- (SO10Z2);
\draw[-latex](SO12a) -- (SO11);
\draw[-latex](SO12b) -- (SO11);
\draw[-latex](SO10) -- (SO9);
\draw[-latex](SO9) -- (SO8);
\draw[-latex](SO9) -- (SO8Z2);
\draw[-latex](SO8) -- (SO7);
\draw[-latex](SO7) -- (G2);
\draw[-latex](G2) -- (SU3);
\draw[-latex](G2) -- (SU3Z2);
\draw[-latex](SU3) -- (Sp1);
\draw[-latex](Sp1) -- (E);
\draw[-latex](SO10Z2) -- (SO8Z2);
\draw[-latex](SO8Z2) -- (SU3Z2);
\draw[-latex](SU3Z2) -- (EZ2);

\draw[-latex](E7) -- (E6);
\draw[-latex](E6) -- (F4);
\draw[-latex](F4) -- (SO8);

\draw[-latex](E7) -- (E6Z2);
\draw[-latex] (E6Z2) -- (SO8Z2);
\draw[-latex](E6) -- (SO8Z3);
\end{tikzpicture}
\caption{Higgsing tree for untwisted/twisted 6d $(1,0)$ SCFTs on with $\fn=1$.
}
\label{higgsingtree1} 
\end{figure}

The Higgsing trees for $\IZ_2$ twisted $\SO$ theories with $\fn=2,3,4$ have been mostly studied in \cite{Kim:2019dqn} using brane webs. We find more examples by our algebraic procedure, and collect the $\fn=4$ Higgsing tree in Figure \ref{higgsingtree4}, $\fn=3$ Higgsing tree in Figure \ref{higgsingtree3} and $\fn=2$ Higgsing tree in Figure \ref{higgsingtree2}. For $\IZ_2$ twisted $\mf{so}$ theories with $\fn=1$, it is not easy to find the brane webs. We use our new method to propose the $\fn=1$ Higgsing tree in Figure \ref{higgsingtree1}.

\begin{figure}[h]
\centering
\hspace{-0.3cm}
\begin{tikzpicture}
\tikzset{block/.style={draw, align=center, inner sep=0.7mm}};

\node (SO11) [block] {
$\,\mf{so}(11)+3\mathbf{V}\!\!$
};

\node (SO12b) [block, above = 0.5cm of SO11] {
$\,\mf{so}(12)+4\mathbf{V}\!\!$
};

\node (SO12bZ2) [block, right = 1.5cm of SO12b] {
$(\mf{so}(12)+4\mathbf{V})/\IZ_2\!\!$
};

\node (SO13) [block, above = 0.5cm of SO12b] {
$\,\mf{so}(13)+5\mathbf{V}\!\!$
};

\node (SOn) [block, above = 0.5cm of SO13] {
$\,\mf{so}(8+m)+m\mathbf{V}\!\!$
};

\node (SO10) [block, below = 0.5cm of SO11] {
$\,\mf{so}(10)+2\mathbf{V}\!\!$
};

\node (SO10Z2) [block, right = 3.5cm of SO10] {
$(\mf{so}(10)+2\mathbf{V})/\IZ_2\!\!$
};

\node (E7) [block, left = 0.9cm of SO11] {
$E_7+2\bf F\!\!$
};

\node (SO9) [block, below = 0.5cm of SO10] {
$\,\mf{so}(9)+\mathbf{V}\!\!$
};

\node (E6) [block, left = 0.5cm of SO10] {
$E_6+2\bf F\!\!$
};

\node (SO8) [block, below = 0.5cm of SO9] {
$\,\mf{so}(8)\!\!$
};

\node (F4) [block, left = 0.3cm of SO9] {
$F_4+{\bf F}\!\!$
};

\node (E6Z2) [block, right = 0.8cm of SO10] {
$(E_6+2{\bf F})/\IZ_2\!\!$
};

\node (SO8Z2) [block, right = 1.75cm of SO8] {
$\,\mf{so}(8)/\IZ_2\!\!$
};

\node (SO8Z3) [block, right = 1.7cm of SO8Z2] {
$\,\mf{so}(8)/\IZ_3\!\!$
};

\draw[-latex](SO11) -- (SO10);
\draw[-latex](SO11) -- (SO10Z2);

\draw[-latex](SO12b) -- (SO11);
\draw[-latex](SO10) -- (SO9);

\draw[-latex](E7) -- (E6);
\draw[-latex](E6) -- (F4);
\draw[-latex](F4) -- (SO8);
\draw[-latex](SO9) -- (SO8);
\draw[-latex](SO9) -- (SO8Z2);

\draw[-latex](E6) -- (SO8Z3);
\draw[-latex](SOn) -- (SO13);
\draw[-latex](SO13) -- (SO12b);
\draw[-latex](SO13) -- (SO12bZ2);
\draw[-latex](SO12bZ2) -- (SO10Z2);
\draw[-latex](SO10Z2) -- (SO8Z2);

\draw[-latex](E6Z2) -- (SO8Z2);
\draw[-latex](E7) -- (E6Z2);
\end{tikzpicture}
\caption{Higgsing tree for untwisted/twisted 6d $(1,0)$ SCFTs on with $\fn=4$.
}
\label{higgsingtree4} 
\end{figure}
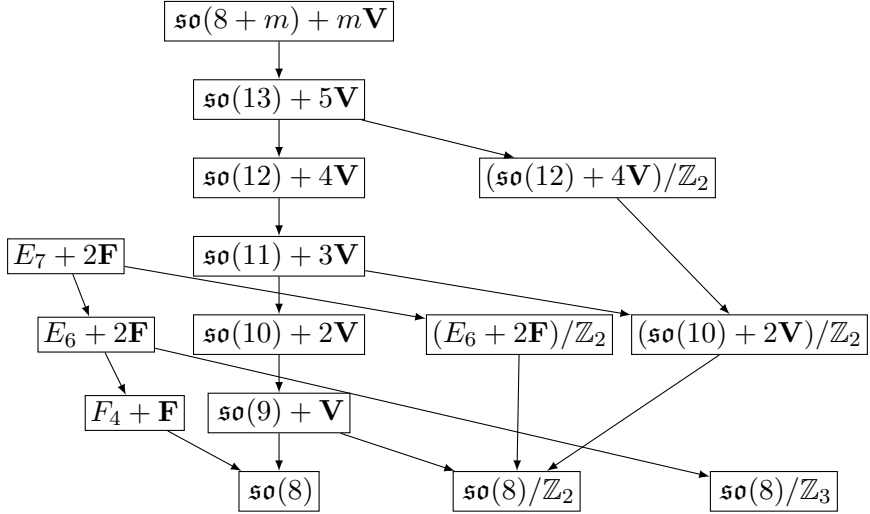

\begin{figure}[]
\centering
\hspace{-0.3cm}
\begin{tikzpicture}
\tikzset{block/.style={draw, align=center, inner sep=0.7mm}};
\node (SO11) [block] {
$\,\mf{so}(11)+4\mathbf{V}+\frac12\bf S\!\!$
};

\node (SO12b) [block, above = 0.5cm of SO11] {
$\,\mf{so}(12)+5\mathbf{V}+\frac12\mathbf{S}\!\!$
};

\node (SO10) [block, below = 0.5cm of SO11] {
$\,\mf{so}(10)+3\mathbf{V}+\bf S\!\!$
};

\node (SO10Z2) [block, right = 3.4cm of SO10] {
$(\mf{so}(10)+3\mathbf{V}+{\bf S})/\IZ_2\!\!$
};

\node (E7) [block, left = 0.9cm of SO11] {
$E_7+\frac52\bf F\!\!$
};

\node (SO9) [block, below = 0.5cm of SO10] {
$\,\so(9)+2\mathbf{V}+\bf S\!\!$
};

\node (E6) [block, left = 0.5cm of SO10] {
$E_6+3\bf F\!\!$
};

\node (SO8) [block, below = 0.5cm of SO9] {
$\,\so(8)\!+\!\mathbf{V}\!+\!{\bf S}\!+\!{\bf C}\!\!$
};

\node (F4) [block, left = 0.3cm of SO9] {
$F_4+2{\bf F}\!\!$
};

\node (E6Z2) [block, right = 0.8cm of SO10] {
$(E_6+3{\bf F})/\IZ_2\!\!$
};

\node (SO8Z2) [block, right = 0.5cm of SO8] {
$(\so(8)\!+\!\mathbf{V}\!+\!{\bf S}\!+\!{\bf C})/\IZ_2\!\!$
};

\node (SO8Z3) [block, right = 0.2cm of SO8Z2] {
$(\so(8)\!+\!\mathbf{V}\!+\!{\bf S}\!+\!{\bf C})/\IZ_3\!\!$
};

\node (SO7) [block, below = 0.5cm of SO8] {
$\,\mf{so}(7)+2\mathbf{S}\!\!$
};

\node (G2) [block, below = 0.5cm of SO7] {
$\,G_2+\mathbf{F}\!\!$
};

\node (SU3) [block, below = 0.5cm of G2] {
$\,\su(3)\!\!$
};

\node (SU3Z2) [block, right = 2.6cm of SU3] {
$\,\su(3)/\IZ_2\!\!$
};

\draw[-latex](SO11) -- (SO10);
\draw[-latex](SO11) -- (SO10Z2);

\draw[-latex](SO12b) -- (SO11);
\draw[-latex](SO10) -- (SO9);

\draw[-latex](E7) -- (E6);
\draw[-latex](E6) -- (F4);
\draw[-latex](F4) -- (SO8);
\draw[-latex](SO9) -- (SO8);
\draw[-latex](SO9) -- (SO8Z2);
\draw[-latex](SO8) -- (SO7);
\draw[-latex](SO7) -- (G2);
%\draw[-latex](SO7) -- (SU4);
%\draw[-latex](SO7) -- (SU4Z2);
\draw[-latex](G2) -- (SU3);
%\draw[-latex](SU4) -- (SU3);
\draw[-latex](G2) -- (SU3Z2);

\draw[-latex](E6) -- (SO8Z3);

%\draw[-latex](SO12bZ2) -- (SO10Z2);
\draw[-latex](SO10Z2) -- (SO8Z2);
\draw[-latex](SO8Z2) --  (SU3Z2);

\draw[-latex](E6Z2) -- (SO8Z2);
\draw[-latex](E7) -- (E6Z2);
\end{tikzpicture}
\caption{Higgsing tree for untwisted/twisted 6d $(1,0)$ SCFTs on with $\fn=3$.
}
\label{higgsingtree3} 
\end{figure}
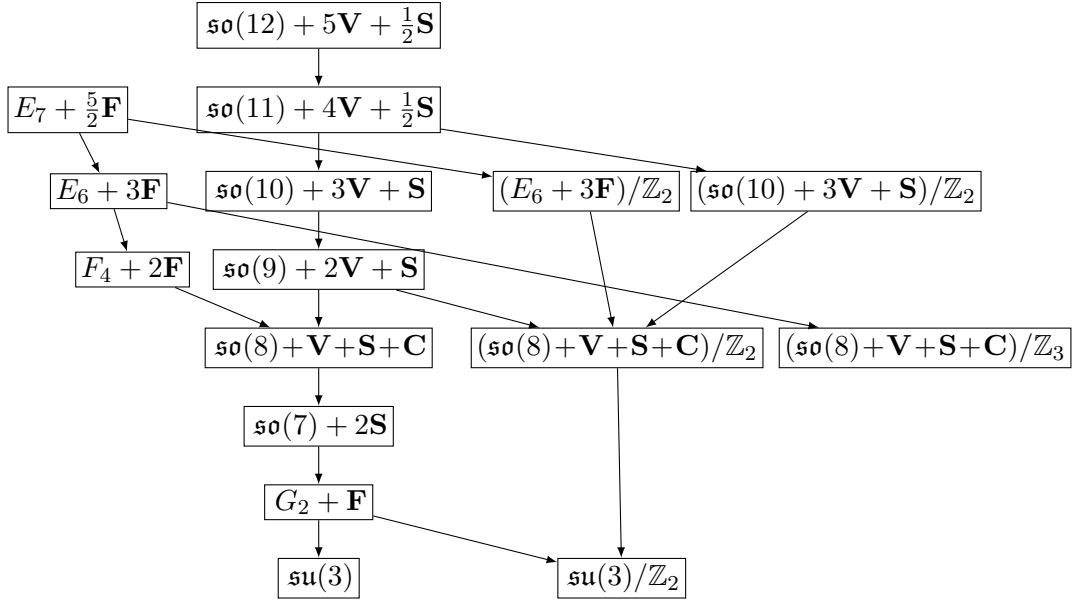

Our method can also be used to rule out some potential Higgsing. For example, we notice that the $\fn=2$, $\mf{su}(N)$ theory cannot be Higgsed to $\fn=2$, $\mf{su}(N-1)^{(2)}$ twisted theory. The reason is that the $\bf Adj$ and $\bf \Lambda^2$ of $\mf{su}(N)$ always have the same KK charge $0$ under representation decompositions, thus can not produce the required one-loop function of the $\mf{su}(N-1)^{(2)}$ twisted theory.

The Higgsing of elliptic blowup equations for 6d $(1,0)$ SCFTs have been discussed in Section 3.3 of \cite{Gu:2020fem}. For the twisted cases, the Higgsing is just similar. The Higgsing conditions can be taken directly in the elliptic genera as well as the (twisted) elliptic blowup equations, at least for the unity ones.  It is easy to check that the characteristic $a$ and value $y$ of elliptic blowup equations are consistent with the Higgsing.

In the following, we pick some interesting examples of twisting from Higgsing to show our algebraic approach and give the precise Higgsing conditions.

\subsection{Higgsing from untwisted to twisted theories}

\subsubsection{$E_7\to E_6^{(2)}$}
A series of interesting examples of twisting from Higgising is the $E_7\to E_6^{(2)}$ type for $\fn=6,5,4,3,2,1$. Let us start with the simplest case $\fn=6$, i.e., Higgsing $E_7+\mathbf{F}\rightarrow E_6^{(2)}.$ 
The flavor algebra of $E_7+\mathbf{F}$ theory is $\mf{so}(2)_{12}$. 
To write down the precise Higgsing condition, we first decompose $E_7\to F_4\times \mf{su}(2)$. It is useful to remark the following representation decompositions
\be
\mathbf{133}\to \bf (1,3)+(26,3)+(52,1),\qquad 56\to (1,4)+(26,2).
\ee
We find the Higgsing can be taken by discretizing the $\mf{su}(2)$ into $\IZ_2$:
\be\label{higgsE7}
m_{\mf{su}(2)}\to \frac{\tau}{2},\quad m_{\mf{so}(2)}\to\epsilon_+-\frac{\tau}{4}.
\ee
It is easy to see  that upon this condition the elliptic genera index of $E_7+\mathbf{F}$ from \eqref{eq:Ed-index}
\be
\IE_d(\eq,\et,m_{E_7},m_{\mf{so}(2)})=-14d\epsilon_+^2+\eq\et(3d^2-2d)+d\left(-3(m,m)_{E_7}+6m_{\mf{so}(2)}^2\right)
\ee
indeed goes to the index of twisted elliptic genera of $E_6^{(2)}$ theory from \eqref{eq:twistedEd-index}
\be
\IE_d(\eq,\et,m_{F_4})=-8d\epsilon_+^2+\eq\et(3d^2-2d)+d(-3(m,m)_{F_4}) .
\ee
To check the one-loop function, we have
\be
\ba
&\phantom{=}\left(-v\chi_{\bf 133}^{E_7}+e^{m_{\mf{so}(2)}}\chi_{\bf 56}^{E_7}\right)\Big|_{Higgsing}\\
&\sim  -v\left(\chi_{\bf 52}^{F_4}+\chi_{\bf 26}^{F_4}(q^{1/2}+1+q^{-1/2}) \right) +vq^{-1/4}\left(\chi_{\bf 26}^{F_4}(q^{1/4}+q^{-1/4}) \right) ,\\
&= -v\left(\chi_{\bf 52}^{F_4}+\chi_{\bf 26}^{F_4}q^{1/2}  \right)    .
\ea
\ee
Here we have droped those gauge singlets.

For $\fn=4$, we can Higgs $E_7+2\mathbf{F}\rightarrow (E_6+2\mathbf{F})/\IZ_2.$ 
The Higgsing condition is precisely \eqref{higgsE7} combined with $m'_{\mf{so}(2)}=m_{\Sp(1)}+\tau/4$.  For the one-loop function, we have
\be
\ba
&\phantom{=}\left(-v\chi_{\bf 133}^{E_7}+e^{m_{\mf{so}(2)}}\chi_{\bf 56}^{E_7}+e^{m'_{\mf{so}(2)}}\chi_{\bf 56}^{E_7}\right)\Big|_{Higgsing}\\
&\sim  -v\left(\chi_{\bf 52}^{F_4}+\chi_{\bf 26}^{F_4}(q^{1/2}+1+q^{-1/2}) \right) +(vq^{-1/4}+e^{m_{\Sp(1)}}q^{1/4})\left(\chi_{\bf 26}^{F_4}(q^{1/4}+q^{-1/4}) \right),\\
&= -v\left(\chi_{\bf 52}^{F_4}+\chi_{\bf 26}^{F_4}q^{1/2}  \right)   +e^{m_{\Sp(1)}}\left(\chi_{\bf 26}^{F_4}(1+q^{1/2}) \right)  .
\ea
\ee
The same applies to all Higgsing
\be
E_7+\frac{8-\fn}{2}\mathbf{F}\rightarrow (E_6+(6-\fn)\mathbf{F})/\IZ_2,\qquad \fn=6,4,2.
\ee
For odd $\fn$, the Higgsing is a bit unorthodox as the twisted theory has unpaired matter content. Under the above procedure, the $\frac12\bf 56$ with zero mass should produce a $\bf 26$ with a  quarter KK-charge and zero mass. 

As a final remark, some 6d $E_6^{(2)}$ theories can also be Higgsed from conformal matter theories. For example, it was found in \cite{Hayashi:2021pcj} that the $\fn=2,E_6^{(2)}$ theory can be obtained from a special Higgsing of the minimal $(E_7,E_7)$ conformal matter theory, while the $\fn=5,E_6^{(2)}$ theory can be obtained from a special Higgsing of extended $(E_7,\und{E_7})$ theory.

\subsubsection{$E_6\to \so(8)^{(3)}$}
One more interesting example of twisting from Higgsing is for the $\IZ_3$ twisted theories. In \cite{Hayashi:2021pcj}, it was found that the $\fn=4,\so(8)^{(3)}$ theory can be obtained from a special Higgsing of the minimal $(E_6,E_6)$ conformal matter theory, while the $\fn=2,\so(8)^{(3)}$ theory can be obtained from a special Higgsing of non-minimal $(E_6,E_6)_2$ conformal matter theory. We notice that in general, the rank-one $\so(8)^{(3)}$ theory can be Higgsed from the rank-one $E_6$ theory with the same tensor coefficient $\fn$. Let us first consider the $\fn=4$ case, i.e. Higgsing $E_6+2\mathbf{F}$ to pure $\so(8)^{(3)}$. To write down the precise Higgsing condition, it is useful to consider decomposition $E_6\to  \mf{su}(3)\times G_2$. Remark that
\be
\mathbf{78}\to \bf (8,1)+(8,7)+(1,14),\qquad 27\to (\bar6,1)+(3,7).
\ee
We then observe that the Higgsing to twisting can be achieved by discretizing the $\mf{su}(3)$ into $\IZ_3$ by the following condition
\be\label{higgsE6}
m_{1,2,3}^{\mf{su}(3)}\to 0,\pm\frac{\tau}{3},\qquad m_{1,2}\to \epsilon_+,\epsilon_+-\frac{\tau}{3}.
\ee
It is easy to see  that upon this condition the elliptic genera index of $E_6+2\mathbf{F}$ from \eqref{eq:Ed-index}
\be
\IE_d(\eq,\et,m_{E_6},m_{1,2})=-14d\epsilon_+^2+\eq\et(3d^2-2d)+d\left(-3(m,m)_{E_6}+6m_{\mf{so}(2)}^2\right)
\ee
indeed goes to the index of twisted elliptic genera of $\so(8)^{(3)}$ theory from \eqref{eq:twistedEd-index}
\be
\IE_d(\eq,\et,m_{G_2})=-8d\epsilon_+^2+\eq\et(3d^2-2d)+d(-3(m,m)_{G_2}) .
\ee
For the one-loop function, we have
\be
\ba
&\phantom{=}\left(-v\chi_{\bf 78}^{E_6}+(e^{m_{1}} +e^{m_{2}})\chi_{\bf 27}^{E_6}\right)\Big|_{Higgsing}\\
&\sim  -v\left(\chi_{\bf 14}^{G_2}+\chi_{\bf 7}^{G_2}(2+2q^{\pm1/3}+q^{\pm2/3}) \right)+(vq^{-1/3} +v)\left(\chi_{\bf 7}^{G_2}(q^{1/3}+1+q^{-1/3}) \right) ,\\
&=  -v\left(\chi_{\bf 14}^{G_2}+\chi_{\bf 7}^{G_2}(q^{1/3} +q^{2/3}) \right)    .
\ea
\ee
This shows that the Higgsing indeed produces the correct one-loop function of pure $\so(8)^{(3)}$ theory. 

For $\fn=3$, we can Higgs $E_6+3\mathbf{F}\rightarrow (\so(8)+\mathbf{V}+\mathbf{S}+\mathbf{C})/\IZ_3.$ 
The Higgsing condition is precisely \eqref{higgsE6} combined with $m_{3}=m_{\Sp(1)}+\tau/3$.  For the one-loop part, we have
\be\nonumber
\ba
&\phantom{=}\left(-v\chi_{\bf 78}^{E_6}+(e^{m_{1}} +e^{m_{2}}+e^{m_{3}})\chi_{\bf 27}^{E_6}\right)\Big|_{Higgsing}\\
&\sim  -v\left(\chi_{\bf 14}^{G_2}+\chi_{\bf 7}^{G_2}(2+2q^{\pm1/3}+q^{\pm2/3}) \right)+(vq^{-1/3} +v+e^{m_{\Sp(1)}}q^{1/3})(q^{1/3}+1+q^{-1/3}) \chi_{\bf 7}^{G_2},\\
&=  -v\left(\chi_{\bf 14}^{G_2}+\chi_{\bf 7}^{G_2}(q^{1/3} +q^{2/3}) \right) +  e^{m_{\Sp(1)}}\chi_{\bf 7}^{G_2}(1+q^{1/3}+q^{2/3})  .
\ea
\ee
The same applies to all Higgsing to twisting
\be
E_6+(6-\fn)\mathbf{F}\rightarrow (\so(8)+(4-\fn)(\mathbf{V}+\mathbf{S}+\mathbf{C}))/\IZ_3,\qquad \fn=4,3,2,1.
\ee

\subsubsection{$\mf{so}(7)\to \mf{su}(4)^{(2)}$}
Consider the $\fn=2$ Higgsing from  $\mf{so}(7)+\bf V+4S$ to the $\IZ_2$ twist of  $\mf{su}(4)+8\bf F$. This twisting from Higgsing has been found in \cite{Kim:2019dqn} from brane webs. Here we adopt a purely algebraic approach. Let us consider the decomposition $\mf{so}(7)\to  \Sp(2)\times \mf{so}(2)$. Regarding the $\so(2)$ as $\mf{u}(1)$. We have the following representation decompositions
\be
\mathbf{21}\to \bf 1_0+5_2+5_{-2}+10_0,\qquad 7\to 1_2+1_{-2}+5_0 ,\qquad 8\to 4_1+4_{-1}.
\ee 
We find the Higgsing to twisting can be achieved by discretizing the $\mf{u}(1)$ by the following condition
\be
m_{\mf{u}(1)}\to \frac{\tau}{4},\quad m_{\bf 7}\to \epsilon_+-\frac{\tau}{2},\quad m_{i,\bf 8}\to m_{i,\bf 4}+\frac{\tau}{4},\ i=1,2,3,4.
\ee
For the one-loop part, we have
\be\nonumber
\ba
&\phantom{=}\bigg(-v\chi_{\bf 21}^{\mf{so}(7)}+e^{m_{\bf 7}}\chi_{\bf 7}^{\mf{so}(7)}+\chi_{\bf 8}^{\mf{so}(7)}\sum_{i=1}^4 e^{m_{i,\bf 8}}\bigg)\bigg|_{Higgsing}\\
&\sim  -v\left(\chi_{\bf 10}^{\Sp(2)}+\chi_{\bf 5}^{\Sp(2)}q^{\pm1/2} \right)+vq^{-1/2}\chi_{\bf 5}^{\Sp(2)}+(q^{-1/4}+q^{1/4}) \chi_{\bf 4}^{\Sp(2)}\sum_{i=1}^4q^{1/4}e^{m_{i,\bf 4}},\\[-2mm]
&=  -v\left(\chi_{\bf 10}^{\Sp(2)}+\chi_{\bf 5}^{\Sp(2)}q^{1/2} \right)+\left(\chi_{\bf 4}^{\Sp(2)}+\chi_{\bf 4}^{\Sp(2)}q^{1/2} \right) \sum_{i=1}^4e^{m_{i,\bf 4}}    .
\ea
\ee
Thus we successfully reach the one-loop function of the $\IZ_2$ twist of $\fn=2$, $\mf{su}(4)+8\bf F$ theory.

\subsection{Higgsing from twisted to twisted theories}
Let us consider a chain of Higgsing $\fn=2$, $\mf{su}(N)^{(2)}\to \mf{su}(N-1)^{(2)}$. From \eqref{eq:twistedEd-index}, we can easily write down the modular indices of the $\mf{su}(N)^{(2)}$ theories. For both $N=2r+1$ and $N=2r$ cases, we have
\be\label{n2sunindex}
\IE_d(\eq,\et,m_{\mf{sp}(r)},m_{\mf{sp}(N)})=-Nd\epsilon_+^2+d^2\eq\et +d\bigg(-(m,m)_{\mf{sp}(r)}+\sum_{i=1}^{N}m_{i,\bf N}^2\bigg).
\ee

Consider the Higgsing from $\mf{su}(2r+1)^{(2)}\to \mf{su}(2r)^{(2)}$. The low energy gauge group $\mf{sp}(r)$ remains the same upon the Higgsing. However, the number of fundamentals gets reduced.  We propose the Higgsing condition to be
\be
m_{2r+1}\to\epsilon_++\frac{\tau}{4},\qquad m_i\to m_i,\quad i=1,2,\dots,2r.
\ee
Obviously, this Higgsing condition gives the correct modular index in \eqref{n2sunindex}. For the one-loop function, we have
\be\nonumber
\ba
&\phantom{=}\bigg(-v\left(\chi_{\bf Adj}^{\mf{sp}(r)}+\chi_{\bf F}^{\mf{sp}(r)}(q^{1/4}+q^{3/4})+\chi_{\bf \Lambda^2}^{\mf{sp}(r)}q^{1/2} \right)  + \chi_{\bf F}^{\mf{sp}(r)}(1+q^{1/2})\sum_{i=1}^{2r+1}e^{m_{i}} \bigg)\Big|_{Higgsing}\\
%&\sim  \\
&=-v\left(\chi_{\bf Adj}^{\mf{sp}(r)}+\chi_{\bf \Lambda^2}^{\mf{sp}(r)}q^{1/2} \right)  + \chi_{\bf F}^{\mf{sp}(r)}(1+q^{1/2})\sum_{i=1}^{2r}e^{m_{i}}  .
\ea
\ee
This gives exactly the one-loop function of $\mf{su}(2r)^{(2)}$ twisted theory.

Now consider the Higgsing from $\mf{su}(2r)^{(2)}\to \mf{su}(2r-1)^{(2)}$. The low energy gauge group reduces from $C_{r}$ to $C_{r-1}$. 
Let us consider the natural decomposition $C_{r}\to C_{r-1}\times \mf{su}(2)$. It is useful to remark the following representation decompositions
\be\nonumber
\bf Adj\to (Adj,1)+(F,2)+(1,3),\quad F\to (F,1)+(1,2),\quad \Lambda^2\to (\Lambda^2,1)+(F,2)+(1,1).
\ee
We find the Higgsing can be taken by discretizing the $\mf{su}(2)$ into $\IZ_2$:
\be\label{higgsSUZ2}
m_{\mf{su}(2)}\to\frac{\tau}{4},\quad m_{2r }\to\epsilon_+-\frac{\tau}{4},\quad m_{i}\to m_i,\ i=1,2,\dots,2r-1.
\ee
Obviously, this Higgsing condition also gives the correct modular index in \eqref{n2sunindex}. For the one-loop function, we have
\be\nonumber
\ba
&\phantom{=}\bigg(-v\left(\chi_{\bf Adj}^{\mf{sp}(r)}+\chi_{\bf \Lambda^2}^{\mf{sp}(r)}q^{1/2} \right)  + \chi_{\bf F}^{\mf{sp}(r)}(1+q^{1/2})\sum_{i=1}^{2r}e^{m_{i}}  \bigg)\bigg|_{Higgsing}\\
&\sim -v\left(\chi_{\bf Adj}^{\mf{sp}(r-1)}+\chi_{\bf F}^{\mf{sp}(r-1)}(q^{-1/4}+2q^{1/4}+q^{3/4})+\chi_{\bf \Lambda^2}^{\mf{sp}(r-1)}q^{1/2} \right)+ \chi_{\bf F}^{\mf{sp}(r)}(1+q^{1/2})vq^{-1/4}\\
&\phantom{\sim}\ + \chi_{\bf F}^{\mf{sp}(r-1)}(1+q^{1/2})\sum_{i=1}^{2r-1}e^{m_{i}}\\[-2mm]
&=-v\left(\chi_{\bf Adj}^{\mf{sp}(r-1)}+\chi_{\bf F}^{\mf{sp}(r-1)}(q^{1/4}+q^{3/4})+\chi_{\bf \Lambda^2}^{\mf{sp}(r-1)}q^{1/2} \right)  + \chi_{\bf F}^{\mf{sp}(r-1)}(1+q^{1/2})\sum_{i=1}^{2r-1}e^{m_{i}}  .
\ea
\ee
This gives exactly the one-loop function of $\mf{su}(2r-1)^{(2)}$ twisted theory.

\section{Summary and outlook}\label{sec:outlook}
In this paper we studied the 2d $(0,4)$ SCFTs of BPS strings associated with the twisted circle compactification of all 6d rank-one $(1,0)$ SCFTs. The twisted elliptic genera of such 2d theories exhibit many extraordinary properties, which above all are classified by twisted affine Lie algebras. We established the functional equations of the twisted elliptic genera, called twisted elliptic blowup equations, which are the generalization of elliptic blowup equations developed in recent years. This powerful tool enables us to calculate twisted elliptic genera for most rank-one theories from twisted compactification. Although we focused on the one-string case, the recursion formula from twisted elliptic blowup equation allows us to compute the twisted elliptic genera of arbitrary number of strings. We also investigated the modular ansatz of the twisted one-string elliptic genera and found there is naturally a $\Gamma_1(N)$ modular structure, where $N=2,3,4$ is just the twist coefficients of the twisted affine Lie algebras. This is directly connected to the underlying $N$-section Calabi-Yau geometries, which are some genus-one fibered Calabi-Yau threefolds, whose blowup equations after converting to gauge theory language are just the twisted elliptic blowup equations. Our blowup equations can also solve the refined BPS invariants of these Calabi-Yau geometries.

We further studied the generalization of spectral flow symmetry proposed in \cite{DelZotto:2016pvm,DelZotto:2018tcj} to the twisted case. We found the spectral flow between R-R and NS-R elliptic genera of one BPS string nicely fits into the twisted situation where the KK charges are naturally shifted by half. In many cases such as the four pure gauge cases, this implies an extra symmetry of the twisted elliptic genera. For theories with 2d localization formulas such as all $\fn=4,\mf{so}(2r+8)^{(2)}$ theories, such spectral flow symmetry can be derived explicitly. 

In \cite{DelZotto:2016pvm}, Del Zotto and Lockhart conjectured a surprising relation between the one-string elliptic genera of 6d $(1,0)$ pure gauge ${G}$ SCFTs and the Schur indices of 4d $\mathcal{N}=2$ $H_{G}$ SCFTs for ${G}=\su(3),\so(8),F_4,E_{6,7,8}$. Though still mysterious, this relation has also been shown to exist to some extent  for multi-string cases \cite{Gu:2019dan}. It is intriguing to consider whether the twisted elliptic genera of the four pure gauge cases discussed in the current work, i.e., $\su(3)^{(2)},\so(8)^{(2)},\so(8)^{(3)},E_6^{(2)}$  are related to the Schur indices of some 4d SCFTs, possibly with defects. This relies on certain decompositions of the twisted elliptic genera as infinite copies of certain simpler functions, similar with those in \cite{DelZotto:2016pvm}. Besides, by SCFT/VOA correspondence, the Schur indices of $H_{G}$ SCFTs are equivalent to the vacuum characters of VOA $(G)_{-\dualCox/6-1}$. Thus we suspect in the twisted case, the VOA $(G^{(n)})_{-\dualCox/6-1}$ will also play a role. Notably, the VOA $(\su(3)^{(2)})_{-3/2}$ has been recently discussed in \cite{kanade}.

One more fascinating structure of the one-string elliptic genera of 6d $(1,0)$ SCFTs with tensor coefficient $\fn$ and gauge group $G$ (possibly with matter content) is an elegant decomposition found in \cite{DelZotto:2018tcj} involving the characters of affine Lie algebra $G^{(1)}$ at negative level $-\fn$, see the Equations (1.6) and (1.7) there and some recent results in \cite{Duque:2022tub}. Presumably, the twisted elliptic genera we investigated in the current paper should have an analogous decomposition involving the characters of twisted affine Lie algebras at negative levels.  We hope to address this issue in the future.

In this paper, we have limited ourselves to the twisted compactification by folding vector multiplets in gauge algebra $G$, which is closely related to twisted affine Lie algebras. As mentioned earlier, it is well-known that there exists another type of twist, that is folding tensor multiplets. This can happen when a higher-rank 6d SCFT exhibits some discrete symmetry in its quiver structure. This type of twisted compactification also gives many interesting 5d KK theories. It should be interesting to study as well the twisted elliptic genera in this setting. In particular, for 6d $(2,0)$ SCFTs, this kind of twisted elliptic genera has been studied in \cite{Duan:2021ges}. Besides, one can even consider the twisted elliptic genera associated to the twisted circle compactification of 6d little string theories  which has drawn some interests recently \cite{DelZotto:2020sop,Bhardwaj:2022ekc,DelZotto:2022xrh}. We expect many methods developed in the current work can be extended to the twisted elliptic genera of little string theories.

We can also study 6d SCFTs with defects. There are two types of defects with codimension two and codimension four, which play important roles in the study of the elliptic quantum Seiberg-Witten curves \cite{Chen:2020jla,Chen:2021ivd,Chen:2021rek}. See \cite{Nazzal:2018brc,Razamat:2018zel,Nazzal:2021tiu} for other approaches to the elliptic quantum Hamiltonians. The blowup equations for Wilson loops/surfaces and codimension four defect partition functions have been proposed in \cite{Kim:2021gyj} for generic 5d/6d theories on $\mathbb{R}^4\times S^1$ and $\mathbb{R}^4\times T^2$. It is also interesting to extend the (elliptic) blowup equation to theories with codimension two defects, and study the modular expressions of the (twisted) elliptic genera. See \cite{Nekrasov:2020qcq,Jeong:2020uxz} for the blowup equation for codimension two defects in 4d. The solution of the defect partition functions would help to bootstrap new elliptic quantum Seiberg-Witten curves that are corresponding to 6d gauge theories with and without twist.

\section*{Acknowledgements}
We would like to thank Lakshya Bhardwaj, Cesar Fierro Cota, Michele Del Zotto, Zhihao Duan, Amir-Kian Kashani-Poor, Hee-Cheol Kim, Minsung Kim, Sung-Soo Kim, Albrecht Klemm, Guglielmo Lockhart, Paul-Konstantin Oehlmann, Thorsten Schimannek, Cumrun Vafa, Haowu Wang and Rui-Dong Zhu for useful discussions. We thank Hee-Cheol Kim for the hospitality in POSTECH where part of this work was done. KS would also like to thank the hospitality of Institut Mittag-Leffler during the workshop ``Enumerative Invariants, Quantum Fields and String Theory Correspondences''. KL, KS and XW are supported by KIAS Grants PG006904, QP081001 and QP079201, respectively. KL is also supported in part by the National Research Foundation of Korea (NRF) Grant funded by the Korea government (MSIT) (No.2017R1D1A1B06034369).

\appendix
\section{Useful formulas}\label{app:A}
We collect some definition formulas used in the twisted elliptic blowup equations \eqref{eq:ebeq}. These are the twisted generalizations of those that appeared in \cite{Gu:2020fem}. 
When computing the contribution of $\cirG$ vector multiplet with KK-charge zero to twisted elliptic blowup equations, we have 
\begin{align}
  A_{{V}}(\tau,\mG,\lG)
  &=
    \prod_{\beta\in\Delta_+}
    \breve{\theta}_V(\beta\cdot\mG,\beta\cdot\lG)
     , \label{eq:AV}
\end{align}
where for $L\in \IZ$, 
\begin{align}
  &\breve{\theta}_V(z,L) :=
    \prod_{\substack{m,n\geq 0\\m+n\leq |L|-1}}
  \frac{\eta}{\theta_1(z+s m\eq+s n\et)}
  \prod_{\substack{m,n\geq 0\\m+n\leq |L|-2}}
  \frac{\eta}{\theta_1(z+s(m+1)\eq+s(n+1)\et)} ,\label{eq:thetaV}
\end{align}
with $s$ the sign of $L$. The contributions of vector multiplets with fractional KK-charges can be defined analogously. For $\cirG$ representation $ R$ with non-zero fractional KK-charge $k=1/2,1/3,2/3,1/4,3/4$,  we have
\begin{align}
  A_{{V}}^{frac}(\tau,\mG,\lG)
  &=
    \prod_{\beta\in R}
    \breve{\theta}_V^{[k]}(\beta\cdot\mG,\beta\cdot\lG)
     , \label{eq:AVfrac}
\end{align}
where $\breve{\theta}_V^{[k]}$ is defined by adding characteristic $k$ for all $\theta_1(\tau,z)$ functions as $\theta_1^{[k]}(\tau,z)$ in \eqref{eq:thetaV}. The Jacobi theta functions with characteristics $a$ are defined as
\begin{align}
\theta_1^{[a]}(\tau,z)=& -i\sum_{k\in\IZ}(-1)^{k+a}q^{(k+1/2+a)^2/2}Q_z^{k+1/2+a},\\
\theta_2^{[a]}(\tau,z)=& \sum_{k\in\IZ}q^{(k+1/2+a)^2/2}Q_z^{k+1/2+a},\\
\theta_3^{[a]}(\tau,z)=&\sum_{k\in\IZ}q^{(k+a)^2/2}Q_z^{k+a},\\
\theta_4^{[a]}(\tau,z)=&\sum_{k\in\IZ}(-1)^{k+a}q^{(k+a)^2/2}Q_z^{k+a},
\end{align}
where $Q_z=e^{2\pi i z}$. It is important to write every component as genuine Jacobi forms, where the following  easy formulas are useful: 
\be
\mathrm{PE}\bigg(\Big(Q_z+\frac{1}{Q_z}\Big)\frac{q^{\frac12}}{1-q}\bigg)=\frac{q^{-\frac{1}{24}}\eta(\tau)}{\theta_4(\tau,z)},
\ee
\be
\mathrm{PE}\bigg(\Big(Q_z+\frac{1}{Q_z}\Big)\frac{q^{\frac13}+q^{\frac23}}{1-q}\bigg)=-\frac{q^{-\frac{1}{6}}\eta(\tau)^2}{\theta_1(\tau,\frac{\tau}{3}+z)\theta_1(\tau,\frac{\tau}{3}-z)}=-\frac{q^{-\frac{1}{18}}\eta(\tau)^2}{\theta_4^{[\frac{5}{6}]}(\tau,z)\theta_4^{[\frac{5}{6}]}(\tau,-z)},
\ee
\be
\mathrm{PE}\bigg(\Big(Q_z+\frac{1}{Q_z}\Big)\frac{q^{\frac14}+q^{\frac34}}{1-q}\bigg)=-\frac{q^{-\frac{1}{12}}\eta(\tau)^2}{\theta_1(\tau,\frac{\tau}{4}+z)\theta_1(\tau,\frac{\tau}{4}-z)}=-\frac{q^{-\frac{1}{48}}\eta(\tau)^2}{\theta_4^{[\frac{3}{4}]}(\tau,z)\theta_4^{[\frac{3}{4}]}(\tau,-z)}.
\ee
The three formulas are used for the twist coefficient $N=2,3,4$ respectively. For example, for $N=2$ theories, we can simply define $\breve{\theta}_V^{[k]}$ by replacing all $\theta_1$ function to $\theta_4$ in \eqref{eq:thetaV}.

The contribution of hypermultiplets of twisted 6d $(1,0)$ theory to the twisted elliptic blowup equations is defined similarly. We have 
\begin{align}
  A^{\cirR}_{{H}}(\tau,\mG,\mF,\lG,\lF)
  &=
    \prod_{\w\in \cirR^{+}}
    \breve{\theta}_H^{[k']}(w_{\cirG}\cdot\mG+w_{\cirF}\cdot\mF, w_{\cirG}\cdot\lG+w_{\cirF}\cdot\lF)  .
    \label{eq:AH}
\end{align}
Here $\cirR^+$ is half of the total weight space of the twisted matter representations. The $k'$ is the KK-charge of each component in the twisted matter content. The $k'$ can be $0,1/2,1/3,2/3,1/4,3/4$. For unity blowup equations, the half weight space can be taken as 
\begin{equation}\label{eq:uR+1}
  \cirR^+ = \{w_{\cirG}\in R_{\cirG},w_{\cirF}\in R_{\cirF}\,|\, w_{\cirF}\cdot \lF = +{1}/{2}\},
\end{equation}
and for vanishing blowup equations as
\begin{equation}\label{eq:uR+}
  \cirR^+ = \{w_{\cirG}\in R_{\cirG},w_{\cirF}\in R_{\cirF}\,|\, w_{\cirG}\cdot\lG+w_{\cirF}\cdot \lF >0\}.
\end{equation}
The $\breve{\theta}_H^{[k']}$ functions are defined as
\begin{align}
  &\breve{\theta}^{[k']}_H(z,L) :=
    \prod_{\substack{m,n\geq 0\\m+n\leq |L|-3/2}}
  \frac{\theta_1^{[k']}(z+s(m+1/2)\eq+s(n+1/2)\et)}{\eta} \ ,
  \quad L\in
  \frac{1}{2}+\IZ .\label{eq:thetaH*}
\end{align}

\section{Modular forms and Jacobi forms}\label{app:B}
The ring of $SL(2,\IZ)$ holomorphic modular forms of even weights is generated by the Eisenstein series $E_4(\tau)$ and $E_6(\tau)$. 
The $SL(2,\IZ)$ weak Jacobi forms of even weights are generated by $E_4(\tau)$, $E_6(\tau)$ and the Eichler-Zagier generators  \cite{eichler1985theory}:
\be
\phi_{-2,1}(\tau,\epsilon) = -\frac{\theta_1(\tau,\epsilon)^2}{\eta(\tau)^6}\quad \text{\emph{and}}\quad
\phi_{0,1}(\tau,\epsilon) = 4\left[\frac{\theta_2(\tau,\epsilon)^2}{\theta_2(\tau)^2}+\frac{\theta_3(\tau,\epsilon)^2}{\theta_3(\tau)^2}+\frac{\theta_4(\tau,\epsilon)^2}{\theta_4(\tau)^2}\right].
\ee
These two weak Jacobi forms have index 1 and weight $-2$ and $0$ respectively. They are frequently used in the modular bootstrap of $SL(2,\IZ)$. We will also use $\phi_{-1,1/2}(\tau,\epsilon):=\phi_{-2,1}(\tau,\epsilon)^{1/2}$ for some occasions.

The principal congruence subgroup of level $N$ in $SL(2,\IZ)$ is defined by
\be
\Gamma(N) =  \left \{ \left({_a \ _b \atop ^c \ ^d}\right) \in SL(2,\IZ) \big| \, a \equiv d \equiv 1\! \pmod{N}, ~b \equiv c \equiv 0\! \pmod{N} \right \} \, .
\ee
The Hecke congruence subgroup of level $N$ is defined by
\be
\Gamma_0(N) =  \left \{ \left({_a \ _b \atop ^c \ ^d}\right) \in SL(2,\IZ) \big| \,  c \equiv 0\! \pmod{N} \right \} \, .
\ee
Another useful congruence subgroup of level $N$ is defined by 
\be
\Gamma_1(N) =  \left \{ \left({_a \ _b \atop ^c \ ^d}\right) \in SL(2,\IZ) \big| \, a \equiv d \equiv 1\! \pmod{N}, ~ c \equiv 0\! \pmod{N} \right \} \, .
\ee
Clearly $\Gamma(N)\subset \Gamma_1(N)\subset \Gamma_0(N)$. Therefore, there are less modular forms of $\Gamma_0(N)$ than those of $\Gamma_1(N)$.
Define weight-two modular forms of level $N$ by
\begin{align}
	E_2^{(N)}(\tau)=-\frac{1}{N-1}\partial_\tau\log\left(\frac{\eta(\tau)}{\eta(N\tau)}\right).
	\label{eqn:gene2}
\end{align}
Then the ring $M_*(N)$ of even-weight modular forms for $\Gamma_0(N)$, $N\in\{2,3,4\}$ is finitely generated by
\begin{align}\label{eq:gammaNgens2}
	M_*(2)&=\langle E_2^{(2)}(\tau),\,E_4(2\tau)\rangle,\\ \label{eq:gammaNgens3}
	M_*(3)&=\langle E_2^{(3)}(\tau),\,E_4(3\tau),\,E_6(3\tau)\rangle,\\
	M_*(4)&=\langle E_2^{(2)}(\tau),\,E_2^{(4)}(\tau),\,E_4(4\tau),\,E_6(4\tau)\rangle.
	\label{eq:gammaNgens4}
\end{align}
The subscripts are the weights of the modular forms.  
Note for $N=3,4$, the generators are not algebraically independent. For example, for $N=3$, it is easy to find
\be\label{relationgen3}
0=E_2^{(3)}(\tau)^4 - 6 E_2^{(3)}(\tau)^2 E_4(3\tau) - 3 E_4(3\tau)^2 + 8 E_2^{(3)}(\tau) E_6(3\tau).
\ee
The rings of modular forms for $\Gamma_1(N)$ usually have more generators. In the case of $N=2$, the two rings are isomorphic. Luckily, for almost all numerators of the modular ansatz in the current work, we only need to use the  $\Gamma_0(N)$ generators. Only in the case of $\fn=1,\so(8)^{(3)}$, we utilize $\Gamma_1(3)$. The ring of $\Gamma_1(3)$ has four generators: $E_2^{(3)}(\tau)$, $E_4(3\tau)$ and two odd ones of weight 3 which are
\be
\ba
E_3^{(b)}(\tau)&= \frac{\eta(3\tau)^9}{\eta(\tau)^3}=q+3 q^2+9 q^3+13 q^4+24 q^5+27 q^6+\dots,\\
E_3^{(a)}(\tau)& = E_1(\tau)^3-18E_3^{(b)}(\tau)=1 + 54q^2 + 72q^3 +\dots .
\ea
\ee
where
\be
E_1(\tau)=\Theta_{A_2}(\tau)=\sum_{n_1,n_2\in\IZ}q^{n_1^2-n_1n_2+n_2^2}=1+6 q+6 q^3+6 q^4+\dots.
\ee
One can also find some algebraic relations like $E_2^{(3)}=E_1^2$ and
\be\label{relationgen3a}
\ba
0&= E_2^{(3)}(\tau)^3-4 E_3^{(a)}(\tau)^2+3 E_4(3\tau) E_2^{(3)}(\tau)-72 E_3^{(a)}(\tau) E_3^{(b)}(\tau)  ,\\
0&=  2 E_3^{(a)}(\tau)^2+E_2^{(3)}(\tau)^3-3 E_4(3\tau) E_2^{(3)}(\tau)-648 E_3^{(b)}(\tau)^2 .
\ea
\ee
For $N=4$, the two generators $E_2^{(2)}(\tau),\,E_2^{(4)}(\tau)$ are sufficient for our purpose.  

For $N>1$, we define the weight-$2N$ cusp forms of $\Gamma_1(N)$ by
\begin{align}
    \Delta_{2N}(\tau)=\frac{1}{q\phi_{-2,1}^N(N\tau,\tau)}=q^{N-1}+\mathcal{O}(q^N).
\end{align}
For $N=2,3,4$, $\Delta_{2N}(\tau)$ are also cusp forms of $\Gamma_0(N)$, they take the form from eta functions or the modular generators as
\begin{align}
	\Delta_{4}(\tau)&=\frac{\eta(2\tau)^{16}}{\eta(\tau)^8}=\frac{1}{192}\left(E_4-(E_2^{(2)})^2\right),\\
\Delta_6(\tau)&=\frac{\eta(3\tau)^{18}}{\eta(\tau)^6}
		=\frac{1}{2^4\cdot 3^6}\left(7(E_2^{(3)})^3-5E_2^{(3)}E_4-2E_6\right),\\
		\Delta_8(\tau)&=\frac{\eta(2\tau)^8\eta(4\tau)^{16}}{\eta(\tau)^8}\\ \nonumber
		&=\frac{1}{2^{17}\cdot 3^2\cdot 17}\left(187 (E_2^{(2)})^4 - 144 (E_2^{(4)})^4- 33 E_4^2 -E_6( 154 E_2^{(2)} - 144 E_2^{(4)} )\right).
\end{align}
These $\Delta_{2N}$ cusp forms are used in the modular ansatz for the twisted elliptic genera.

\section{More on $\su(2r+1)^{(2)}$ theories} \label{app:C}
In Section \ref{sec:n2su2r1}  we have discussed the $\IZ_2$ twist of 6d $(1,0)$ $\su(2r+1)+2(2r+1)\bf F$ theories where we choose the twisted matter content  as $(2r+1)(\mathbf{F}_0+\mathbf{F}_{1/2})$. Here we choose a different assignment of the KK-charges of the twisted matter content as $(2r+1)(\mathbf{F}_{1/4}+\mathbf{F}_{3/4})$. We notice this choice results in some quite interesting twisted elliptic genera, thus we record our results here. This choice also makes the tensor parameter has integral $B$ field. 
With this matter content, the circle reduction gives 5d $\mathcal{N}=1$ pure $\Sp(r)$ gauge theories. Interestingly, we find there exist two possible solutions for the twisted elliptic blowup equations by choosing different characteristics $a$.  The two solutions have circle reduction as 5d  pure $\Sp(r)_0$ and $\Sp(r)_\pi$ theories respectively. 
Since the matter content is still invariant under the KK-charge shift by $1/2$, we can expect some spectral flow symmetry. 
Most interestingly, we find the two twisted elliptic genera are dual to each other under spectral flow transformation! Besides, both twisted elliptic genera allow perfect modular ansatz. We have fixed the modular ansatz for both elliptic genera for $r=1,2,3$ against the results from blowup equations and find perfect agreements. We summarize the relevant data of the modular ansatz in Table \ref{tb:mainMA1}. In the following we show some detailed results for $r=1,2$.

\paragraph{\underline{$\fn=2,\su(3)^{(2)}$}}
With twisted matter content $3(\mathbf{F}_{1/4}+\mathbf{F}_{3/4})$, we find there are two possible low energy gauge groups $\Sp(1)_0$ or $\Sp(1)_{\pi}$. For each case, we solve the twisted one-string elliptic genus from twisted unity elliptic blowup equations to $q^4$ order, and collect the Fourier coefficients in $(q,v)$ expansion in Tables \ref{tab:bijsu3z2new} and \ref{tab:bijsu3z2new2}. For $\theta=0$ in Table \ref{tab:bijsu3z2new}, we recognize that the circle reduction of the twisted R-R sector colored orange gives exactly the one $\Sp(1)_0$ instanton Hilbert series, while the circle reduction of the twisted NS-R sector colored red gives exactly the one $\Sp(1)_{\pi}$ instanton Hilbert series. For $\theta=\pi$ in Table \ref{tab:bijsu3z2new2}, it is exactly the opposite. It is interesting to see that the two elliptic genera are dual to each other under spectral flow!

\begin{table}[h]
\begin{center}
$\begin{tabu}{c|cccccccccc}
q\backslash v& & 1 & & 2 & &3 & & 4 & & 5  \\ \hline
0& \ 0 & \color{orange}1 & 0 & 0 & 0 & \color{orange}3 & 0 & 0 & 0 & \color{orange}5 \\
1/4& \ 0 & 0 & 2 & 0 & -12 & 0 & 6 & 0 & -24 & 0 \\
1/2& \ 0 & -6 & 0 & 18 & 0 & -24 & 0 & 54 & 0 & -48 \\
3/4& \ \color{red}2 & 0 & -12 & 0 & 38 & 0 & -88 & 0 & 108 & 0 \\
1& \ 0 & 4 & 0 & -44 & 0 & 113 & 0 & -176 & 0 & 327 \\
5/4& \ 0 & 0 & 36 & 0 & -100 & 0 & 250 & 0 & -516 & 0 \\
3/2& \ 0 & -18 & 0 & 72 & 0 & -282 & 0 & 626 & 0 & -1002 \\
7/4& \ \color{red}4 & 0 & -36 & 0 & 218 & 0 & -580 & 0 & 1254 & 0 \\
2& \ 0 & 8 & 0 & -132 & 0 & 431 & 0 & -1300 & 0 & 2669 \\
9/4& \ 0 & 0 & 72 & 0 & -264 & 0 & 1114 & 0 & -2588 & 0 \\
\end{tabu}
$
\caption{The coefficients $b_{ij}$ for the  $\IE_1$ of 6d $\su(3)^{(2)}+3(\mathbf{F}_{1/4}+\mathbf{F}_{3/4})$ theory with $\theta=0$.}
\label{tab:bijsu3z2new}
\end{center}
\end{table}

\begin{table}[h]
\begin{center}
$\begin{tabu}{c|cccccccccc}
q\backslash v& & 1& & 2 & &3 & & 4 & & 5   \\ \hline
0& \ 0 & 0 & 0 & \color{orange}2 & 0 & 0 & 0 & \color{orange}4 & 0 & 0 \\
1/4& \ \color{red}1 & 0 & -6 & 0 & 4 & 0 & -18 & 0 & 8 & 0 \\
1/2& \ 0 & 2 & 0 & -12 & 0 & 36 & 0 & -36 & 0 & 72 \\
3/4& \ 0 & 0 & 18 & 0 & -44 & 0 & 72 & 0 & -132 & 0 \\
1& \ 0 & -12 & 0 & 38 & 0 & -100 & 0 & 218 & 0 & -264 \\
5/4& \ \color{red}3 & 0 & -24 & 0 & 113 & 0 & -282 & 0 & 431 & 0 \\
3/2& \ 0 & 6 & 0 & -88 & 0 & 250 & 0 & -580 & 0 & 1114 \\
7/4& \ 0 & 0 & 54 & 0 & -176 & 0 & 626 & 0 & -1300 & 0 \\
2& \ 0 & -24 & 0 & 108 & 0 & -516 & 0 & 1254 & 0 & -2588 \\
9/4& \ \color{red}5 & 0 & -48 & 0 & 327 & 0 & -1002 & 0 & 2669 & 0 \\
\end{tabu}
$
\caption{The coefficients $b_{ij}$ for the  $\IE_1$ of 6d $\su(3)^{(2)}+3(\mathbf{F}_{1/4}+\mathbf{F}_{3/4})$ theory with $\theta=\pi$.}
\label{tab:bijsu3z2new2}
\end{center}
\end{table}

\paragraph{\underline{$\fn=2,\su(5)^{(2)}$}}
With twisted matter content $5(\mathbf{F}_{1/4}+\mathbf{F}_{3/4})$, we find there are two possible low energy gauge groups $\Sp(2)_0$ or $\Sp(2)_{\pi}$. For each case, we solve the twisted one-string elliptic genus from twisted unity elliptic blowup equations to $q^4$ order, and collect the Fourier coefficients in $(q,v)$ expansion in Tables \ref{tab:bijsu5z2new} and \ref{tab:bijsu5z2new2}. For $\theta=0$ in Table \ref{tab:bijsu5z2new}, we recognize that the circle reduction of the twisted R-R sector colored orange gives exactly the one $\Sp(2)_0$ instanton Hilbert series, while the circle reduction of the twisted NS-R sector colored red gives exactly the one $\Sp(2)_{\pi}$ instanton Hilbert series. For $\theta=\pi$ in Table \ref{tab:bijsu5z2new2}, the situation is exactly the opposite. Again we observe that the two elliptic genera are dual to each other under spectral flow. 

\begin{table}[h]
\begin{center}
$\begin{tabu}{c|cccccccccccc}
q\backslash v& & 2 & &3 & & 4 & & 5 & & 6  \\ \hline
0&\ 0 & \color{orange}1 & 0 & 0 & 0 &\color{orange} 10 & 0 & 0 & 0 & \color{orange}35  \\
1/4&\ 0 & 0 & 4 & 0 & -40 & 0 & 24 & 0 & -200 & 0   \\
1/2&\ 0 & -10 & 0 & 60 & 0 & -110 & 0 & 536 & 0 & -450  \\
3/4&\ \color{red}4 & 0 & -40 & 0 & 224 & 0 & -920 & 0 & 1280 & 0   \\
1&\ 0 & 11 & 0 & -280 & 0 & 1112 & 0 & -2440 & 0 & 7536   \\
5/4&\ 0 & 0 & 220 & 0 & -960 & 0 & 3528 & 0 & -10928 & 0   \\
3/2&\ 0 & -100 & 0 & 596 & 0 & -3932 & 0 & 12656 & 0 & -27432 \\
7/4&\ \color{red}20 & 0 & -240 & 0 & 3328 & 0 & -11648 & 0 & 35348 & 0  \\
2&\ 0 & 45 & 0 & -2160 & 0 & 8602 & 0 & -37680 & 0 & 105345  \\
9/4&\ 0 & 0 & 1060 & 0 & -5120 & 0 & 33104 & 0 & -100168 & 0 \\
%5/2& 0 & -350 & 0 & 2380 & 0 & -23960 & 0 & 80712 & 0 & -279644 & 0 & 709700 \\
%11/4& 56 & 0 & -760 & 0 & 14400 & 0 & -55168 & 0 & 250860 & 0 & -685816 & 0 \\
\end{tabu}
$
\caption{The coefficients $b_{ij}$ for the  $\IE_1$ of 6d $\su(5)^{(2)}+5(\mathbf{F}_{1/4}+\mathbf{F}_{3/4})$ theory with $\theta=0$.}
\label{tab:bijsu5z2new}
\end{center}
\end{table}

\begin{table}[h]
\begin{center}
$\begin{tabu}{c|cccccccccccc}
q\backslash v&  & 2 & &3 & & 4 & & 5 & & 6   \\ \hline
0&\ 0 & 0 & 0 & \color{orange}4 & 0 & 0 & 0 & \color{orange}20 & 0 & 0   \\
1/4&\ \color{red}1 & 0 & -10 & 0 & 11 & 0 & -100 & 0 & 45 & 0   \\
1/2&\ 0 & 4 & 0 & -40 & 0 & 220 & 0 & -240 & 0 & 1060   \\
3/4&\ 0 & 0 & 60 & 0 & -280 & 0 & 596 & 0 & -2160 & 0   \\
1& 0\ & -40 & 0 & 224 & 0 & -960 & 0 & 3328 & 0 & -5120   \\
5/4&\ \color{red}10 & 0 & -110 & 0 & 1112 & 0 & -3932 & 0 & 8602 & 0   \\
3/2&\ 0 & 24 & 0 & -920 & 0 & 3528 & 0 & -11648 & 0 & 33104  \\
7/4&\ 0 & 0 & 536 & 0 & -2440 & 0 & 12656 & 0 & -37680 & 0   \\
2&\ 0 & -200 & 0 & 1280 & 0 & -10928 & 0 & 35348 & 0 & -100168  \\
9/4&\ \color{red}35 & 0 & -450 & 0 & 7536 & 0 & -27432 & 0 & 105345 & 0   \\
%5/2& 0 & 76 & 0 & -4200 & 0 & 17648 & 0 & -93680 & 0 & 268360 & 0 & -685816 \\
%11/4& 0 & 0 & 1844 & 0 & -9400 & 0 & 70496 & 0 & -222272 & 0 & 709700 & 0 \\
\end{tabu}
$
\caption{The coefficients $b_{ij}$ for the  $\IE_1$ of 6d $\su(5)^{(2)}+5(\mathbf{F}_{1/4}+\mathbf{F}_{3/4})$ theory with $\theta=\pi$.}
\label{tab:bijsu5z2new2}
\end{center}
\end{table}

\section{Cases with unpaired matter content}\label{app:D}
We have focused on the twisted compactification of 6d $(1,0)$ SCFTs with \emph{paired} matter content in the main context. The unpaired cases are trickier and have been  discussed in \cite{Bhardwaj:2020kim} including the Calabi-Yau geometries and 5d low energy limits. The twist in this situation always involves zero mass of some hypermulitplet, thus there should exist no unity elliptic blowup equation or recursion formula for the twisted elliptic genera.  However, we notice that sometimes it is possible to pretend the hyper has a free mass for blowup equations and then turn the mass off after solving the twisted elliptic genera. We find that for some twisted theories, this procedure allows consistent solutions of twisted elliptic genera. Remarkably, the solutions have perfect modular ansatz, spectral flow symmetry and required behavior in the $(q,v)$ expansion. It is intriguing to consider whether the solutions are indeed physical. We give one example in the following.

Consider the $\IZ_2$ twist of $E_6+\bf F$ theory with $\fn=5$. It was determined in \cite{Bhardwaj:2020kim} that the 5d low energy limit should be a pure $F_4$ theory. In other words, the twisted matter context should be $\mathring{R}={\bf 1}_0+\mathbf{26}_{1/2}$. Let us pretend the fundamental hyper $\mathbf{26}_{1/2}$ of $\cirG=F_4$ has a free mass $m$. This mass can have a shift in blowup equations such that there exist unity twisted elliptic blowup equations. Then from the recursion formula \eqref{Z1m}, we compute the twisted one-string elliptic genus to $q$ order $5$. Let us denote 
\be
\IE_1(q,v)=q^{-\frac{4}{3}}\sum_{i,j=0}^{\infty}b_{ij}q^{{i}} v^{-2i+j}.
\ee
We obtain the  coefficients $b_{ij}$ in Table \ref{tab:bijn5E6z2a}. We recognize the leading order of R-R elliptic genus colored orange gives exactly the one-instanton partition function of 5d pure $F_4$ theory, while the leading order of NS-R elliptic genus colored red gives exactly the one-instanton partition function of 5d $F_4+\bf F$ theory.
\begin{table}[h]
\begin{center}
$\begin{tabu}{c|ccccccccc}
q\backslash v  &5 &6 &7 & 8 & 9 & 10 & 11 & 12 & 13\\ \hline
0& 0 & 0 & 0 & \color{orange}1 & 0 & \color{orange}52 & 0 & \color{orange}1053 & 0 \\
1/2& 0 & 0 & 0 & 0 & 26 & -52 & 1079 & -2106 & 18954 \\
1& 0 & 0 & 0 & 0 & -2 & 381 & -1300 & 15209 & -41910 \\
3/2& 0 & 0 & 0 & 0 & -3 & -46 & 4235 & -18670 & 164753 \\
2& 0 & 0 & 0 & 0 & -2 & -73 & -610 & 38768 & -201290 \\
5/2& \color{red}1 & -2 & 1 & 0 & -1 & 8 & -1100 & -6134 & 308968 \\
3& \color{red}4 & 23 & -50 & 22 & 2 & 4 & 578 & -12778 & -50036 \\
7/2& \color{red}-78 & 154 & 307 & -766 & 441 & -112 & -167 & 10866 & -122955 \\
4& \color{red}754 & -2197 & 2610 & 3113 & -8788 & 6143 & -2582 & -4314 & 131968 \\
9/2& \color{red}-4433 & 17214 & -33127 & 31736 & 25627 & -82494 & 68509 & -36674 & -65257 \\
\end{tabu}$
\caption{The coefficient matrix $b_{ij}$ for the $\IE_1$ of 6d $E_6^{(2)}$ theory with matter ${\bf 1}_0+\mathbf{26}_{1/2}$.}
\label{tab:bijn5E6z2a}
\end{center}
\end{table}

\begin{table}[h]
\begin{center}
$\begin{tabu}{c|cccccccccc}
q\backslash v  &2 &3  &4 &5 &6 &7 & 8 & 9 & 10 & 11 \\ \hline
0& 0 & 0 & 0 & 0 & 0 & \color{orange}1 & \color{orange}4 & \color{orange}-78 & \color{orange}754 & \color{orange}-4433 \\
1/2& 0 & 0 & 0 & 0 & 0 & -2 & 23 & 154 & -2197 & 17214 \\
1& 0 & 0 & 0 & 0 & 0 & 1 & -50 & 307 & 2610 & -33127 \\
3/2& \color{red}1 & 0 & 0 & 0 & 0 & 0 & 22 & -766 & 3113 & 31736 \\
2& 0 & 26 & -2 & -3 & -2 & -1 & 2 & 441 & -8788 & 25627 \\
5/2& \color{red}52 & -52 & 381 & -46 & -73 & 8 & 4 & -112 & 6143 & -82494 \\
3& 0 & 1079 & -1300 & 4235 & -610 & -1100 & 578 & -167 & -2582 & 68509 \\
7/2& \color{red}1053 & -2106 & 15209 & -18670 & 38768 & -6134 & -12778 & 10866 & -4314 & -36674 \\
4& 0 & 18954 & -41910 & 164753 & -201290 & 308968 & -50036 & -122955 & 131968 & -65257 \\
\end{tabu}$
\caption{The coefficient matrix $b_{ij}$ for the $\IE_1$ of 6d $E_6^{(2)}$ theory with matter ${\bf 26}_0+\mathbf{1}_{1/2}$.}
\label{tab:bijn5E6z2b}
\end{center}
\end{table}

Interestingly, we find that even for twisted matter content
{$\mathring{R}={\bf 26}_0+\mathbf{1}_{1/2}$}, elliptic blowup equations still allow perfectly reasonable solution as long as we pretend the hyper has free mass. From the recursion formula \eqref{Z1m}, we also compute the twisted one-string elliptic genus of this case to $q$ order $5$. Let us denote 
\be
\IE_1(q,v)=q^{-\frac{7}{12}}\sum_{i,j=0}^{\infty}b_{ij}q^{{i}} v^{-2i+j}.
\ee
We obtain the coefficients $b_{ij}$ in Table \ref{tab:bijn5E6z2b}. 
We recognize the leading order of R-R elliptic genus colored orange gives the one-instanton partition function of 5d $F_4+\bf F$ theory, while the leading order of NS-R elliptic genus colored red gives the one-instanton partition function of 5d pure $F_4$ theory. In fact, we find the twisted elliptic genera of the two cases are exactly spectral dual to each other. We also compute the twisted elliptic genera for both matter contents with the mass parameter turned on to $q$ order $5$ and find the spectral flow relation still holds perfectly. Moreover, we determine the modular ansatz  for both cases and find complete consistency with the results from the recursion formula \eqref{Z1}. We have summarized the relevant data of the two modular ansatz in Table \ref{tb:mainMA1}.

\bibliographystyle{JHEP}     {\small{\bibliography{main}}}

\end{document}